%% file: main.tex
\documentclass[10pt,aps,prl,twocolumn,superscriptaddress,nofootinbib,tightenlines,floatfix,titlepage]{revtex4-2}

\usepackage{bm}
\usepackage{graphicx}
\usepackage{rotating}
\usepackage{float}
\usepackage[pdftex]{color}
\usepackage{amsmath, amssymb}
\usepackage{array}
\usepackage{rotating}
\usepackage{mathtools}
\usepackage{multirow}
\usepackage{changepage}
\usepackage[normalem]{ulem}
\usepackage{verbatim}
\usepackage{physics}
\usepackage{tabularx}
\usepackage[ragged]{sidecap}
\usepackage{placeins}

\usepackage[sort&compress]{natbib}

\usepackage[hyperfootnotes=false]{hyperref}
\hypersetup{
    colorlinks=true,     
    linkcolor=blue,      
    citecolor=blue,      
    filecolor=blue,      
    urlcolor=blue,        
    linktoc=page
}

\usepackage[capitalise]{cleveref}

\crefname{section}{Sec.}{Secs.}
\Crefname{section}{Section}{Sections}
\crefrangeformat{equation}{Eqs.~(#3#1#4)--(#5#2#6)}
\Crefrangeformat{equation}{Equations~(#3#1#4)--(#5#2#6)}
\crefrangeformat{figure}{Figs.~#3#1#4--#5#2#6}
\Crefrangeformat{figure}{Figures~#3#1#4--#5#2#6}

\newcolumntype{C}{>{\centering\arraybackslash}X}
\newcolumntype{L}{>{\raggedright\arraybackslash}X}
\newcolumntype{R}{>{\raggedleft\arraybackslash}X}

\newcommand{\panel}[1]{\textit{{#1}}}

\newcommand{\bi}{\begin{itemize}}
\newcommand{\ei}{\end{itemize}}

\newcommand{\ben}{\begin{enumerate}}
\newcommand{\een}{\end{enumerate}} 

\newcommand{\be}{\begin{equation}}
\newcommand{\ee}{\end{equation}}

\newcommand{\bea}{\begin{eqnarray}}
\newcommand{\eea}{\end{eqnarray}}

\newcommand{\nn}{\nonumber}

\newcommand{\chpt}{$\chi$PT}

\newcommand{\cov}{\operatorname{Cov}}

\newcommand{\amu}{a_\mu}
\newcommand{\amuHVP}{a_\mu^{\mathrm{HVP,LO}}}

\newcommand{\amuLFull}{a^{ll, \mathrm{Full}}_{\mu}(\mathrm{conn.})}

\newcommand{\amuL}{a^{ll}_{\mu}(\mathrm{conn.})}

\newcommand{\amuLW}{a^{ll,\,{\mathrm W}}_{\mu}(\mathrm{conn.})}

\newcommand{\amuLSD}{a^{ll,\,{\mathrm {SD}}}_{\mu}(\mathrm{conn.})}

\newcommand{\amuLLD}{a^{ll,\,{\mathrm {LD}}}_{\mu}(\mathrm{conn.})}

\newcommand{\amuLSDRes}{a^{ll,\,{\mathrm{SD}}}_{\mu}(\mathrm{conn.}) &= 48.139(11)(91)[92]\times 10^{-10}}

\newcommand{\amuLWRes}{a^{ll,\,{\mathrm{W}}}_{\mu}(\mathrm{conn.}) &= 206.90(14)(62)[63] \times 10^{-10}}

\newcommand{\amuLLDRes}{a^{ll,\,{\mathrm{LD}}}_{\mu}(\mathrm{conn.}) = 400.2(2.3)(3.7)[4.3] \times 10^{-10}}

\newcommand{\amuLLDResABS}{a^{ll,\,{\mathrm{LD}}}_{\mu}(\mathrm{conn.}) = 400.2(2.3)_{\mathrm{stat}}(3.7)_{\mathrm{syst}}[4.3]_{\mathrm{total}} \times 10^{-10}}

\newcommand{\amuLFullRes}{a^{ll,\,{\mathrm{Full}}}_{\mu}(\mathrm{conn.}) = 654.2(2.4)(4.7)[5.3] \times 10^{-10}}

\newcommand{\amuLFullResFPI}{a^{ll,\,{\mathrm{Full}}}_{\mu}(\mathrm{conn.}) = 650.5(2.4)(4.4)[5.0] \times 10^{-10}}

\newcommand{\amuLRes}{a^{ll}_{\mu}(\mathrm{conn.}) = 655.2(2.3)(3.9)[4.5] \times 10^{-10}}

\newcommand{\amuLResABS}{a^{ll}_{\mu}(\mathrm{conn.}) = 655.2(2.3)_{\mathrm{stat}}(3.9)_{\mathrm{syst}}[4.5]_{\mathrm{total}} \times 10^{-10}}

\newcommand{\gmtwo}{$g-2$} 

\newcommand{\pr}{\operatorname{pr}}

\newcommand{\coloaf}{Department of Physics, University of Colorado, Boulder, Colorado 80309, USA}
\newcommand{\fnalaf}{Theory Division, Fermi National Accelerator Laboratory, Batavia, Illinois, 60510, USA}
\newcommand{\iuaf}{Department of Physics, Indiana University, Bloomington, Indiana 47405, USA}

\newcommand{\msuaf}{Department of Computational Mathematics, Science and Engineering, and Department of Physics and Astronomy, Michigan State University, East Lansing, Michigan 48824, USA}
\newcommand{\ugraf}{CAFPE and Departamento de Física Teórica y del Cosmos, Universidad de Granada, \\E-18071 Granada, Spain}
\newcommand{\uiucaf}{Department of Physics, University of Illinois Urbana-Champaign, Urbana, Illinois, 61801, USA}
\newcommand{\icasuuiaf}{Illinois Center for Advanced Studies of the Universe, University of Illinois Urbana-Champaign, \\ Urbana, Illinois, 61801, USA}
\newcommand{\unizar}{Departmento de Física Teórica, Universidad de Zaragoza, 50009 Zaragoza, Spain}
\newcommand{\utahaf}{Department of Physics and Astronomy, University of Utah, Salt Lake City, Utah 84112, USA}
\newcommand{\glasaf}{SUPA, School of Physics and Astronomy, University of Glasgow, \\ Glasgow G12 8QQ, United Kingdom}
\newcommand{\cornaf}{Laboratory for Elementary-Particle Physics, Cornell University, Ithaca, New York 14853, USA}
\newcommand{\plyaf}{Centre for Mathematical Sciences, University of Plymouth, Plymouth PL4 8AA, United Kingdom}
\newcommand{\syracuseaf}{Department of Physics, Syracuse University, Syracuse, New York 13244, USA}
\newcommand{\csuaf}{Department of Physics, Colorado State University, Fort Collins, Colorado 80523, USA}
\newcommand{\washuaf}{Department of Physics, Washington University, St.~Louis, Missouri 63130, USA}
\newcommand{\capa}{Center for Astroparticles and High Energy Physics (CAPA), Calle Pedro Cerbuna 12, 50009 Zaragoza, Spain}

\definecolor{red}{rgb}{0.8,0.0,0.0}
\definecolor{green}{rgb}{0.0,0.6,0.0}
\definecolor{darkblue}{rgb}{0.0,0.1,0.7}
\definecolor{brown}{rgb}{0.6,0.1,0.0}
\definecolor{gray}{rgb}{0.6,0.6,0.6}
\definecolor{darkgreen}{rgb}{0.0, 0.545098, 0.0}
\definecolor{purple}{rgb}{0.5,0.0,0.5}
\definecolor{applegreen}{rgb}{0.55, 0.71, 0.0}
\definecolor{babypink} {rgb}{0.64, 0.44, 0.44}
\definecolor{orange}{rgb}{0.9,0.4,0.0}
\definecolor{skyblue}{rgb}{0.53,0.81,0.92}
\definecolor{teal}{rgb}{0.0, 0.52, 0.52}

\begin{document}

\preprint{FERMILAB-PUB-24-0957-T}

\title{Hadronic vacuum polarization for the muon \texorpdfstring{\boldmath\gmtwo}{g-2} from lattice QCD: long-distance and full light-quark connected contribution}

\author{Alexei~Bazavov}\affiliation{\msuaf}
\author{Claude~W.~Bernard}\affiliation{\washuaf}
\author{David~A.~Clarke}\affiliation{\utahaf}
\author{Christine Davies}\affiliation{\glasaf}
\author{Carleton~DeTar}\affiliation{\utahaf}
\author{Aida~X.~El-Khadra}\affiliation{\uiucaf}\affiliation{\icasuuiaf}
\author{Elvira~G\'amiz}\affiliation{\ugraf}
\author{Steven~Gottlieb}\affiliation{\iuaf}
\author{Anthony~V.~Grebe}\affiliation{\fnalaf}
\author{Leon~Hostetler}\affiliation{\iuaf}
\author{William~I.~Jay}\affiliation{\csuaf}
\author{Hwancheol~Jeong}\affiliation{\iuaf}
\author{Andreas~S.~Kronfeld}\affiliation{\fnalaf}
\author{Shaun~Lahert}\email{shaun.lahert@gmail.com}\affiliation{\utahaf}
\author{Jack~Laiho}\affiliation{\syracuseaf}
\author{G.~Peter~Lepage}\affiliation{\cornaf}
\author{Michael~Lynch}\email{ml11@illinois.edu}\affiliation{\uiucaf}\affiliation{\icasuuiaf}
\author{Andrew~T.~Lytle}\affiliation{\uiucaf}\affiliation{\icasuuiaf}
\author{Craig~McNeile}\affiliation{\plyaf}
\author{Ethan~T.~Neil}\affiliation{\coloaf}
\author{Curtis~T.~Peterson}\affiliation{\msuaf}
\author{James~N.~Simone}\affiliation{\fnalaf}
\author{Jacob~W.~Sitison}\email{jacob.sitison@colorado.edu}\affiliation{\coloaf}
\author{Ruth~S.~\surname{Van~de~Water}}\affiliation{\fnalaf}
\author{Alejandro~Vaquero}\affiliation{\utahaf}\affiliation{\unizar}\affiliation{\capa}

\collaboration{Fermilab Lattice, HPQCD, and MILC Collaborations}
\noaffiliation

\date{\today}

\begin{abstract}

We present results for the dominant light-quark connected contribution to the long-distance window (LD) of the hadronic vacuum polarization contribution (HVP) to the muon $g-2$ from lattice quantum chromodynamics (QCD). Specifically, with a new determination of the lattice scale on MILC's physical-mass HISQ ensembles, using the $\Omega^-$ baryon mass, we obtain a result of $\amuLLDResABS$. Summing this result with our recent determinations of the light-quark connected contributions to the short- (SD) and intermediate-distance (W) windows, we obtain a sub-percent precision determination of the light-quark-connected contribution to HVP of $\amuLResABS$. Finally, as a consistency check, we verify that an independent analysis of the full contribution is in agreement with the sum of individual windows. We discuss our future plans for improvements of our HVP calculations to meet the target precision of the Fermilab $g-2$ experiment.

\end{abstract}

\maketitle

\raggedbottom
\allowdisplaybreaks

\paragraph{Introduction ---}
The muon's anomalous magnetic moment or ``muon $(g-2)$'' is one of the most sensitive probes of the Standard Model of particle physics, with both the experimental results \cite{Muong-2:2021ojo,Muong-2:2023cdq} and the Standard Model prediction \cite{Aoyama:2020ynm} reaching precisions well below one part per million. Tensions between theory and experiment in this quantity have long captured the attention of particle physicists since the results of the E821 experiment at Brookhaven \cite{Muong-2:2006rrc}. With forthcoming experimental results from Fermilab \cite{Muong-2:2023cdq} and J-PARC \cite{Abe:2019thb,E34webpage} expected to improve further the accuracy of the muon $(g-2)$ measurement, concurrent improvements in the Standard Model theoretical calculation are urgently needed.

Although they are a relatively small part of the total prediction, the contributions to the muon $(g-2)$ from hadronic physics are among the most challenging to calculate due to strong-coupling effects. The muon does not feel the strong force directly, so these contributions arise from corrections to either the photon two-point function (hadronic vacuum polarization, or HVP), or the photon four-point function (hadronic light-by-light scattering, or HLbL). As described in detail in Ref.~\cite{Aoyama:2020ynm}, both HVP and HLbL can be estimated either from dispersive theory relying on experimental measurements of hadronic processes or calculated \emph{ab initio} using lattice QCD.

Of these quantities, HLbL represents a smaller contribution to muon $(g-2)$ but is also more difficult to calculate precisely. While previous summary reports from the Muon $g-2$ Theory Initiative \cite{Aoyama:2020ynm,Colangelo:2022jxc} already found good agreement between dispersive evaluations
\cite{Melnikov:2003xd,Masjuan:2017tvw,Colangelo:2017fiz,Hoferichter:2018kwz,Gerardin:2019vio,Bijnens:2019ghy,Colangelo:2019uex,Pauk:2014rta,Danilkin:2016hnh,Jegerlehner:2017gek,Knecht:2018sci,Eichmann:2019bqf,Roig:2019reh,Leutgeb:2022lqw}
 and lattice QCD calculations \cite{Blum:2019ugy,Chao:2021tvp}, more recent lattice and dispersive results \cite{Bijnens:2020xnl,Ludtke:2020moa,Bijnens:2021jqo,Hoferichter:2020lap,Leutgeb:2021mpu,Zanke:2021wiq,Danilkin:2021icn,Colangelo:2021nkr,Cappiello:2021vzi,Bijnens:2022itw,Hoferichter:2023tgp,Ludtke:2023hvz,Colangelo:2023een,Hoferichter:2024fsj,Hoferichter:2024vbu,Hoferichter:2024bae,Bijnens:2024jgh,Holz:2022hwz,Holz:2024lom,Leutgeb:2024rfs,Colangelo:2024xfh,Estrada:2024rfi,Miramontes:2024fgo,Estrada:2024cfy,Asmussen:2022oql,Chao:2022xzg,Blum:2023vlm,Gerardin:2023naa,ExtendedTwistedMass:2023hin,Lin:2024khg,Fodor:2024jyn} are rapidly pushing this quantity towards the precision goals needed for the final muon $(g-2)$ experimental results.

For HVP, rapid progress on the lattice QCD side \cite{Borsanyi:2020mff,Aubin:2022hgm,Alexandrou:2022amy,Ce:2022kxy,FermilabLatticeHPQCD:2023jof,Blum:2023qou,Boccaletti:2024guq,Kuberski:2024bcj,RBC:2024fic,Spiegel:2024dec,Djukanovic:2024cmq,ExtendedTwistedMassCollaborationETMC:2024xdf,FermilabLattice:2024yho} has revealed tensions between lattice HVP results and data-driven dispersive estimates \cite{Colangelo:2022jxc,Benton:2024kwp}. Efforts to understand these tensions, as well as to cross-check systematic effects in lattice calculations from different groups, have led to widespread study of HVP restricted to Euclidean windows \cite{RBC:2018dos}, which isolate the contribution from a particular temporal range. Corresponding dispersive estimates of HVP restricted to the same windows \cite{Colangelo:2022vok,Benton:2023dci,Davier:2023cyp,Benton:2023fcv,Benton:2024kwp} can be studied and compared with lattice results, an approach that can help shed light on these tensions. 

In this work, we present results for the dominant long-distance window (LD) contribution to HVP, specifically the connected contributions from light (up and down) quarks in the isospin-symmetric limit, which comprise about 90\% of the total HVP. LD quantities suffer from large statistical noise and are sensitive to systematic effects from finite lattice volume and scale-setting \cite{DellaMorte:2017dyu} uncertainties. This work employs the high-statistics lattice datasets that were used to study the short- and intermediate-distance contributions in Ref.~\cite{FermilabLattice:2024yho}, a companion paper to this work. As in that work, we set the lattice scale at high precision using the mass of the $\Omega^-$ baryon \cite{Bazavov:2024dov,Grebe:2025}.
The combination of high statistics and improved scale setting allows us to achieve greatly improved precision for the LD HVP as well as for the full light-quark connected HVP by summing our LD result with the other window observables calculated in Ref.~\cite{FermilabLattice:2024yho}, and updated here with the improved scale setting.

\paragraph{Definitions ---}
The leading-order hadronic-vacuum-polarization contribution to the muon anomalous magnetic moment $\amu \equiv (g-2)_\mu/2$ is obtained via
\begin{equation}
    \amuHVP = 4\alpha^{2} \int_{0}^{\infty} \dd{t} C(t)  \tilde{K}(t),
    \label{eq:amuTint}
\end{equation}
where $\alpha$ is the fine-structure constant, the kernel $\tilde{K}$ stems from quantum electrodynamics (QED) \cite{Blum:2002ii,Bernecker:2011gh}, and 
$C(t) = \frac{1}{3} \sum_{k=1}^3 \sum_{\bm{x}}\left\langle J^{k}(\bm{x}, t) J^{k}(0)\right\rangle$ is the Euclidean-time two-point correlation function of the electromagnetic current $J^{\mu}(x)= \sum_{f} q_{f} \bar{\psi}_{f}(x) \gamma^{\mu} \psi_{f}(x)$, summed over all quark flavors $f\in\{u,d,s,c,b,t\}$ of electric charge $q_f$. 
Lattice-QCD calculations of $\amuHVP$ separately compute the contributions from each quark flavor and from connected and disconnected Wick contractions. Additionally, these calculations are performed in the isospin-symmetric limit in pure QCD where the up- and down-quark masses are taken to be equal and QED effects are removed. The effects of isospin-breaking corrections are studied in Ref.~\cite{FermilabLattice:2024yho,QEDpaper}. In this work, we focus on the dominant isospin-symmetric, light-quark connected (LQC) contribution $\amuL$.

\begin{figure}
\centering
\includegraphics[width=\linewidth]{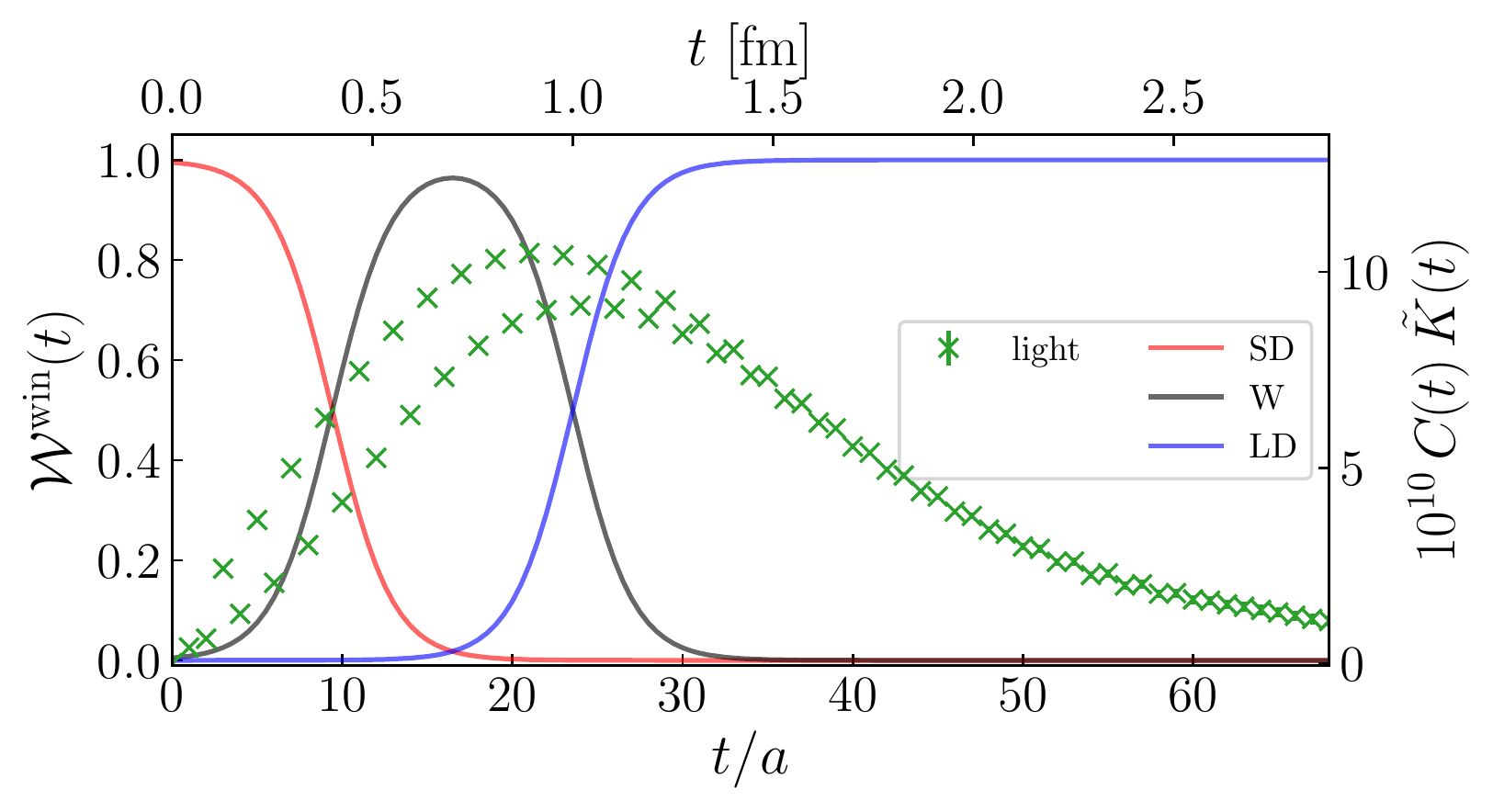}
\vspace{-5mm}
\caption{The SD (red), W (black), and LD (blue) window functions as defined in the text, overlaid with raw lattice data for the LQC vector correlator multiplied by the integration kernel $C(t) \tilde{K}(t)$ (green crosses) on our 0.06~fm ensemble. Note that the data points show the characteristic oscillating behavior of staggered fermions due to parity mixing at finite lattice spacing.}
\label{fig:window06Data}
\end{figure}

We also consider window observables \cite{RBC:2018dos} that restrict the Euclidean-time region over which $C(t)$ is integrated,
\begin{equation}
a_\mu^{\mathrm{win}} = 4 \alpha^{2} \int_{0}^{\infty} \dd{t} C(t)\tilde{K}(t) \mathcal{W}^{\mathrm{win}}\left(t\right). \label{eq:amuTintWin}
\end{equation}
Here we focus on the ``long-distance'' window defined by
\begin{align}
    \mathcal{W}^{\mathrm{LD}}(t) &= \frac{1}{2}\left[1 + \tanh \left(\frac{t-t_1}{\Delta}\right)\right] , \label{eq:windowLD}
\end{align}
with $t_1 = 1$ fm, and $\Delta = 0.15$ fm. This window is complemented by the ``short'' and ``intermediate-distance'' windows, defined in Eqs.~(2.9)--(2.12) of Ref.~\cite{FermilabLattice:2024yho} and studied there in detail, to cover the full integration region. The functions for the three complementary windows (SD, W, and LD) are shown in \cref{fig:window06Data}. The LD window covers the tail of the integrand, also shown in \cref{fig:window06Data}, where statistical errors increase significantly. Our strategies for alleviating this issue are discussed in detail below and in the Supplemental Material \cite{supplement}\nocite{Budapest-Marseille-Wuppertal:2017okr,Lepage:2001ym,Bouchard:2014ypa,FermilabLattice:2016ipl}.

\paragraph{Physical inputs ---}
This calculation makes use of ensembles of four-flavor gauge-field configurations. Converting the lattice results into physical units requires the selection of a set of physical inputs that determine the quark masses and the lattice spacing (or scale). The masses of the $u$, $d$, $s$, and $c$ quarks are determined by the masses of the $\pi^+, K^0$, $K^+$, and $D_s^+$ mesons. Following our previous work \cite{FermilabLattice:2024yho}, we define the physical point in the isospin-symmetric limit of pure QCD with $M_{\pi^+} =$ 135.0 MeV, $M_{K} = $ 494.6 MeV, and $M_{D_s} = 1967$ MeV \cite{FlavourLatticeAveragingGroupFLAG:2024oxs}. In this prescription, we work with the physical light-quark mass $m_l \equiv (m_u + m_d) / 2$. 

We set the lattice scale using the $\Omega^-$ baryon mass \cite{Bazavov:2024dov,Grebe:2025}. To fix the scale to the pure-QCD world, we use the value of $M_\Omega= 1.67126(32)$~GeV \cite{Bazavov:2024dov,Grebe:2025}, obtained by subtracting the QED effects. The error includes statistical, correlator-fit-systematic, and continuum-extrapolation uncertainties. Systematic errors from finite volume, strong isospin breaking, and charged sea-quark effects are sub-leading. We also consider a secondary scale setting based on the pion decay constant $f_\pi$ \cite{FermilabLattice:2014tsy,FermilabLattice:2017wgj,Bernard:2025} using the physical value $f_{\pi}=0.13050(13)$~GeV \cite{ParticleDataGroup:2022pth,FlavourLatticeAveragingGroupFLAG:2024oxs}, allowing a consistency check between the two most commonly used scale-setting observables for $\amuL$. Values for $aM_\Omega$ and $af_\pi$ on the ensembles used in this work are given in \cref{table:ensParams}.

\paragraph{Lattice ensembles and correlation functions ---}
\begin{table*}[t]
\centering
\caption{Ensemble parameters used in this work. The first column lists the approximate lattice spacings in~fm using $M_{\Omega}$ scale setting. The second column gives the spatial length $L$ of the lattices in~fm. The third column lists the volumes of the lattices in number of space-time points. The fourth column gives the sea-quark masses in lattice-spacing units. The fifth column lists the $\Omega^-$ baryon mass in lattice units, $aM_\Omega$ \cite{Bazavov:2024dov,Grebe:2025}. The sixth column lists the values of $af_\pi$.}
\label{table:ensParams}
\begin{tabularx}{\linewidth}{llCCRR}
\hline \hline
$\approx a/\mathrm{fm}$ & $L_{M_{\Omega}}/\mathrm{fm}$ & $N_s^3 \times N_t$ & $a m_{l}^\text{sea} / a m_{s}^\text{sea} / a m_{c}^\text{sea}$ & $aM_{\Omega}$  &  $af_{\pi}$  \\ 
\hline
$0.15$ & 5.00 & $ 32^3 \times 48$ & 0.002426/0.0673/0.8447  & 1.3246(26) & 0.100015(39)\\ 
$0.12$ & 5.95 & $48^3 \times 64$ & 0.001907/0.05252/0.6382  & 1.0494(17)  & 0.080290(62)\\ 
$0.09$ & 5.70 & $64^3 \times 96$ & 0.001326/0.03636/0.4313 &  0.75372(97) & 0.058145(63)\\ 
$0.06$ & 5.48 & $96^3\times128$ & 0.0008/0.022/0.260  & 0.4834(11)  & 0.037526(32)\\ 
\hline
\hline
\end{tabularx}
\end{table*}

The lattice ensembles and correlation function datasets employed in this work are described in detail in our companion paper, Ref.~\cite{FermilabLattice:2024yho}. Briefly, this work employs gauge field ensembles generated by the MILC collaboration \cite{MILC:2010pul,MILC:2012znn,Bazavov:2017lyh,MILCConfigsGitHub} with light, strange, and charm quarks in the sea (all tuned to their physical values) spanning the range of lattice spacings of $a\approx 0.15$--0.06~fm, as summarized in \cref{table:ensParams}.\footnote{The $0.09$~fm ensemble was, in-part, generated by the CalLat collaboration \cite{Miller:2020evg} using retuned values of the quark masses determined by MILC \cite{FermilabLattice:2014tsy}.} Further details on the quark masses and corresponding meson masses are given in Table II of Ref.~\cite{FermilabLattice:2024yho}. The vector-current two-point correlation function datasets are described in Table IV of Ref.~\cite{FermilabLattice:2024yho} and the corresponding renormalization factors $Z_V$ are taken from Refs.~\cite{Hatton:2019gha,Hatton:2020qhk} and listed in Table III of Ref.~\cite{FermilabLattice:2024yho}. We use the ``local'' and ``one-link'' current datasets, which correspond to two different discretizations of the vector current, on the $a\approx 0.12, 0.09, 0.06$~fm ensembles, all of which were generated using low-mode averaging (LMA) \cite{DeGrand:2004qw,Giusti:2004yp,Blum:2012uh}.

\paragraph{Analysis strategy ---}
Our overall strategy for HVP observables is described in our companion paper, Ref.~\cite{FermilabLattice:2024yho}. In particular, the lattice corrections, continuum fit functions and strategy for performing the continuum extrapolations, the Bayesian Model Averaging (BMA), and systematic error estimates follow the detailed description of Ref.~\cite{FermilabLattice:2024yho}. As in Ref.~\cite{FermilabLattice:2024yho}, each individual HVP observable was multiplied with an unknown blinding factor, which was removed only after the analysis was finalized.

Long-distance HVP observables are affected by the well-known, rapid growth of statistical errors in the vector-current correlation functions \cite{Lepage:1989hd}. While the exact low-modes used in the generation of the correlators help to reduce the statistical errors at large Euclidean times, these errors are still a significant source of uncertainty. We address this issue with a correlator-reconstruction strategy to replace $C(t)$ at large times. In Ref.~\cite{Lahert:2024vvu}, a precise reconstruction of the vector-current correlator at large times, obtained from an exclusive channel study of the two-pion contributions, was used to test the fit method (described below). The fit was found to provide a reliable, albeit less precise, description of the vector-current correlator in the long time region. Hence, our results are obtained with this approach; we also employ the bounding method \cite{RBC:2018dos} as a cross-check (see Supplemental Material \cite{supplement} and references).

\paragraph{Fit method ---}
In this approach, the correlation function data are fit over the range $t_\textrm{min} \leq t \leq t_\textrm{max}$ to a function matching the expected spectral decomposition. The noisy correlation function data is then replaced with the fit reconstruction at large Euclidean distances, $t \geq t^\star$, where $t^\star$ is a hyperparameter that we optimize by minimizing the overall variance of $\amu$. Each fit makes use of the full sample covariance matrix over the fit range, which is well determined for all lattice spacings and currents.

The spectral decompositions of the staggered correlation functions employed in this analysis contain two towers of states, one of which has an oscillating phase, yielding the following fit function:
\begin{align}
        C_{\textrm{fit}}(t) &= \textrm{const.} +  \sum^{N_{\textrm{states}}}_{n=0}\left[ Z^2_n \left(e^{-E_n t} + e^{-E_n (T-t)}\right)\right] \nn \\
        &+ (-1)^t \sum^{M_{\textrm{states}}}_{m=0} \left[ Z^2_{m,\textrm{osc}} \left(e^{-E_{m,\textrm{osc}} t} + e^{-E_{m,\textrm{osc}} (T-t)}\right)\right], 
        \label{eq:corrfitfunc}
\end{align}
where the energies ($E_n, E_{m,\textrm{osc}}$), amplitudes ($Z_n, Z_{m,\textrm{osc}}$), and constant term are the parameters obtained from the fit. Here, $T = a N_t$ is the temporal extent of the lattice, and the term that is constant in time separation $t$ incorporates the leading additional finite-$T$ effects \cite{Chakraborty:2016mwy,Lahert:2024vvu}. The vector-current correlators are too noisy to resolve precisely the low-lying two-pion states that contribute to the correlator at large times. Hence, the low-lying spectrum obtained with the fit method is only approximate, representing a mix of the $\rho$ meson and two-pion states. Tests of the fidelity of the fit in reproducing the correlator are given in the Supplemental Material.

Rather than using a single fit Ansatz, we carry out a model average over multiple values of $t_\textrm{min}$ using the Bayesian Akaike information criterion (BAIC) with the data-subset penalty \cite{Neil:2022joj,Neil:2023pgt}, holding $t_{\textrm{max}}$ fixed at a value beyond which the noise-to-signal rises above 85\%. The value of $t_{\textrm{min}}$ is allowed to vary over a restricted range. Log-normal priors are employed for the energies and amplitudes, following Ref.~\cite{FermilabLattice:2019ugu}; the prior for the constant term is set based on the energy and amplitude priors to match the leading constant contribution of $Z_0^2 e^{-E_\pi T}$. Choices of fit hyperparameters and results for ground-state energies are given in the Supplemental Material.

Our final $C(t)$, which enters into \cref{eq:amuTint,eq:amuTintWin}, is constructed by taking the raw data $C(t)$ for $t < t^\star$ and $C_{\textrm{recon.}}(t)$ for $t\geq t^\star$; the latter corresponds to the prediction of the fit model \cref{eq:corrfitfunc} evaluated in the limit $T \rightarrow \infty$. Other variations of the analysis choices, such as changing the number of states, have been tested and have negligible impact on the results.

Following the correlator reconstruction, to calculate $\amu$, we extend the independent time range of our correlation function dataset from $N_t/2+1$ to $2N_t$ using the infinite-$T$ correlator-reconstruction and then integrate \cref{eq:amuTint,eq:amuTintWin} using the trapezoidal rule. All error propagation is carried out using $\textsc{gvar}$ \cite{gvarGitHub}, cross-checked using jackknife resampling.

\paragraph{Lattice corrections ---}
We perform explicit corrections for finite volume (FV), pion-mass ($M_{\pi}$) mistuning, and (optionally) taste breaking (TB) in that order, using the effective field theory (EFT) based correction schemes described in Ref.~\cite{FermilabLattice:2024yho}; details are shown in the Supplemental Material. Finite-time corrections, computed using next-to-leading order chiral perturbation theory (NLO \chpt) \cite{Borsanyi:2020mff}, on the contribution from $t<t^\star$ are negligible at the current level of precision.

\paragraph{Continuum extrapolations ---}
The continuum extrapolations of our $\amuLLD$ and $\amuL$ data are based on the form $a_{\mu}(a, M_A)= a_{\mu}\left[1 + F^{{a}} (a) + F^M (M_A)\right]$ where $F^M(M_A)$ accounts for residual sea-quark-mass--mistuning effects and is given in Eq.~(3.7) of Ref.~\cite{FermilabLattice:2024yho} and 
\begin{align}
    F^{{a}} (a) &= C_{a^{2},n}(a\Lambda)^{2} \alpha_s^{n} + \sum^{m}_{k=2} C_{a^{2k}} (a\Lambda)^{2k}, \label{eq:discfunc}
\end{align}
where $n = 1,2$ ($n=0$) for the local (one-link) current, and $m=2,3$. As described in Ref.~\cite{FermilabLattice:2024yho}, the Gaussian prior $C_{a^{2m}}=0(2)$ is imposed on the highest-order terms. In joint fits to data obtained from the local and one-link currents, the parameters in $F^M(M_A)$ and $a_\mu$ are shared. 

We first employ the empirical Bayes procedure discussed in Sec.~III.C of Ref.~\cite{FermilabLattice:2024yho} 
to obtain guidance for the choice of scale $\Lambda$ and the relevant terms in the continuum fit function. The procedure is performed for both currents independently and jointly, and separately for data corrected and not corrected for TB effects. After varying the correction schemes and windows, as well as correlator-reconstruction methods, we find $\Lambda \approx 0.5$~GeV for data corrected for TB effects and $\Lambda \approx 1.0$~GeV for uncorrected data, reflecting the fact that discretization effects are larger in the uncorrected $\amuL$ data. The empirical Bayes analysis further reveals that the data corrected for TB is sensitive to discretization terms up to $a^4$ whereas the uncorrected data is sensitive to terms up to $a^6$. Hence, we consider variations of the functions $F^a(a)$ where $m=2$ (quadratic) or $m=3$ (cubic).  For the quadratic fits, we consider a variation where the 0.15~fm data point is dropped. For datasets not corrected for TB effects, we perform only joint fits and fix $n=2$ for the local current, to account for the dominant TB effects~\cite{FermilabLatticeHPQCD:2023jof}.

\paragraph{Continuum extrapolation BMA ---}
Systematic uncertainties are estimated using BMA~\cite{Jay:2020jkz,Neil:2022joj}, following the procedure and formulae detailed in Refs.~\cite{FermilabLatticeHPQCD:2023jof,FermilabLattice:2024yho}. In the BMA, we include variations for all FV, $M_{\pi}$-mistuning and TB correction schemes discussed in the Supplemental Material, namely NNLO \chpt, chiral model (CM), and Meyer-Lellouch-L\"{u}scher-Gounaris-Sakurai (MLLGS). We always correct both currents using the same scheme. Extrapolations using MLLGS exclude the coarsest data point ($0.15$~fm) as discussed in Ref.~\cite{FermilabLattice:2024yho}. FV and $M_{\pi}$-mistuning corrections are always included, along with a 10\% associated uncertainty to capture differences between EFT-based and data-driven corrections \cite{Borsanyi:2020mff,RBC:2024fic,Djukanovic:2024cmq}. Variations with and without TB corrections are considered. 
\begin{figure}
\centering
\includegraphics[width=\linewidth]{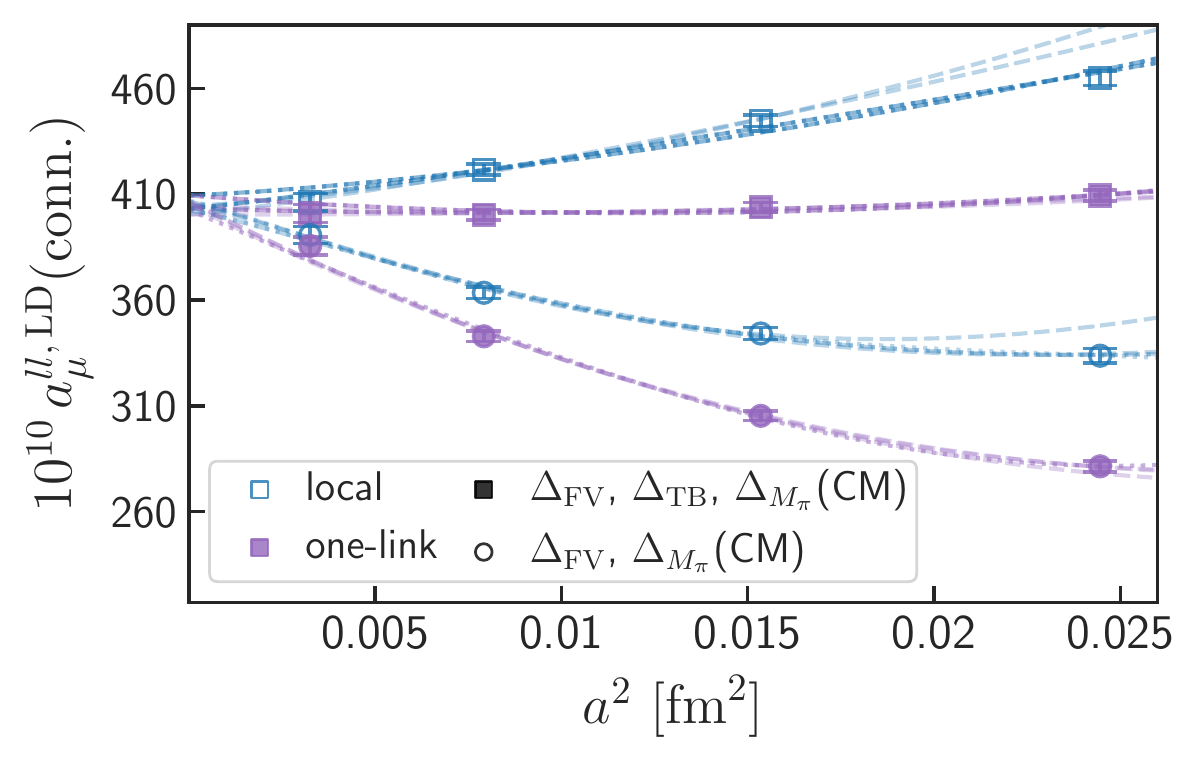}
\vspace{-5mm}
\caption{
Continuum extrapolations of the long-distance window observable $\amuLLD$. Data points shown are corrected for FV and $M_\pi$ mistuning using the CM scheme. Blue unfilled (purple filled) points are from the local (one-link) current. Squares (circles) represent data corrected (uncorrected) for TB using the CM scheme. Different extrapolations are obtained from variations of the fit functions and ensembles included.
}\label{fig:BMALDFitCont}
\end{figure}

We perform continuum extrapolations using all fit function variations described above, including fits that exclude one of the two currents entirely \cite{FermilabLattice:2024yho}. Sample individual results from the continuum extrapolation BMA for $a^{ll,\,{\mathrm {LD}}}_{\mu}(\mathrm{conn.})$ are shown in \cref{fig:BMALDFitCont}. Additional results and plots are shown in the supplemental material.

\begin{table}[t]
\centering
\caption{Approximate absolute error budgets (in units of $10^{-10}$) for the uncertainties reported in\cref{eq:SDRes,eq:WRes,eq:LDRes,eq:SumRes}. 
From left to right, the contributions to the error are Monte Carlo statistics and $t_{\min}$ variations in the correlator reconstructions for $a^{ll,\,{\mathrm {LD}}}_{\mu}$ and $a^{ll}_{\mu}$, continuum extrapolation and TB corrections, FV and $M_{\pi}$-mistuning corrections, scale setting (due to the uncertainties in $aM_\Omega$ and in the input $M_\Omega$ mass in physical units) and current renormalization. }
\label{table:IndividualUncertaintyMomega}
\vspace{1mm}
\begin{tabularx}{\linewidth}{LCLLLLll}
\hline \hline
 &  stat., & $a\to0$, &  $\Delta_{\textrm{FV}}$, &  &  &  \\ 
Contrib. &  $t_{\min}$ &  $\Delta_{\mathrm{TB}}$   &   $\Delta_{M_{\pi}}$ &   $a$  & $Z_V$ &  Total\\ 
\hline
$a^{ll,\,{\mathrm {SD}}}_{\mu}$ & 0.011 & 0.060 & --- & 0.028 & 0.063 & 0.092 \\
$a^{ll,\,{\mathrm {W}}}_{\mu}$ & 0.14 & 0.37 & 0.34 & 0.31 & 0.18 & 0.63 \\
$a^{ll,\,{\mathrm {LD}}}_{\mu}$ & 2.3 & 2.8 & 1.4 & 1.9 & 0.2 & 4.3 \\
$a^{ll}_{\mu}$ & 2.3 & 2.9 & 1.7 & 1.8 & 0.3 & 4.5 \\
\hline \hline
\end{tabularx}
\end{table}

\paragraph{Results ---}
Here we report our main results, which use $M_{\Omega}$ scale setting and the fit method for correlator reconstructions. Error budgets for all listed quantities below are given in \cref{table:IndividualUncertaintyMomega}. Additional results for $f_{\pi}$ scale setting or the bounding method are given in the End Matter. Note that \cref{table:IndividualUncertaintyMomega,table:IndividualUncertaintyPi} list the scale error, which is useful for comparisons of results for scheme-dependent flavor-specific quantities from different lattice groups. Finally, the Supplemental Material (Table II) contains comparisons of all isospin-symmetric observables obtained here and in Ref.~\cite{FermilabLattice:2024yho} with $M_\Omega$, $f_\pi$, and $w_0$ scale setting,\footnote{We follow the Muon g-2 Theory Initiative’s White-Paper scheme (WP25) and take $w_0$ from Ref.~\cite{Borsanyi:2020mff} as a fixed value, $w_0 = 0.17236$~fm.} including their correlated differences.

First, we take our results for $\amuLSD$ and $\amuLW$ from Ref.~\cite{FermilabLattice:2024yho}, which we restate here for convenience,
\begin{align}
    \amuLSDRes, \label{eq:SDRes}\\
    \amuLWRes. \label{eq:WRes}
\end{align}
The uncertainties are statistical (second column of \cref{table:IndividualUncertaintyMomega}), systematic (third to sixth columns of \cref{table:IndividualUncertaintyMomega}) and total.
For the long-distance window we obtain,
\begin{align}
    \amuLLDRes. \label{eq:LDRes}
\end{align}
Summing the LQC contribution in the three windows, \cref{eq:SDRes,eq:WRes,eq:LDRes}, with the correlation matrix given in \cref{table:correlationsMomegafpi}, yields
\begin{align}
    \amuLRes \label{eq:SumRes}.
\end{align}
Finally, from the independent analysis of the full integration region we obtain $\amuLFullRes$, a result that is completely consistent with \cref{eq:SumRes} with a correlated difference of $1.0(5.5)\times 10^{-10}$.

\paragraph{Summary and outlook ---}
As part of our ongoing lattice QCD project to calculate the complete HVP with few-permille precision, we have calculated $\amuLLD$ at $1.1\%$ precision, while we obtain $\amuL$ with a total uncertainty of $0.69\%$, the most precise determination of this important quantity from lattice QCD to date.\footnote{Our results for $\amuLLD$ and $\amuL$ obtained from $f_\pi$ scale setting are at $1.0\%$ and $0.63\%$ precision, respectively.} These results are made possible by the high-statistics correlation function data set we have generated to date \cite{FermilabLattice:2024yho}, as well as the new high-precision scale-setting using the $\Omega^-$ baryon mass \cite{Bazavov:2024dov,Grebe:2025}. 

\begin{figure}
    \centering
    \includegraphics[width=\linewidth]{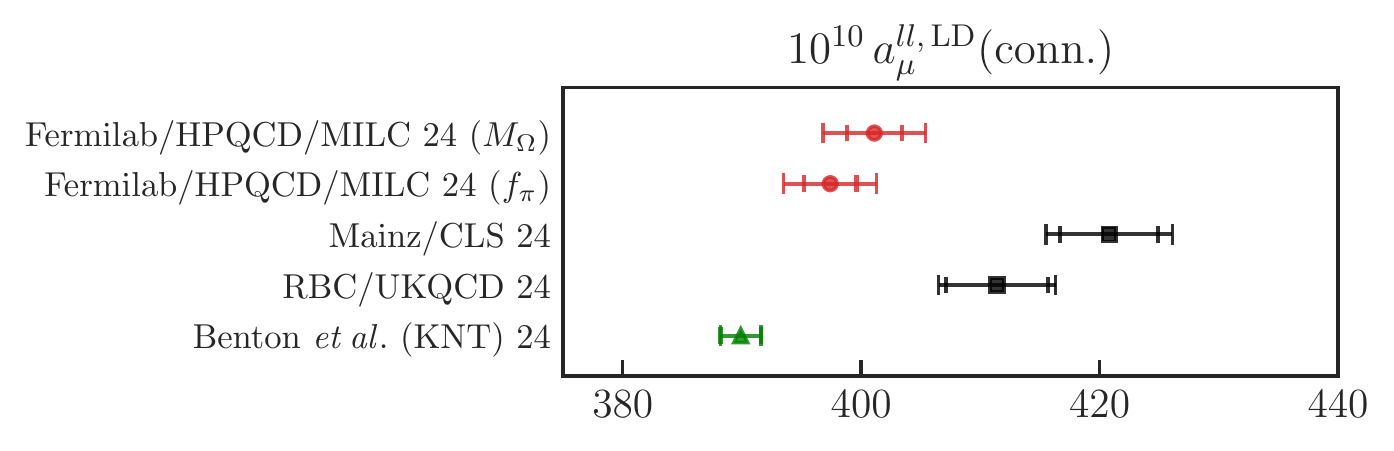}
    \vspace{-5mm}
    \caption{Comparison of our lattice determinations for $\amuLLD$ (filled red circles) labeled ``Fermilab/HPQCD/MILC~24'' with $n_f=2+1$ (black squares) lattice-QCD calculations by Mainz/CLS~24 \cite{Djukanovic:2024cmq}, RBC/UKQCD~24 \cite{RBC:2024fic}. The inner error bar shown for our result is from Monte Carlo statistics. Also shown is a data-driven evaluation of $\amuLLD$ using $e^+e^-$ cross section data (green triangle) by Benton {\it et al.}~24 \cite{Benton:2024kwp}.}
    \label{fig:LDIndividualCompare}
\end{figure}

\begin{figure}
    \centering
    \includegraphics[width=\linewidth]{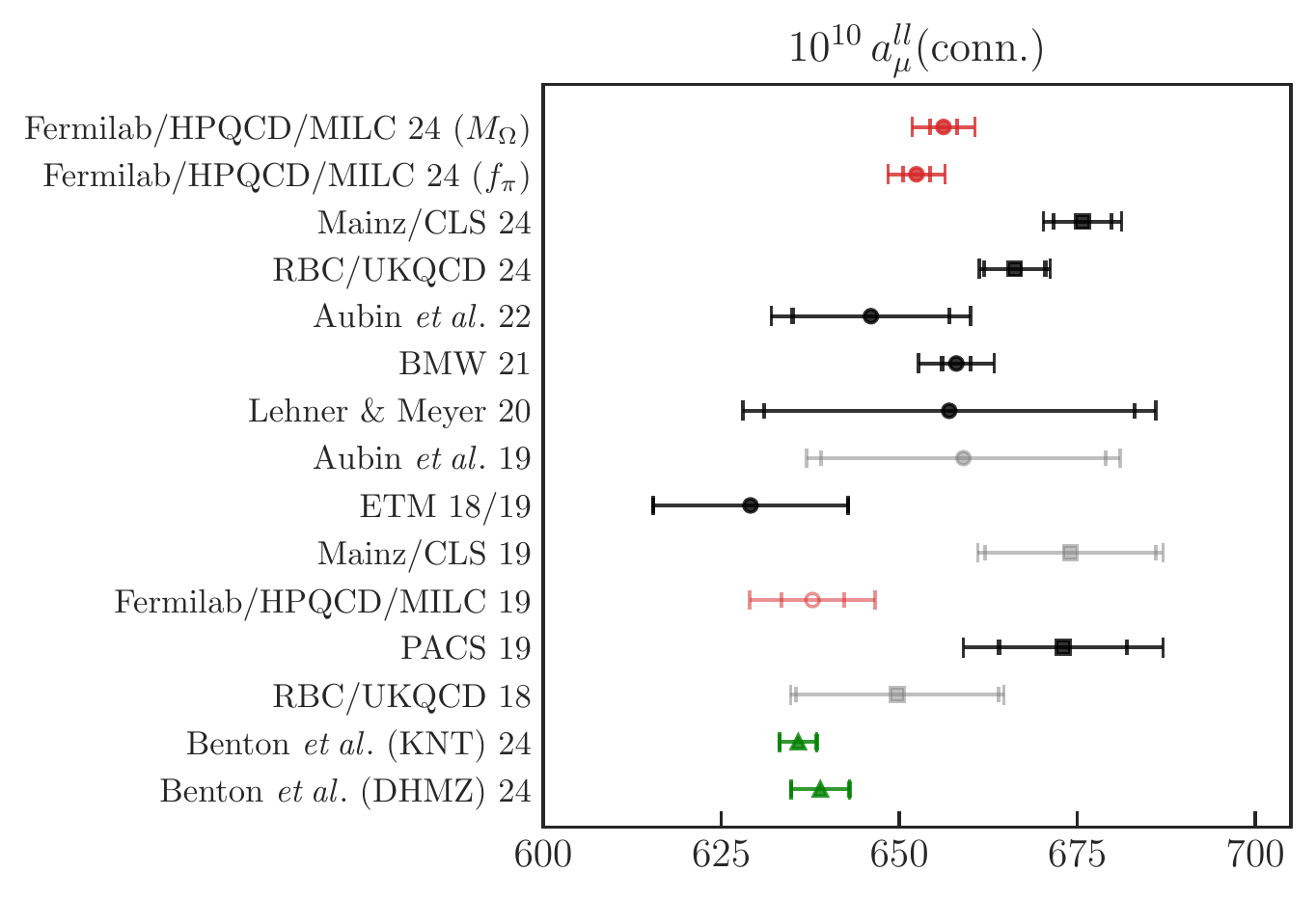}
    \vspace{-5mm}
    \caption{Comparison of our lattice determinations for $\amuL$ (filled red circles) labeled ``Fermilab/HPQCD/MILC~24'' with $n_f=2+1+1$ (black circles) and $n_f=2+1$ (black squares) lattice-QCD calculations by Mainz/CLS~24 \cite{Djukanovic:2024cmq}, RBC/UKQCD~24 \cite{RBC:2024fic}, Aubin {\it et al.}~22 \cite{Aubin:2022hgm}, BMW~21 \cite{Borsanyi:2020mff}, Lehner \& Meyer~20 \cite{Lehner:2020crt}, Aubin {\it et al.}~19 \cite{Aubin:2019usy}, ETM~18/19 \cite{Giusti:2018mdh}, Mainz/CLS~19 \cite{Gerardin:2019rua}, PACS~19 \cite{Shintani:2019wai}, and RBC/UKQCD~18 \cite{RBC:2018dos}. Our previous determination, Fermilab/HPQCD/MILC~19 \cite{FermilabLattice:2019dbx}, is shown with an unfilled red circle.
    Inner error bars indicate the reported statistical uncertainty.
    Also shown are data-driven evaluations of $\amuL$ using $e^+e^-$ cross section data (green triangles) by Benton {\it et al.}~24 \cite{Benton:2024kwp}.}
    \label{fig:FullIndividualCompare}
\end{figure}

A comparison of our results for $\amuLLD$, using both $M_\Omega$ and $f_\pi$ scale settings, with those of the RBC/UKQCD \cite{RBC:2024fic} and Mainz \cite{Djukanovic:2024cmq} collaborations as well as a data-driven evaluation \cite{Benton:2024kwp} is shown in \cref{fig:LDIndividualCompare}. Using the same scale setting in the comparisons, we find that our result is lower than RBC/UKQCD 24 ($M_\Omega$) and Mainz 24 ($f_\pi$) by $1.7\sigma$ and $3.6\sigma$, respectively, while being higher than the data-driven evaluation of Ref.~\cite{Benton:2024kwp} by $2.2\sigma$, if using $M_\Omega$. 
In \cref{fig:FullIndividualCompare} we compare our result for $\amuL$ with previous lattice-QCD calculations as well as the data-driven evaluations of Ref.~\cite{Benton:2024kwp}. Focusing on the three most precise lattice-QCD results, we find significances in the differences of $0.4\sigma$ with BMW~21 ($M_\Omega$), $1.6\sigma$ with RBC/UKQCD 24 ($M_\Omega$), and $3.5\sigma$ with Mainz 24 ($f_\pi$). We find a difference of $1.4\sigma$ with our previous result of Ref.~\cite{FermilabLattice:2019ugu}, if using the $f_\pi$ scale. For a more direct comparison with Ref.~\cite{FermilabLattice:2019ugu} we take our result from the direct analysis of the full integration region obtained with $f_\pi$ scale setting of $\amuLFullResFPI$, which yields a significance of $1.3\sigma$. Finally, we find that our result in \cref{eq:SumRes} differs from the data-driven evaluation of Ref.~\cite{Benton:2024kwp} by $3.8\sigma$ ($2.7\sigma$) if using the KNT (DHMZ) compilations, confirming the tensions between lattice-QCD and data-driven evaluations seen previously and in windowed HVP observables. More precise lattice-QCD calculations are clearly needed.   

Looking towards future improvements of our HVP calculations, we are generating correlators on ensembles with both a finer lattice spacing $a \approx 0.04$ fm and with a larger spatial length $L \approx 11$ fm at $a \approx 0.09$ fm, both of which will help to address significant sources of uncertainty (as shown in \cref{table:IndividualUncertaintyMomega}.) We also plan to add exclusive two-pion channel correlator data, which is expected to greatly improve the statistical precision for LD HVP, even for staggered fermions \cite{Lahert:2024vvu}, as well as parallel improved calculations of subleading contributions due to quark-disconnected diagrams and isospin-breaking effects \cite{FermilabLattice:2024yho}. Taken all together, we expect that these improvements will allow us to meet the target precision set by the Fermilab $g-2$ experiment.

\acknowledgments

\hfill
\paragraph{Acknowledgments ---}
We are grateful to Yin Lin, Aaron Meyer, and Michael Wagman for their contributions to the $\Omega^-$ scale-setting project, the results of which are used in this work.
We thank Urs Heller for his contributions to our projects on gauge flow and the pion decay constant, also used in scale setting.

Computations for this work were carried out in part with computing and long-term storage resources provided by the USQCD Collaboration, the National Energy Research Scientific Computing Center (Edison, Cori, Perlmutter), the Argonne Leadership Computing Facility (Mira, Theta) under the INCITE program, and the Oak Ridge Leadership Computing Facility (Summit, Frontier) under the Innovative and Novel Computational Impact on Theory and Experiment (INCITE) and the ASCR Leadership Computing Challenge (ALCC) programs.
The Argonne Leadership Computing Facility (ALCF), is supported by the U.S. Department of Energy (USDOE) under contract DE-AC02-06CH11357.
The National Energy Research Scientific Computing Center (NERSC) is located at Lawrence Berkeley National Laboratory and operated under USDOE 
Contract No. DE-AC02-05CH11231.  (Our awards are HEP-ERCAP0032813, HEP-ERCAP0031326, and possible predecessors.)
The Oak Ridge Leadership Computing Facility (OLCF) at the Oak 
Ridge National Laboratory is supported by the USDOE under Contract No. DE-AC05-00OR22725.
ALCF, NERSC, and OLCF are all USDOE Office of Science User Facilities.
This work used the Extreme Science and Engineering Discovery Environment (XSEDE) supercomputer Stampede 2 at the Texas Advanced Computing Center (TACC) through allocation TG-MCA93S002. The XSEDE program is supported by the National Science Foundation under grant number ACI-1548562.
Computations on the Big Red II+, Big Red 3, and Big Red 200 supercomputers were supported in part by Lilly Endowment, Inc., through its support for the Indiana University Pervasive Technology Institute.
The parallel file system employed by Big Red II+ was supported by the National Science Foundation under Grant No.~CNS-0521433.
This work utilized the RMACC Summit supercomputer, which is supported by the National Science Foundation (awards ACI-1532235 and ACI-1532236), the University of Colorado Boulder, and Colorado State University. The Summit supercomputer is a joint effort of the University of Colorado Boulder and Colorado State University.
Some of the computations were done using the Blue Waters sustained-petascale computer, which was supported by the National Science Foundation (awards OCI-0725070 and ACI-1238993) and the state of Illinois. Blue Waters was a joint effort of the University of Illinois at Urbana-Champaign and its National Center for Supercomputing Applications.

This work was supported in part by the U.S.~Department of Energy, Office of Science, under Awards
No.~DE-SC0010005 (E.T.N. and J.W.S.),
No.~DE-SC0010120 (S.G.), 
No.~DE-SC0015655 (A.X.K., S.L., M.L., A.T.L.),
No.~DE-SC0009998 (J.L.),
the ``High Energy Physics Computing Traineeship for Lattice Gauge Theory'' No.~DE-SC0024053 (J.W.S.),
and the Funding Opportunity Announcement Scientific Discovery through Advanced Computing: High Energy Physics, LAB 22-2580 (D.A.C., C.T.P., L.H., M.L., S.L.); by the Exascale Computing Project (17-SC-20-SC), a collaborative effort of the U.S. Department of Energy Office of Science and the National Nuclear Security Administration (H.J.); 
by the National Science Foundation under Grants Nos.~PHY20-13064 and PHY23-10571 (C.D., D.A.C., S.L., A.V.), PHY23-09946 (A.B.), Grant No. 2139536 for Characteristic Science Applications for the Leadership Class Computing Facility (L.H., H.J.);
by the Simons Foundation under their Simons Fellows in Theoretical Physics program (A.X.K.); 
by the Universities Research Association Visiting Scholarship awards 20-S-12 and 21-S-05 (S.L.);  
by MICIU/AEI/10.13039/501100011033 and FEDER (EU) under Grant PID2022-140440NB-C21 (E.G.);
by Consejeria de Universidad, Investigaci\'on e Innovaci\'on and Gobierno de Espa\~na and EU--NextGenerationEU, under Grant AST22 8.4 (E.G.);
by AEI (Spain) under Grant No.\ RYC2020-030244-I / AEI / 10.13039/501100011033 (A.V.);
and by U.K. Science and Technology Facilities Council under Grant ST/T000945/1 (C.T.H.D).
A.X.K. and E.T.N. are grateful to the Pauli Center for Theoretical Studies and the ETH Z\"urich for support and hospitality. A.X.K, A.S.K, and E.T.N are grateful to the Kavli Institute for Theoretical Physics (KITP) for hospitality and support during the ``What is Particle Theory?'' program. The KITP is supported in part by the National Science Foundation under Grant PHY-2309135.
This document was prepared by the Fermilab Lattice, HPQCD, and MILC Collaborations using the resources of the Fermi National Accelerator Laboratory (Fermilab), a U.S. Department of Energy, Office of Science, HEP User Facility.
Fermilab is managed by Fermi Research Alliance, LLC (FRA), acting under Contract No.~DE-AC02-07CH11359.

\bibliographystyle{apsrev4-2}
\bibliography{refs}

\section*{End Matter}

\begin{figure*}[htb]
\centering
\includegraphics[width=0.48\linewidth]{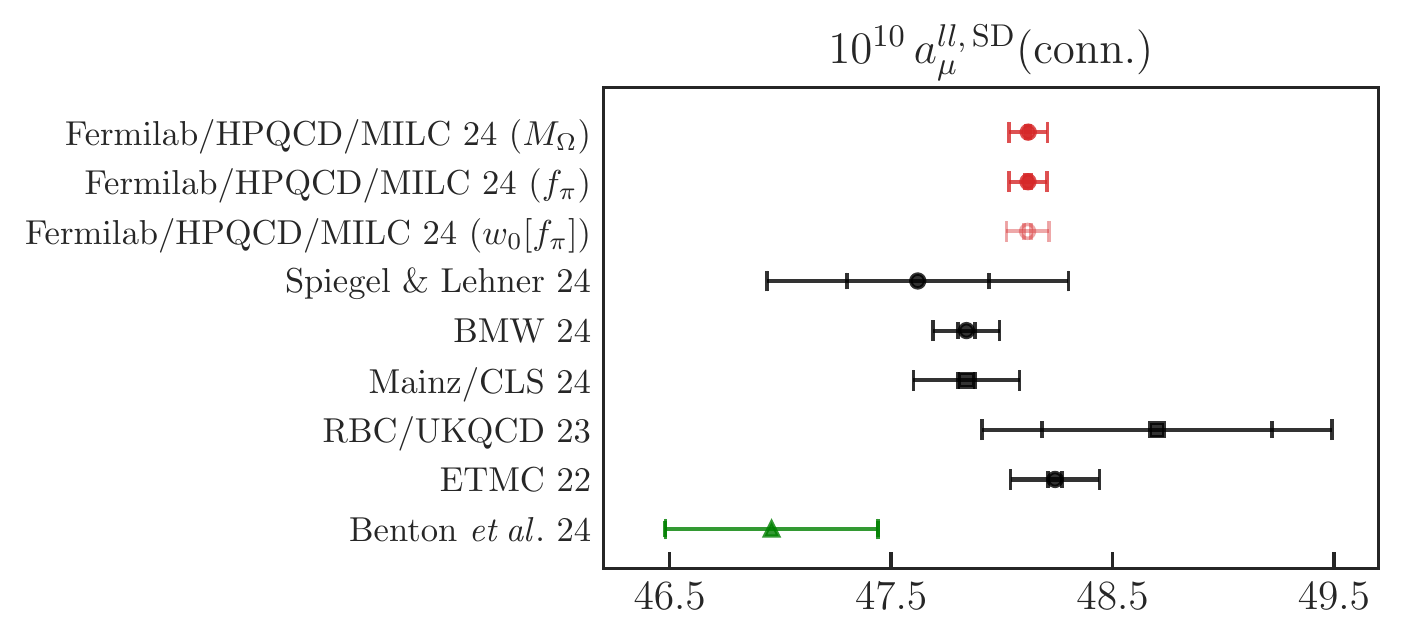}
\hfill
\includegraphics[width=0.48\linewidth]{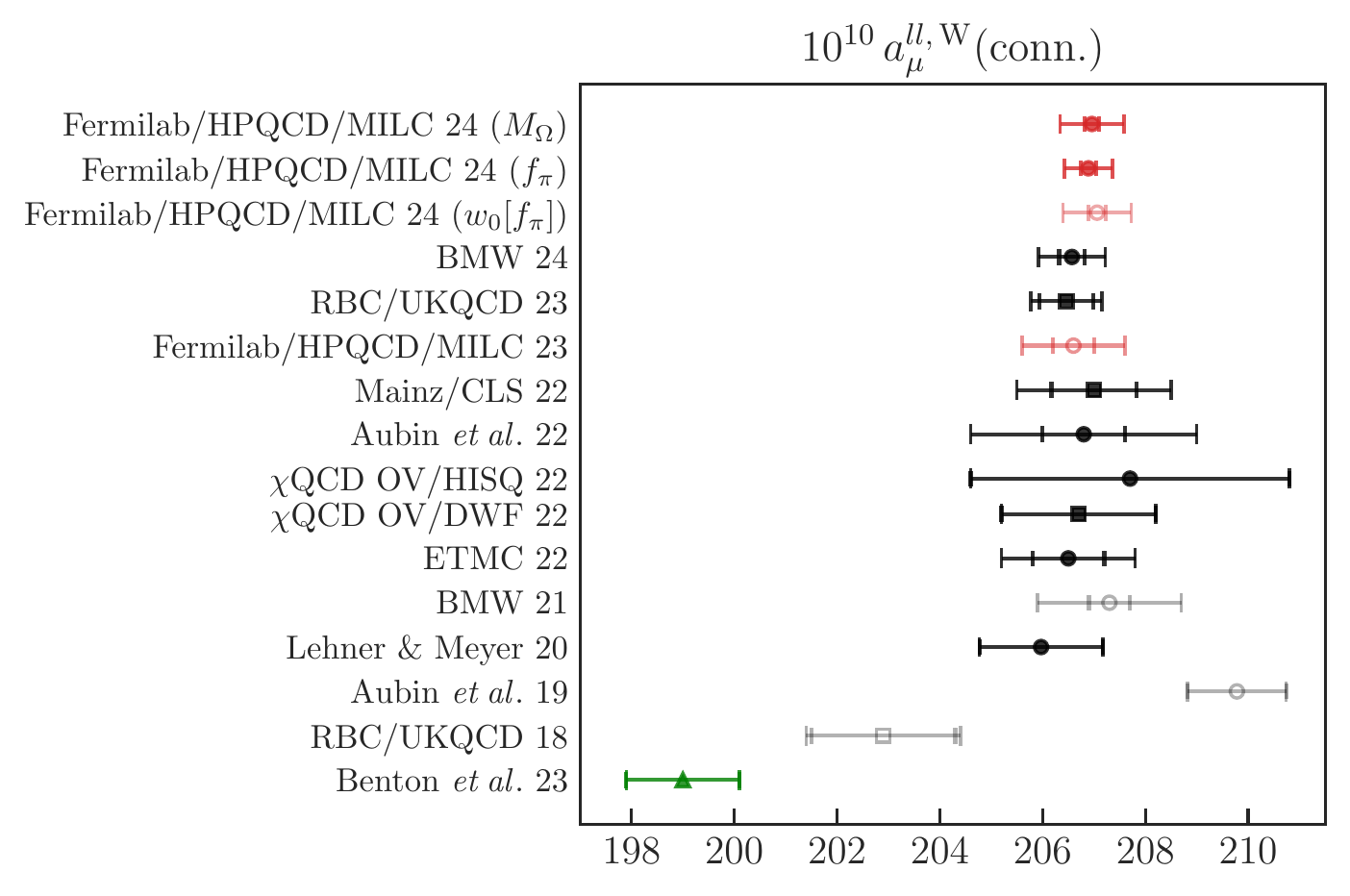}
\caption{Comparison of our lattice determinations for $\amuLSD$ (left) and $\amuLW$ (right) (red circles) labeled ``Fermilab/HPQCD/MILC~24~($M_{\Omega}$)'' and ``Fermilab/HPQCD/MILC~24~($f_{\pi}$)'' to $n_f=2+1+1$ (black circles) and $n_f=2+1$ (black squares) lattice-QCD calculations by 
Spiegel \& Lehner~24 \cite{Spiegel:2024dec},
BMW~24 \cite{Boccaletti:2024guq},
Mainz/CLS~24 \cite{Kuberski:2024bcj},
RBC/UKQCD~23 \cite{Blum:2023qou},
Mainz/CLS~22 \cite{Ce:2022kxy},
Aubin {\it et al.}~22 \cite{Aubin:2022hgm},
$\chi$QCD 22 \cite{Wang:2022lkq},
ETMC~22 \cite{Alexandrou:2022amy}, and
Lehner \& Meyer~20 \cite{Lehner:2020crt}.
Our previous result, Fermilab/HPQCD/MILC 23, is shown in light red. BMW~21 \cite{Borsanyi:2020mff},  Aubin {\it et al.}~19 \cite{Aubin:2019usy} and RBC/UKQCD~18 \cite{RBC:2018dos}, shown in gray, have been superseded.
The inner error bar shown for our result is from Monte Carlo statistics.  Also shown is a data-driven evaluation of $\amuLSD$ and $\amuLW$ using $e^+e^-$ cross section data (green triangles) by
Benton {\it et al.}~24 \cite{Benton:2024kwp} and Benton {\it et al.}~23 \cite{Benton:2023dci,Benton:2024kwp}, respectively.
}\label{fig:SDWIndividualCompare}
\end{figure*}

Here we collect additional results for the various scale-setting schemes and correlator-reconstruction strategies considered. We compare our updated results for $a^{ll,\,{\mathrm {SD}}}_{\mu}(\mathrm{conn.})$ and $a^{ll,\,{\mathrm {W}}}_{\mu}(\mathrm{conn.})$ to previous determinations in \cref{fig:SDWIndividualCompare}. Final results for the scale-setting and correlator-reconstruction variations considered are listed in \cref{tab:altResults}. For the correlated differences in \cref{tab:altResults}, we take the BMA systematic contributions to the uncertainties to be 100\% correlated for a given window observable across each scale. For the fit method, we give the correlation matrices for these results using $M_{\Omega}$ and $f_{\pi}$ scale setting in \cref{table:correlationsMomegafpi}, and error budgets for results using $f_{\pi}$ are given in \cref{table:IndividualUncertaintyPi}.

\begin{table*}
\centering
\caption{Final results for $\amu$ (in units of $10^{-10}$). The first column lists the specific LQC contribution. Columns two and three give results using the fit method (or just raw data in the case of $\amuLSD$ and $\amuLW$) based on scale setting via $M_\Omega$ and $f_\pi$, respectively. The correlated difference between these values is given in column four.  The final two columns list the results corresponding to columns two and three using the bounding method. The small differences in continuum-extrapolated results between the fit and bounding methods are predominantly driven by a fluctuation in the finest $a\sim 0.06$ fm data, which is discussed further in the Supplemental Material.}
\label{tab:altResults}
\begin{tabularx}{\linewidth}{lCCCCC}
\hline
\hline
 & \multicolumn{3}{c}{Fit (or raw data)} & \multicolumn{2}{c}{Bounding} \\
Contribution & $M_\Omega$ & $f_\pi$ & $\Delta_{M_\Omega, f_\pi}$ & $M_\Omega$ & $f_\pi$ \\
\hline
$a^{ll,\,\mathrm{SD}}_{\mu}$ & 48.139(11)(91)[92] & 48.126(15)(100)[101] & 0.013(78) & --- & --- \\
$a^{ll,\,\mathrm{W}}_{\mu}$ & 206.90(14)(61)[63] & 206.94(15)(46)[49] & $-0.03(43)$ & --- & --- \\  
$a^{ll,\,\mathrm{LD}}_{\mu}$ & 400.2(2.3)(3.7)[4.3] & $396.6(2.2)(3.3)[4.0]$ &  $3.6(3.9)$ & $405.1(3.6)(4.0)[5.3]$ & $401.0(3.4)(3.6)[5.0]$ \\
$a^{ll,\,\mathrm{Full}}_{\mu}$&  $654.2(2.4)(4.7)[5.3]$ & $650.5(2.4)(4.4)[5.0]$ & $3.7(4.5)$ &$659.5(3.8)(5.3)[6.5]$ & $655.3(3.7)(5.2)[6.4]$\\ 
$a^{ll}_{\mu}$ & $655.2(2.3)(3.9)[4.5]$ & $651.7(2.2)(3.5)[4.1]$ & $3.5(4.0)$ & $660.1(3.1)(4.5)[5.4]$ & $656.0(3.1)(4.0)[5.0]$ \\
\hline
\hline
\end{tabularx}
\end{table*}

\begin{table*}
\centering
\caption{Correlation matrices for the LQC contributions to the window observables using $M_{\Omega}$ (left) and $f_{\pi}$ (right) scale setting and (where applicable) the fit method correlator-reconstruction strategy.}
\label{table:correlationsMomegafpi}
\begin{tabularx}{0.49\linewidth}{LCCCC}
\hline\hline
{}           &  $a^{ll,\,{\mathrm {SD}}}_{\mu}$ & $a^{ll,\,{\mathrm {W}}}_{\mu}$ & $a^{ll,\,{\mathrm {LD}}}_{\mu}$ &  $a^{ll,\,{\mathrm {Full}}}_{\mu}$ \\
\hline
$a^{ll,\,{\mathrm {SD}}}_{\mu}$   &  1.00 &  0.09  &  0.01  &  0.01  \\
$a^{ll,\,{\mathrm {W}}}_{\mu}$ & 0.09 & 1.00  & 0.18 & 0.18  \\
$a^{ll,\,{\mathrm {LD}}}_{\mu}$ & 0.01  & 0.18 & 1.00 &   0.33  \\
$a^{ll,\,{\mathrm {Full}}}_{\mu}$ & 0.01 & 0.18 & 0.33 & 1.00 \\
\hline
\hline
\end{tabularx}
\hfill
\begin{tabularx}{0.49\linewidth}{LCCCC}
\hline\hline
{}           &  $a^{ll,\,{\mathrm {SD}}}_{\mu}$ & $a^{ll,\,{\mathrm {W}}}_{\mu}$ & $a^{ll,\,{\mathrm {LD}}}_{\mu}$ &  $a^{ll,\,{\mathrm {Full}}}_{\mu}$ \\[-0.82em]
\hline
$a^{ll,\,{\mathrm {SD}}}_{\mu}$    &   1.00    & 0.19   & 0.02   & 0.02\\
$a^{ll,\,{\mathrm {W}}}_{\mu}$     &   0.19    & 1.00     & 0.16   & 0.15\\
$a^{ll,\,{\mathrm {LD}}}_{\mu}$    &   0.02     & 0.16   & 1.00      & 0.31\\
$a^{ll,\,{\mathrm {Full}}}_{\mu}$  &   0.02     & 0.15   & 0.31     & 1.00\\       
\hline
\hline
\end{tabularx}
\end{table*}

\begin{table*}
\centering
\caption{Approximate absolute error budgets (in units of $10^{-10}$) for the uncertainties reported in \cref{tab:altResults} for the $f_\pi$ results using the fit method. From left to right, the contributions to the error are Monte Carlo statistics and $t_{\min}$ variation in the correlator fits (where applicable), continuum extrapolation and TB corrections, FV and $M_{\pi}$-mistuning corrections, scale setting (due to the uncertainties in $af_\pi$ and the input $f_\pi$ value in physical units) , and current renormalization.}
\label{table:IndividualUncertaintyPi}
\vspace{1mm}
\begin{tabularx}{\linewidth}{LCLLLLll}
\hline \hline
 Contribution &  statistics, $t_{\min}$ & $a\to0$, $\Delta_{\mathrm{TB}}$   &  $\Delta_{\textrm{FV}}$, $\Delta_{M_{\pi}}$ &   $a$  & $Z_V$ &  Total\\ 
\hline
$a^{ll,\,{\mathrm {SD}}}_{\mu}$ & 0.015 & 0.072 & --- & 0.009 & 0.069 & 0.101 \\
$a^{ll,\,{\mathrm {W}}}_{\mu}$ & 0.15 & 0.21 & 0.34 & 0.14 & 0.19 & 0.49 \\
$a^{ll,\,{\mathrm {LD}}}_{\mu}$ & 2.2 & 2.7 & 1.4 & 1.4 & 0.2 & 4.0 \\
$a^{ll}_{\mu}$ & 2.2 & 2.7 & 1.7 & 1.4 & 0.3 & 4.1\\
\hline \hline
\end{tabularx}
\end{table*}

\clearpage
\newpage

\onecolumngrid
\include{main_sup}

\end{document}

%% file: main_sup.tex
\begin{center}
  {\LARGE\bfseries Supplementary Materials}\\[1ex]
\end{center}

\section{Additional fit method results}

\begin{figure*}
\centering
\includegraphics[scale=0.9]{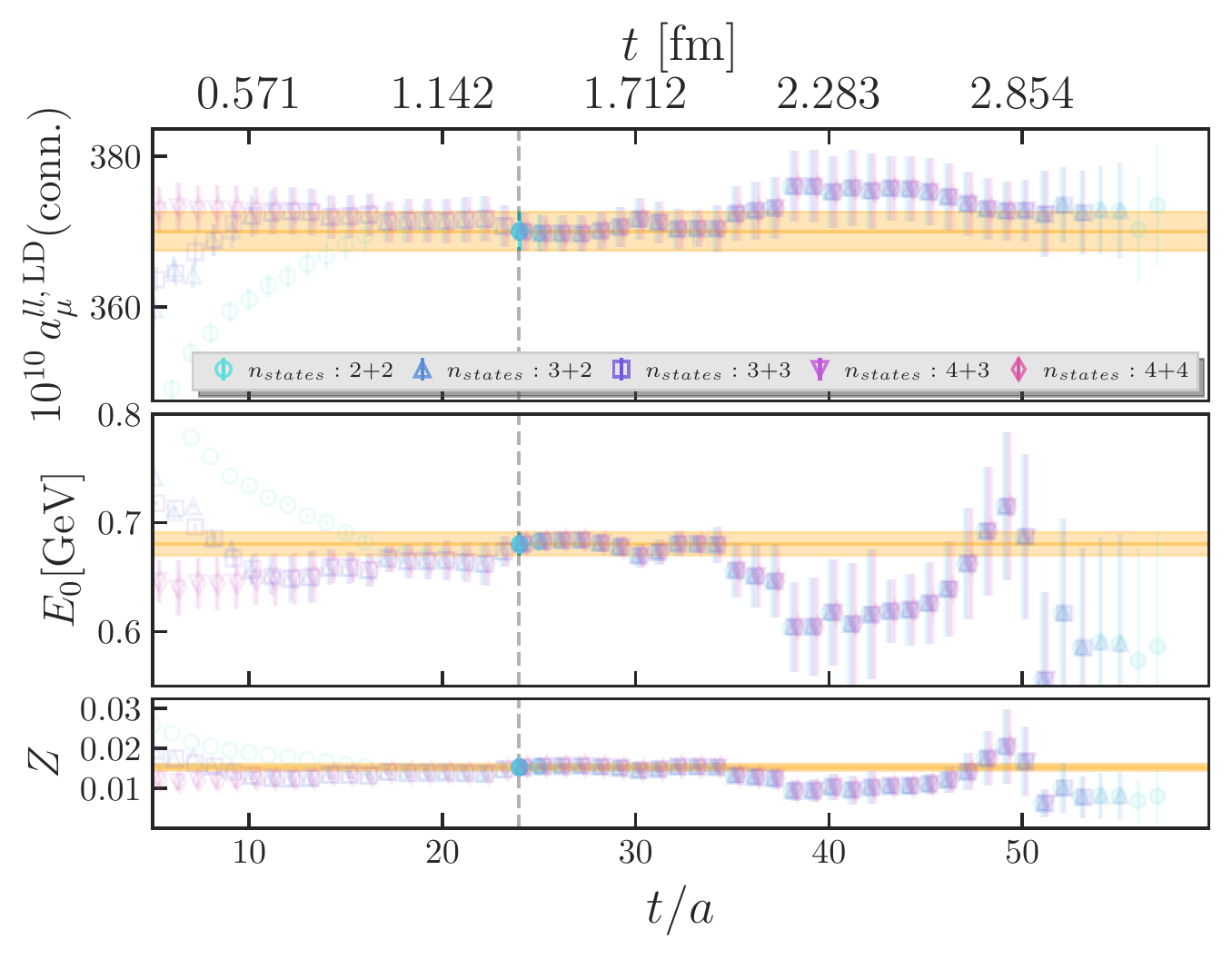}
\vspace{-5mm}
\caption{Fit stability of $a_\mu$ (top panel), the ground state energy (second panel) and amplitude (third panel) on the 0.06~fm ensemble using the local vector current operator. We perform a BMA over the set of points corresponding to $2+2$ states for all $t_\textrm{min}$ after the dashed line. The model probabilities for the BMA procedure are given in the bottom panel. The BMA result is given by the yellow band.}
\label{fig:fit06}
\end{figure*}

 \Cref{fig:fit06} shows individual fit results for the ground state parameters as a function of $t_{\textrm{min}}/a$, compared with the corresponding Bayesian model averaging (BMA) results. The allowed range of $t_{\textrm{min}}$ is $[t_\textrm{min:min}, t_\textrm{min:max}]$, where $t_{\textrm{min:min}}$ is selected as the first time slice in the plateau in the ground-state fit parameters, and $t_\textrm{min:max}$ is obtained by requiring that at least nine data points are present to ensure that there are enough degrees of freedom to determine all fit parameters. To illustrate the fit reconstruction fidelity, \cref{fig:DeltaReconn06} provides an example of single fit reconstructions on the 0.06~fm ensemble using the local vector current operator with $t_{\min}=t_{\mathrm{min:min}}$ and $n_{\mathrm{states}} = 4+2,3+2,2+2$, the best fits in these cases as determined by the BAIC, compared with the raw correlator data. We see that $C_{\mathrm{fit}}(t)$ (defined in Eq.~(4) of the main text) describes the data very well in the fit region, particularly for $t < t^\star$, beyond which the data rapidly becomes noisy. As expected, $C_{\mathrm{fit}}(t)$ loses fidelity as $t\to 0$. At $t> t_{\rm max}$, the correlator data exhibit large statistical fluctuations, which are examined in more detail in \cref{fig:bound06,fig:boundEff06}. Choices of fit hyperparameters and results for ground-state energies are tabulated in \cref{table:NRparams}. Note the convergence of $E_{0,\textrm{fit}}$ between the two different currents seen as $a \rightarrow 0$, reflecting the vanishing of TB effects as the continuum limit is approached.

\begin{figure*}
\centering
\includegraphics[width=0.8\textwidth]{l96192f211b672m0008m022m260LocalCompareDataRecon.pdf}
\vspace{-5mm}
\caption{Comparison of $C_{\mathrm{fit}}(t)$ for  $t_{\min}=t_{\mathrm{min:min}}$ with the raw correlator data $C_{\mathrm{data}}(t)$ on the 0.06~fm ensemble using the local vector current operator.  The gray band corresponds to the range $[t_{\mathrm{min:min}},t_{\mathrm{min:max}}]$, the black dotted line to $t_{\max}$, and the black dashed line to $t^*$ (note that $t^*<t_{\max}$ in this example).  Marker colors and symbols correspond to the number of states used in the fit, among which there is little distinction for $t\geq t_{\mathrm{min:min}}$.  The black crosses correspond to the raw correlator data.
\label{fig:DeltaReconn06}
}
\end{figure*}

\begin{table*}[t]
\centering
\caption{Description of operator measurements, hyperparameters, and ground-state energies used in the two correlator-reconstruction approaches employed in this work. The range of data points used to compute the bounding method result is from $t_c$ to $t_\textrm{max}$.}
\label{table:NRparams}
\begin{tabularx}{\linewidth}{lCCCCCCC}
\hline \hline
$\approx a/\mathrm{fm}$ &  $J$ op. & $N_{\mathrm{conf}}$ & $t_\textrm{min}/\mathrm{fm}$ range & $t_\textrm{max}/\mathrm{fm}$ & $t^\star/\mathrm{fm}$ & $E_{0, \textrm{fit}}/\mathrm{GeV}$ &  $t_\textrm{c}/\mathrm{fm}$ \\ 
\hline
\multirow{2}{*}{$0.15$}  
 & local  & 957 & [1.1, 1.7]  & 3.0  & 2.7 & 0.720(14) & 2.8  \\
 & one-link & 957 & [0.9, 1.9]  & 3.1  & 2.3 & 0.7560(85) & 2.7  \\
\multirow{2}{*}{$0.12$}
& local & 1060  & [1.4, 2.4] & 3.3 & 3.2 & 0.685(27) & 3.0 \\
& one-link & 1060 & [1.2, 2.4] & 3.3 & 2.7 & 0.715(12) & 3.0 \\
\multirow{2}{*}{$0.09$}
& local& 993   & [1.3, 2.8] & 3.5 & 2.8 & 0.692(13) & 3.0 \\
& one-link& 993  & [1.2, 2.7] & 3.4 & 2.8 & 0.704(12) & 3.1 \\
\multirow{2}{*}{$0.06$}
& local& 1009   & [1.4, 3.1] & 3.6 & 2.5 & 0.684(13) & 3.1 \\
& one-link& 900  & [1.4, 3.1] & 3.6 & 2.6 & 0.685(12) & 3.1 \\
\hline
\hline
\end{tabularx}
\end{table*}

\section{Bounding method}

As an alternative to fitting the data, one can use an \emph{Ansatz} for replacing data in the region where a single exponential dominates the spectral decomposition of Eq. (4) in the main text. This enables bounds to be imposed on $C(t)$ from observing that the true ground state in $C(t)$ is an interacting two-pion state with vector quantum numbers, which has lower energy than the mass of the $\rho$ meson.  A lower bound on this energy (and thus an \emph{upper} bound on the exponentially decaying $C(t)$) comes from $E_{\pi \pi, \textrm{free}} = 2\sqrt{(2\pi/L)^2 + M_\pi^2}$, the energy of two P-wave non-interacting pions in a finite volume.\footnote{In principle, one should instead use the interacting energy; however, the approximate use of the free energy here is well within the current precision of $a_{\mu}^{ll}(\mathrm{conn}.)$.  See Refs.~\cite{Lahert:2024vvu,Budapest-Marseille-Wuppertal:2017okr} for further discussion.}  An upper bound on the true ground-state energy (giving a \emph{lower} bound on $C(t)$) is provided by the ground-state energy $E_{0,\textrm{fit}}$ taken from the fit method and corresponding BMA over $t_{\min}$. As can be seen in the middle panels of \cref{fig:boundEff06}, this energy coincides with effective mass of the correlator in the region where it is well determined. The bounding functions are defined from the correlator data point $C(t_c)$ by the relations
\begin{align}
C_\textrm{lower}(t) e^{-E_{0,\textrm{fit}} t_c} &= C(t_c) e^{-E_{0,\textrm{fit}} t}, \\
C_\textrm{upper}(t) C_{\pi \pi}^{0}(t_c, T) &= C(t_c) e^{-E_{\pi \pi} t},
\end{align}
where, for the upper bound, $C^0_{\pi \pi}(t_c, T) \equiv e^{-E_{\pi \pi}t_c} + e^{-E_{\pi \pi}(T-t_c)} + 2e^{-E_{\pi \pi}T/2}$ includes the leading corrections from the finite temporal extent $T$.

For both bounds, a time value $t_c$ is selected such that the data are replaced with the bound for $t \geq t_c$. The value of $t_c$ is determined for each $C(t)$ as the point where the upper and lower bounds meet, which we define as the time $t_c$ where the absolute differences between upper and lower bound values at $t_c/a$ and $t_c/a + 1$ are less than the standard deviation of the correlated average $(C_\textrm{upper} + C_\textrm{lower})/2$ at each of the two points.  The resulting $t_c$ values are tabulated in \cref{table:NRparams}. As for the fit method, finite-time corrections for $t<t_c$ are found to be negligible from NLO $\chi$PT \cite{Borsanyi:2020mff}.

\begin{figure*}
\centering
\includegraphics[height=0.28\textwidth]{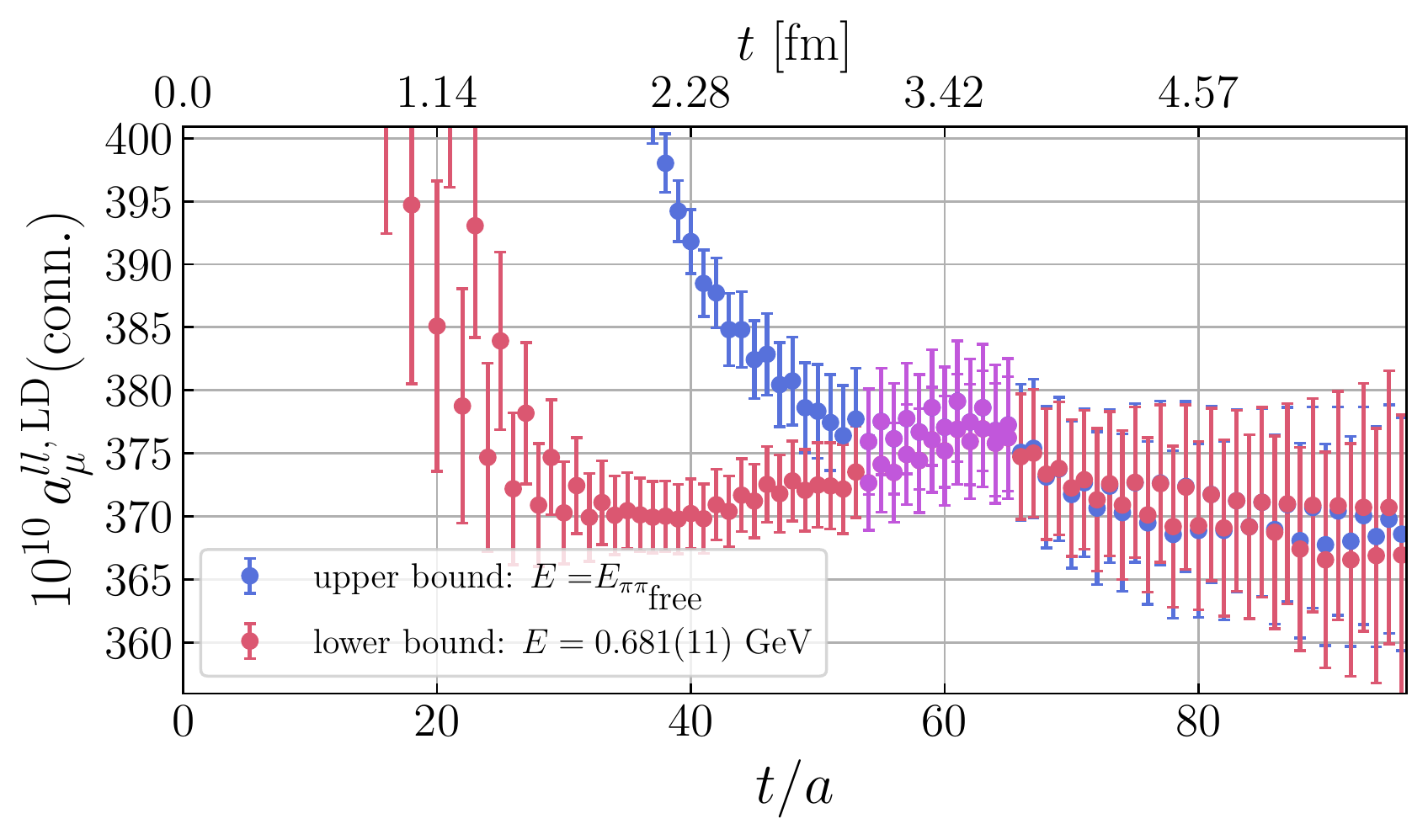}
\hspace{2mm}\includegraphics[height=0.28\textwidth]{l6496f211b630m001326m03636m4313LMIbounding.pdf}
\vspace{-5mm}
\caption{The bounding method applied to the local vector current on the 0.06~fm ensemble (left) and 0.09~fm ensemble (right). The range of data that are averaged to obtain the resultant $a_\mu$, given by $[t_c,t_{\max}]$, is shown in purple.
}
\label{fig:bound06}
\end{figure*}

Once $t_c$ is determined, the bound is used to replace the data for $t\geq t_c$.  To mitigate the effect of statistical fluctuations, this replacement is done by averaging over the (even) range of points from $t_c$ to $t_{\max}$, inclusively, where $t_{\textrm{max}}$ is the same as the value used in the fit method above; if the number of time slices in this range is odd, the value of $t_c/a$ is increased by $1$.  In addition, a systematic error is added equal to the difference in central values from repeating this procedure on the left and right (even, {\it i.e}, possibly overlapping) halves of the full range from $t_c$ to $t_\textrm{max}$ \cite{Budapest-Marseille-Wuppertal:2017okr}. Examples of results from this procedure for the local current on the 0.06~fm (left) and 0.09~fm (right) ensembles are shown in \cref{fig:bound06}.

\begin{figure}
\centering
\includegraphics[width=0.7\linewidth]{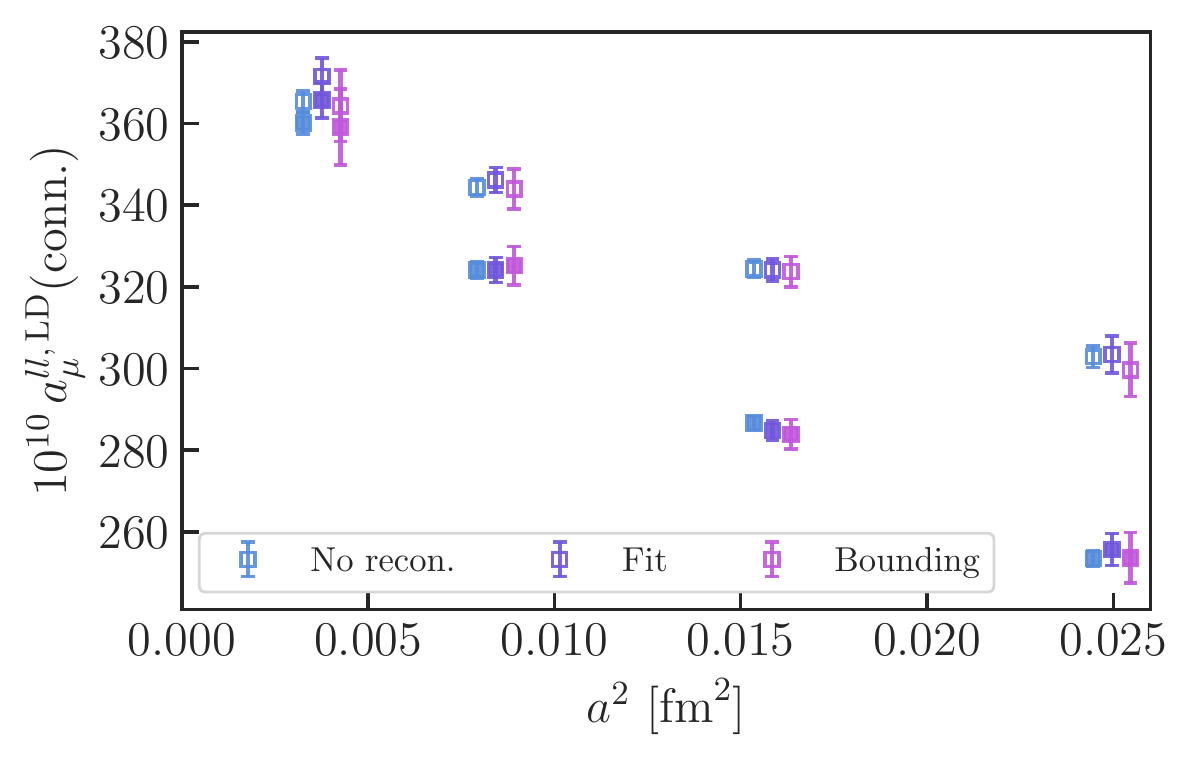}
\vspace{-5mm}
 \caption{Comparison of $\amuLLD$ obtained from the two correlator-reconstruction approaches, the fit method (blue) and the bounding method (magenta), for the ensembles and currents used in this work. Also included is the result of using no correlator-reconstruction strategy (``No recon.,'' yellow). The unfilled symbols correspond to the local current and the filled to the one-link.}
\label{fig:NRcompare}
\end{figure}

\begin{figure*}
\centering
\includegraphics[width=\textwidth]{LocalMEff.pdf} 
\vspace{-5mm}
\caption{Further analysis of the bounding method efficacy for the 0.06~fm (left panels) and 0.09~fm (right panels) ensembles using the local vector current operator.
\textit{Top panels}: Raw correlator data (gray points) compared with a single-exponential decay (green dashed curve) where, for purposes of illustration, the latter is given by the ground state fit parameters.
\textit{Middle left/bottom right panels}:  Effective mass stability compared with fit method ground state $E_{0,\mathrm{fit}}$ (green dashed line) and the energy lower bound $E_{\pi\pi}\approx E_{\pi\pi,\mathrm{free}}$ using two definitions, $M_{\mathrm{eff}}^{\mathrm{arccosh}}(t)$ (blue points) and $M_{\mathrm{eff}}^{\mathrm{log}}(t)$ (red points).   Due to noise, $M_{\mathrm{eff}}^{\mathrm{arccosh}}(t)$ is not defined for $t/a\in[61,63]$ in the local 0.06~fm data.
\textit{Bottom left panel}: Upper bound on $\amuLLD$ versus $t_c$ for two choices of energy lower bound: the theoretical lower bound $E=E_{\pi\pi}\approx E_{\pi\pi,\mathrm{free}}$ (solid curves) corresponding to the blue points in the left plot of \cref{fig:bound06}, and approximately 60\% of said theoretical lower bound (dashed curves with small rightward offset).   
Colors distinguish even (blue) and odd (green) times.
In all panels, the purple band corresponds to the $t$ range where the correlator upper and lower bounds overlap (cf. \cref{fig:bound06}).
}
\label{fig:boundEff06}
\end{figure*}

\section{Tests of the bounding and fit method}

\subsection{Analysis of ground-state stability}

We compare the results of the two correlator-reconstruction approaches in \cref{fig:NRcompare}. We find close agreement between the two approaches in most cases. However, on our finest ensemble ($a \approx$ 0.06 fm), we observe that the bounding method approach does not handle significant correlated fluctuations in late time slices, which result in the tail of the correlator not exhibiting single-exponential behavior. \Cref{fig:boundEff06} examines the statistical fluctuation that leads to this tension. In particular, the two panels compare the correlators for 0.06~fm and 0.09~fm. In the case of 0.06~fm, we observe the deviation from the single-exponential \emph{Ansatz} in the region of bounding (purple band). This fluctuation can be studied in more detail by examining the effective mass in this region (middle panels). We employ two definitions of the effective mass, a finite- and infinite-time version, both taking into account the effects of staggering: 
\begin{align}
M_{\mathrm{eff}}^{\mathrm{arccosh}}(t) &\equiv\frac{1}{2}\mathrm{arccosh\left[\frac{C(t-2)+C(t+2)}{2C(t)}\right]}, \\
M_{\mathrm{eff}}^{\mathrm{log}}(t) & \equiv-\frac{1}{2}\mathrm{ln\left[\frac{C(t+2)}{C(t)}\right]}.
\end{align}
The first accounts for periodic boundary conditions when $t\approx T$, whereas the second is more stable for $t\ll T$.
For 0.09~fm, we observe an effective mass that is consistent with $E_{0,\textrm{fit}}$  for all times. On the other hand, for 0.06~fm we observe the effective mass falling to an unphysical energy below $E_{\pi\pi,\mathrm{free}}$ before rising back up above $E_{0,\textrm{fit}}$. We take this as evidence that this is a statistical fluctuation mimicking a spurious state in our data. To further illustrate this point, the bottom left panel of \cref{fig:boundEff06} shows the upper bound for the 0.06~fm correlator for two choices for energy lower bound. With $E_{\pi\pi,\mathrm{free}}$, the correlator bound is not monotonic.  For the correlator upper bound to fall monotonically (up to oscillations from staggering), the bounding energy must be approximately $60\%$ of $E_{\pi\pi,\mathrm{free}}$, which is well below any possible physical state.

\subsection{Fake-data tests}

To further test the two approaches, we examine both the fit and bounding methods using lattice correlators that are constructed from EFT-based models. For this purpose we use the CM and MLLGS schemes described in Ref.~\cite{FermilabLattice:2024yho}, which are also used to obtain lattice corrections for finite volume, taste-breaking and pion-mass mistuning. These models are understood to reproduce the low-energy two-pion spectrum in the long-distance tail region where the noise-reduction strategies are applied.

\begin{figure}
\centering
\includegraphics[width=0.9\linewidth]{modelSpectruml96192f211b672m0008m022m260LMI.pdf}
\vspace{-5mm}
 \caption{
 \textit{Top panel}: Finite-volume, taste-broken spectra (energies $E_i$ and amplitudes $Z_i$) for the 0.06~fm HISQ ensemble predicted by CM (green) and MLLGS (orange). Two-state fit reconstruction to this data is shown in pink (CM) and gray (MLLGS). \textit{Bottom Panel}: The resulting contribution to $\amuLLD$ from the spectra in the top panel.
 }
\label{fig:fakeDataSpectra}
\end{figure}

The MLLGS scheme allows us to extract this information directly as in Eqs.~(120)--(124) of Ref.~\cite{Borsanyi:2020mff}. For the CM scheme, this information can be extracted by numerically determining the poles and residues of Eq.~(B33) in Ref.~\cite{Chakraborty:2016mwy}. The resulting spectrum (energies and amplitudes) for the 0.06~fm local current from CM (green) and MLLGS (orange) are shown in the top panel of \cref{fig:fakeDataSpectra}. We observe good consistency between the two spectra, demonstrating that both schemes describe the same two-pion physics.
The small differences in the fine structure are due to differences in the implementation of taste-breaking effects in the two models. The model means for the lattice correlation function $\overline{C(t)}$ are then obtained from the model spectra via Eq.~(4) of the main text in the infinite-$T$ limit. For the oscillating contribution, we use a single $h_1$ state, which is the expected opposite-parity partner ground state. The energy for this state is taken from the PDG \cite{ParticleDataGroup:2022pth}, and the amplitude is taken to be in line with the fit result to the true lattice correlator across all lattice spacings. Using the data covariance matrix $\Sigma_{\mathrm{data}}$, a multivariate normal distribution is constructed,
\begin{align}
    \operatorname{pr} \left( C(t) \right) \propto \exp \left[-\frac{1}{2N_{\mathrm{conf}}} \sum_{t, t^\prime} \left(C(t) - \overline{C(t)}\right) \left(\Sigma^{-1}_{\mathrm{data}}\right)_{t, t^\prime}   \left(C(t^\prime) - \overline{C(t^\prime)}\right) \right],\label{eq:fakeDataDist}
\end{align}
for each lattice spacing and current, and both models. $N_{\mathrm{conf}}$ is the number of configurations associated with the original datasets, as given in \cref{table:NRparams}.
\begin{figure}
\centering
\includegraphics[width=0.9\linewidth]{modelTestCompareTruth135353.pdf}
\vspace{-5mm}
 \caption{
 Difference in $\amuLLD$ between reconstructions and ``truth" for the fake model data set for the local current, based on CM (top panel) and MLLGS (bottom panel). Compared are fit-method and bounding-method reconstructions with different fake sample sizes drawn from \cref{eq:fakeDataDist}.} 
\label{fig:fakeDataTest}
\end{figure}

Sampling correlators $C(t)$ from this distribution provides the datasets on which the tests of the bounding and fit method are performed. For both methods, we use the same hyperparameters as in our analyses of the corresponding correlator data and then apply them 
to fake data sets of size: $N_{\mathrm{conf}}$, $2N_{\mathrm{conf}}$, $4N_{\mathrm{conf}}$ and  $8N_{\mathrm{conf}}$  and evaluate how the methods approach the ``truth'', {\it i.e.}, $\amuLLD$ computed from the true means $\overline{C(t)}$. The results of these tests are shown in \cref{fig:fakeDataTest}. We observe that both methods describe the truth quite well, with improved accuracy as the sample sizes are increased. We repeat this test multiple times using different random seeds to draw the data from \cref{eq:fakeDataDist} and also for the one-link current, and observe consistency across all tests. 

In some cases, due to large, correlated statistical fluctuations in the tail in the sampled fake data, as in the top left of \cref{fig:boundEff06}, there are occasional small deviations at the one $\sigma$ level with both methods, particularly for the sample of size $N_{\mathrm{conf}}$, which is not unexpected. However, as illustrated in \cref{fig:fakeDataTest}, the results from the bounding method, more often than those from the fit method, show deviations outside one $\sigma$ level. Hence we find that the bounding method appears less robust than the fit method under such fluctuations. For the fit method, we find that the differences with the truth are generally $\lesssim 1\sigma$ for all sample sizes, and when sample sizes are $\geq 2N_{\rm conf}$, we find that central value differences are $< 2\times 10^{-10}$. In summary, our fake data analysis shows that the fit method, as employed in our analysis, provides a robust estimate of the noisy tail and hence a reliable determination of $a_\mu$ from our correlator data and we adopt it for our main result.

\section{Comparison with our result from 2019}

As noted in the main text, the results of the present analysis differ from our previous determination in Ref.~\cite{FermilabLattice:2019ugu} by $1.4\sigma$. Here we elucidate this tension.

A comparison of $\amuLFull$ for each lattice spacing using the local vector current operator between the two analyses is given in \cref{fig:2019compare}.  Each set of results uses the correlator reconstruction strategy of each respective work.  Corrections for lattice artifacts (FV, $M_{\pi}$-mistuning, and TB) are not included (are included) in the bottom (top) sets of data points.
Uncertainties in \cref{fig:2019compare} correspond to statistical errors ({\it i.e.}, without parametric uncertainty from scale setting and current renormalization), except for the 0.06~fm ensemble from Ref.~\cite{FermilabLattice:2019ugu} where statistical uncertainty dominates. The statistical correlation coefficients 
between the two data sets are $\rho < 0.05$, {\it i.e.} negligible. The reason is that the 2025 data set is generated with exact low modes and uses different random-wall sources. The statistical errors in $\amu$ are largely driven by the noise in the correlators at large times, which is where the two methods differ. For the comparison at $a \approx 0.15$~fm, we include the data on the same ensemble from  Ref.~\cite{FermilabLattice:2019ugu} as in the present work.

To sharpen the comparison, we make several modifications to the 2025 data set in \cref{fig:2019compare}. We replace the data set at 0.09 fm obtained on a newer, better tuned ensemble, with an LMA data set on the same ensemble as used in Ref.~\cite{FermilabLattice:2019ugu}, so that the two data sets are on the same ensembles at 0.06, 0.09, and 0.15~fm. We don't have an LMA data on the 0.12~fm ensemble used in \cite{FermilabLattice:2019ugu}. Furthermore, the comparison is performed in the gradient-flow scale $w_0$ scale \cite{BMW:2012hcm} used in Ref.~\cite{FermilabLattice:2019ugu}, which was determined in Ref.~\cite{Dowdall:2013rya} using $f_{\pi}$ to fix the overall scale; values for the relative scale $w_0/a$ are taken from Ref.~\cite{FermilabLattice:2019ugu}, except for the newer 0.12~fm ensemble for which we use $w_0/a=1.41060(28)$ from arXiv v1 of Ref.~\cite{Bazavov:2024dov}. Additionally, we use the current renormalization factors from Ref.~\cite{FermilabLattice:2019ugu}, which come from the form-factor method of Ref.~\cite{Chakraborty:2017hry}.  Lastly, for the corrected data shown in \cref{fig:2019compare}, we employ the CM from Ref.~\cite{FermilabLattice:2019ugu} to correct for lattice artifacts (FV, $M_{\pi}$-mistuning, and TB). 

\begin{figure*}
\centering
\includegraphics[width=0.48\linewidth]{2019Comparison.pdf}
\hspace{2mm}\includegraphics[width=0.48\linewidth]{2019ComparisonIntegrand.pdf}
\caption{\textit{Left plot}: Comparison of $\amuLFull$ for each lattice spacing using the local vector current operator between the analyses of this work (blue) and Ref.~\cite{FermilabLattice:2019ugu} (red) as described in the text.  Circles are not corrected for lattice artifacts, whereas squares include FV, $M_{\pi}$-mistuning, and TB corrections using CM from Ref.~\cite{FermilabLattice:2019ugu}.
\textit{Right plot}: Comparison of integrands from Eqn. (1) of the main text using LMA (blue points) and TSM (red points) using the local vector current operator on the 0.06~fm ensemble.
 }
\label{fig:2019compare}
\end{figure*}

As shown in \cref{fig:2019compare} (left), the two data sets are consistent at $\leq 1 \sigma$ on the shared ensembles (at 0.06, 0.09, and 0.15~fm). However, the 2025 data points lie consistently above the 2019 ones. As illustrated in \cref{fig:2019compare} (right), at least part of the effect appears to be due to the noisy tail in the TSM data. The large noise in the TSM data at large times also impacts the resolution of the low-energy spectrum in the fit method, which could further contribute to the difference. This statistical effect is the dominant source of the tension between Ref.~\cite{FermilabLattice:2019ugu} and the present analysis. However, as can be seen in \cref{fig:2019compare} (left), the  improved statistics in the 2025 data set increases the sensitivity to higher-order terms in the continuum extrapolations. We note that the 2025 data set uses different (better-tuned) ensembles at two lattice spacings and includes a second current discretization. In addition, the analysis strategy in our current work includes changes, such as new scale setting and current renormalizations, the inclusions of a broad range of correction schemes, and a systematic error analysis based on BMA. While all these changes yield a much more robust analysis of the continuum limit than was possible in Ref.~\cite{FermilabLattice:2019ugu}, they also all appear to further increase the difference from $\lesssim 1 \sigma$ to $1.4\sigma$.

\section{Lattice correction results}

\begin{figure*}
\centering
\includegraphics[height=0.55\linewidth]{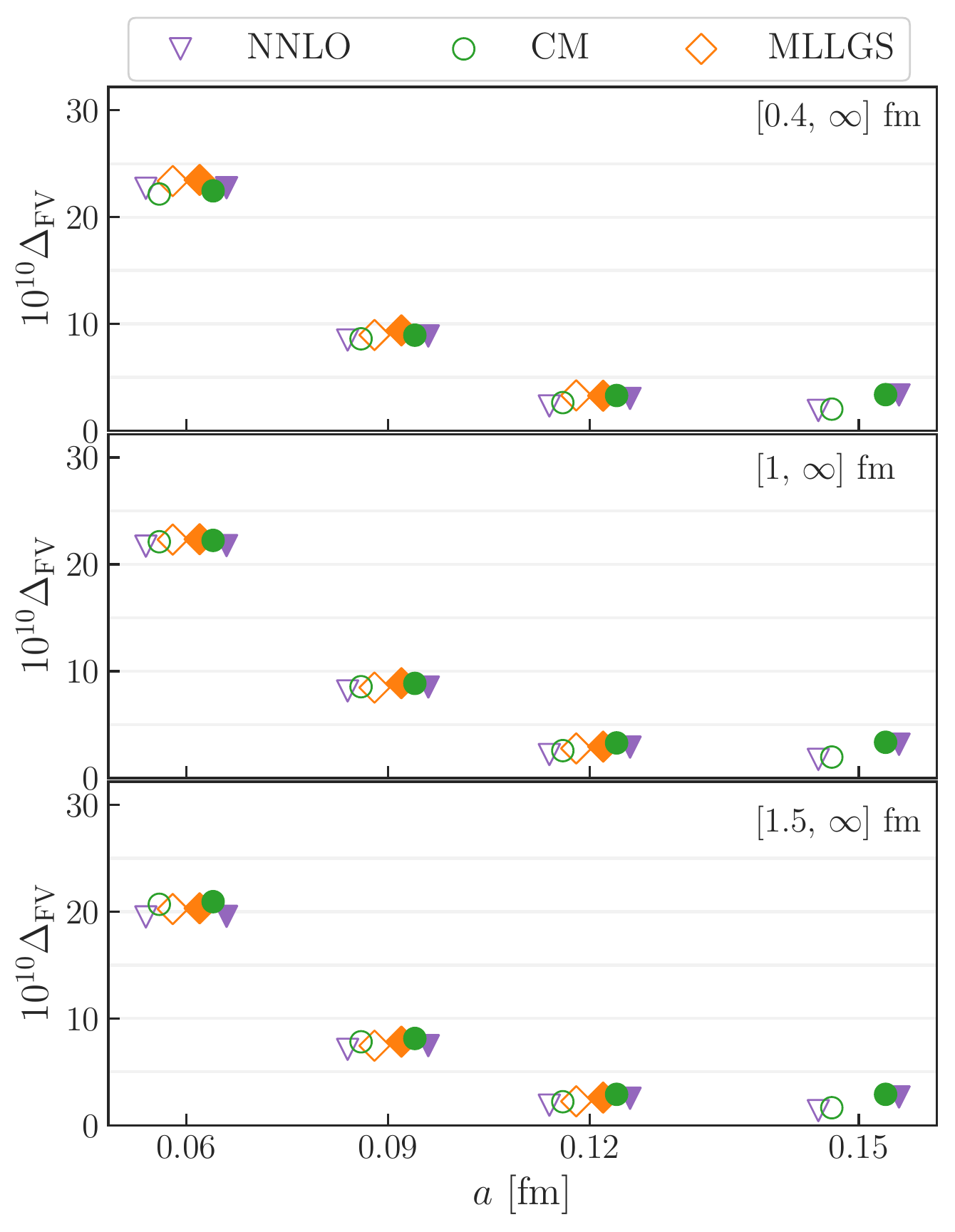}\hspace{5mm}\includegraphics[height=0.55\linewidth]{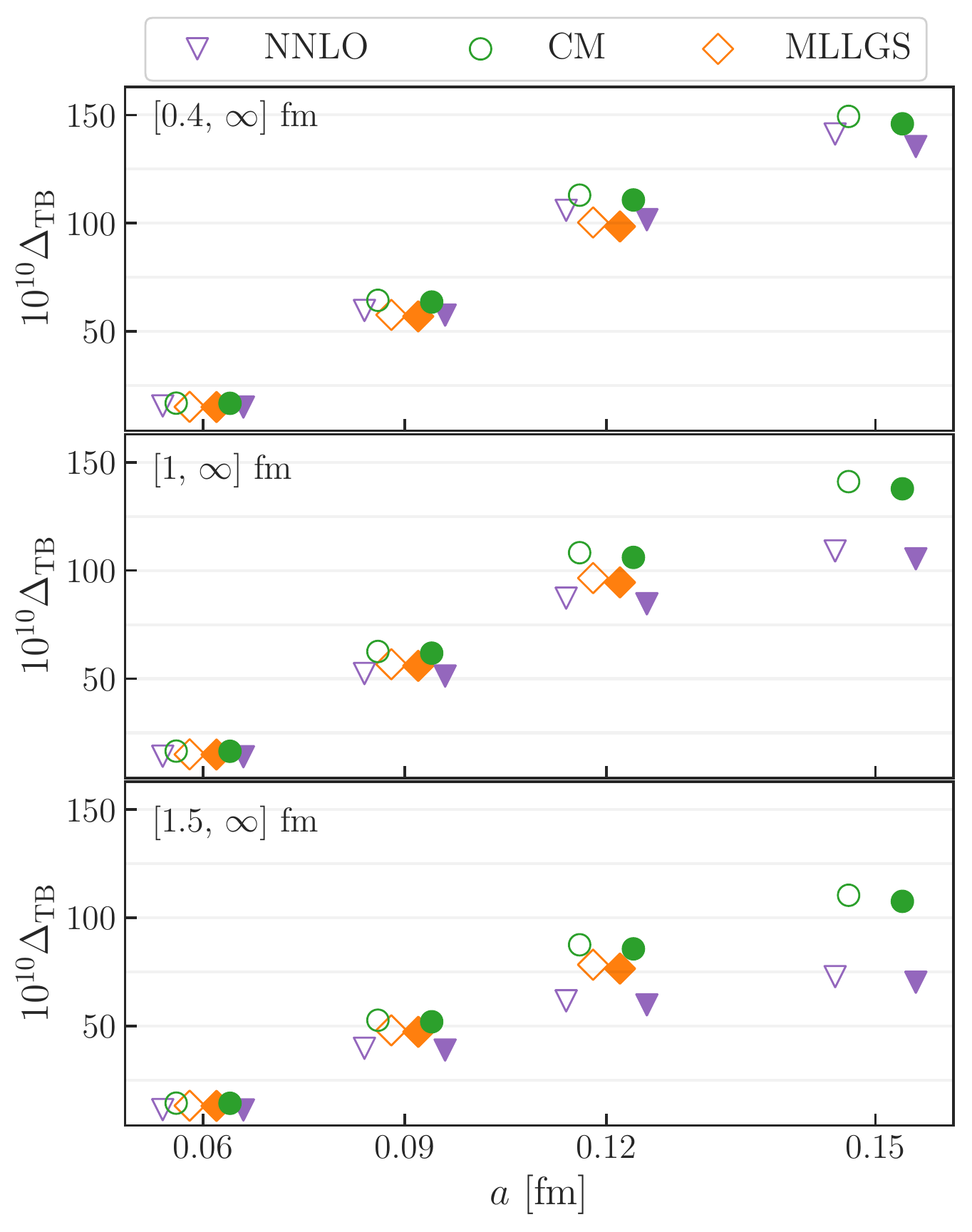}
\vspace{-5mm}
\caption{FV (left) and TB (right) corrections for each ensemble in this work using $M_{\Omega}$ scale setting. The unfilled symbols correspond to the local current and the filled to the one-link.}\label{fig:FVCorr}
\end{figure*}

Results for FV and TB corrections using the $M_{\Omega}$ scale are shown in \cref{fig:FVCorr}. $M_{\pi}$-mistuning corrections are small due to the accurate tuning of the ensemble masses and hence are not shown. We consider three sub-windows of Eq.~(3) in the main text from which we compute the corrections, where the boundaries are given by $t_1=0.4,1,1.5$ fm. These choices of $t_1$ are motivated by boundaries of the W, W2 and LD windows and the regions of validity of the various EFT-based correction schemes \cite{Aubin:2022hgm}. For a more detailed discussion of this approach see Sec.~X and Sec.~III.C of Refs.~\cite{Aubin:2022hgm,FermilabLatticeHPQCD:2023jof}. \Cref{fig:FVCorr} shows that the expected dependencies on the lattice spacing, lattice volume, pion mass, and correction region are largely followed.

\section{Details of BMA scheme}

We use BMA to estimate systematic uncertainties associated with the various analysis choices that are included.  We perform model averages at two stages in the analysis: over $t_{\min}$ variations to obtain correlator reconstructions, and then over the continuum extrapolation and lattice correction models.  The mean and covariance from the BMA are given by
\begin{align}
\left\langle A\right\rangle =& \sum_{n=1}^{N_{M}}\left\langle A\right\rangle_{n} \pr\left(M_{n} \mid D\right), \label{eq:BMAMean} \\
\cov[A,B]=& \sum_{n=1}^{N_{M}}\cov_n[A,B]\pr\left(M_{n} \mid D\right)\notag\\
    &+\sum_{n=1}^{N_{M}}\left\langle A\right\rangle_{n}\left\langle B\right\rangle_{n} \pr\left(M_{n} \mid D\right)-\left\langle A\right\rangle\left\langle B\right\rangle, \label{eq:BMACov} 
\end{align}
where $A$ and $B$ are each functions of the parameters of each model $M_n$, $N_M$ is the number of models, and the probability of model $M$ given the data $D$ is defined through the BAIC weight~\cite{Neil:2022joj},
\begin{equation}
    \pr(M \mid D) \equiv \pr(M) \exp \left[-\frac{1}{2}\left(\chi_{\rm data}^{2}\left(\mathbf{a}^{\star}\right)+2 k+2 N_{\mathrm{cut}}\right)\right]. \label{eq:modelProb}
\end{equation}
$\chi_{\rm data}^{2}$ is the standard chi-squared function not including the priors, $\mathbf{a}^{\star}$ is the posterior mode ({\it i.e.}, the set of parameters that minimizes the augmented chi-squared function \cite{Lepage:2001ym}), $k$ is the number of parameters in each model $M$, and $N_{\mathrm{cut}}$ is the number of data points cut from the dataset ($t_{\min}$ for correlator fits and the number of omitted ensembles for each current for continuum extrapolations).  The factor $\pr(M)$ is the prior probability of a given $M$, which we take to be uniform for correlator fits and as in Ref.~\cite{FermilabLattice:2024yho} for continuum extrapolations.

The probability weights defined in \cref{eq:modelProb} can be used to assess the relative weight of specific analysis choices in the BMA. Comparison of these weights can identify whether a particular correction scheme or fit-function variation is preferred or suppressed by the averaging procedure. This is achieved by computing the ``subset probability'' of the model subset $S$ by the relative posterior probability of the variations contained in $S$:
\begin{equation}
    \pr(S | D) = \sum_{M_i \in S} \pr(M_i | D). \label{eqn:ssProb}
\end{equation}
The subset probability encapsulates the relative weight of the models in a given subset compared with the whole model space as informed by the data (see, {\it e.g.}, the pie charts in \cref{fig:BMALDFitOmega}).

The first line of \cref{eq:BMACov} gives the statistical and parametric contributions to the covariance, and the second line gives the systematic contribution from the spread of the different model variations in the average; we refer to the latter as ``systematic covariance'' here.  In principle, the exact treatment of the systematic covariance requires including every combination of analysis choices through every stage of the analysis, {\it e.g.}, every $t_{\min}$ variation for every lattice spacing and current and every continuum extrapolation model.  The number of such combinations is computationally infeasible.  For the present analysis, we neglect the systematic covariances between the correlator fits on each lattice spacing and current; this assumption is equivalent to assuming the choice of $t_{\min}$ is independent between each lattice spacing and current.  We also neglect this systematic covariance between the choices of $t_{\min}$ and the different continuum extrapolation models. These assumptions are further justified because the size of the systematic variance (and therefore any neglected covariance) due to variation of $t_{\rm min}$ is small relative to other sources of error in our analysis. Therefore, we are able to perform independent model averages and compose their results in series.  For emphasis, the systematic contributions to the variances after each model average are not neglected, merely the systematic covariance between separate model averages.

In order to propagate statistical and parametric correlations though the series of model averages, it is useful to consider \cref{eq:BMACov} in the case where one of the arguments is constant across the model space, {\it e.g.}, $\left\langle B\right\rangle_{n}=\left\langle B\right\rangle$.  This will be true for scale-setting parameters, current renormalization factors, raw correlator data, and the results of other systematically-independent model averages.  In the example above, it follows that 
\begin{align}
\cov[A,\overline{B}
]=& \sum_{n=1}^{N_{M}}\cov_n[A,\overline{B}]\pr\left(M_{n} \mid D\right),\label{eq:BMACovConst} 
\end{align}
where the bar denotes the parameter that is constant over the models $M_n$.  A similar result is derived in Ref.~\cite{FermilabLattice:2024yho}.

The above outlines the procedure used to obtain statistical and parametric covariances for continuum values of $\amu^{ll}(\mathrm{conn.})$ on each window.  We also conservatively account for systematic correlations due to the shared use of lattice correction models on each window following the procedure in Ref.~\cite{FermilabLattice:2024yho}.  Additional systematic correlations due to common continuum model variations between windows, which were not relevant in Ref.~\cite{FermilabLattice:2024yho}, were considered and found to be negligible.  These correlations will be reconsidered in future works when determining other full HVP observables beyond the LQC contribution.

Our procedure for obtaining complete error budgets from the respective BMA analyses is described in Refs.~\cite{FermilabLatticeHPQCD:2023jof,FermilabLattice:2024yho}. In particular, we decompose the final uncertainty into its various contributions as described in Appendix A of Ref.~\cite{Bouchard:2014ypa}; the statistical and parametric contributions are computed approximately by using the best correlator fit, as determined by the BAIC, at each lattice spacing and current.

\section{Additional scale-setting variations}

In this section we give results for all isospin-symmetric quantities, from this work and Ref.~\cite{FermilabLattice:2024yho}, using three different observables to set the scale. They are $M_\Omega=1.67126(32)$ \cite{Bazavov:2024dov}, $f_\pi = 0.13050(13)$ \cite{FermilabLattice:2014tsy,FermilabLattice:2017wgj}, and $w_0 = 0.17236$~fm \cite{Borsanyi:2020mff}, (which together with the meson-mass inputs given Ref.~\cite{FermilabLattice:2024yho}) defines the Muon g-2 Theory Initiative's White-Paper scheme (WP25). The values for $w_0/a$ come from arXiv v1 of Ref.~\cite{FermilabLattice:2024yho}. They are: $w_0/a = 1.13227(18), 1.41060(28), 1.95021(57), 3.01838(92)$ for $a \approx 0.15, 0.12, 0.09, 0.06$ fm respectively.  All quark masses are fixed using the scheme described Ref.~\cite{FermilabLattice:2024yho}.

\begin{table*}[th]
\centering
\caption{Results for $\amu$ (in units of $10^{-10}$). The first column lists the specific $\amuHVP$ contribution. Columns two through four give results based on scale setting via $M_\Omega$, $f_\pi$, or WP25 $w_0$~fm, respectively.  Columns four through six are the correlated differences between these results. The $a_{\mu}(\textrm{iso})$ results contain the contribution from bottom quarks which is given in Eqns.~(5.7) and (5.5) in Ref.~\cite{FermilabLattice:2024yho}.}
\label{tab:altResultsSup}
\begin{tabularx}{\linewidth}{lCCCccc}
\hline
\hline
 & \multicolumn{3}{c}{Results} & \multicolumn{3}{c}{Correlated differences} \\
Contribution & $M_\Omega$ & $f_\pi$ & WP25 $w_0$ & $\Delta_{M_\Omega, f_\pi}$ & $\Delta_{M_\Omega, \textrm{WP25}}$ & $\Delta_{f_\pi, \textrm{WP25} }$ \\
\hline  
\\
 \multicolumn{7}{c}{SD} \\
 
$a^{ll,\,\mathrm{SD}}_{\mu}$& $48.139(11)(91)[92]$ & $48.126(15)(100)[101]$ & $48.115(15)(108)[109]$ & $0.013(78)$ & $0.024(83)$ & $0.011(80)$ \\
$a^{ss,\,\mathrm{SD}}_{\mu}$& $9.111(3)(16)[17]$ & $9.105(3)(17)[17]$ & $9.113(3)(17)[18]$ & $0.007(15)$ & $-0.002(15)$ & $-0.009(14)$ \\
$a^{cc,\,\mathrm{SD}}_{\mu}$ & $11.46(0)(17)[17]$ & $11.43(0)(13)[13]$ & $11.57(0)(13)[13]$ & $0.04(17)$ & $-0.10(17)$ & $-0.14(14)$ \\ 
$a^{\mathrm{SD}}_{\mu}(\textrm{disc.})$ & $-0.0019(4)(26)[26]$ & $-0.0029(4)(26)[26]$ & $-0.0023(4)(26)[26]$ & $0.00106(16)$ & $0.00042(13)$ & $-0.00064(9)$ \\ 
$a^{\mathrm{SD}}_{\mu}(\textrm{iso})$ & $69.01(1)(21)[21]$ & $68.95(2)(19)[19]$ & $69.09(2)(19)[19]$ & $0.06(19)$ & $-0.08(19)$ & $-0.14(16)$ \\ 
\hline  
\\
 \multicolumn{7}{c}{W} \\

$a^{ll,\,\mathrm{W}}_{\mu}$ & $206.90(14)(61)[63]$ & $206.94(15)(46)[49]$ & $207.40(15)(39)[42]$ & $-0.03(43)$ & $-0.50(42)$ & $-0.47(27)$ \\
$a^{ss,\,\mathrm{W}}_{\mu}$&$27.20(1)(13)[13]$ & $27.10(2)(8)[8]$ & $27.32(2)(5)[6]$ & $0.10(14)$ & $-0.12(13)$ & $-0.220(82)$ \\
$a^{cc,\,\mathrm{W}}_{\mu}$ & $2.624(0)(87)[87]$ & $2.626(0)(56)[56]$ & $2.717(0)(42)[42]$ & $-0.002(79)$ & $-0.093(73)$ & $-0.091(49)$   \\
$a^{\mathrm{W}}_{\mu}(\textrm{disc.})$ & $-0.85(5)(19)[20]$ & $-0.85(6)(19)[20]$ & $-0.85(6)(19)[20]$ & $0.000(99)$ & $0.000(75)$ & $0.00(11)$  \\
$a^{\mathrm{W}}_{\mu}(\textrm{iso})$ &$235.89(15)(75)[76]$ & $235.82(16)(56)[58]$ & $236.60(16)(47)[50]$ & $0.07(57)$ & $-0.71(54)$ & $-0.78(35)$\\
\hline  
\\
 \multicolumn{7}{c}{LD} \\

$a^{ll,\,\mathrm{LD}}_{\mu}$ & $400.2(2.3)(3.7)[4.3]$ & $396.6(2.2)(3.3)[4.0]$ & $401.3(2.3)(3.1)[3.8]$ & $3.6(3.9)$ & $-1.2(3.6)$ & $-4.7(3.3)$ \\
\hline  
\\
 \multicolumn{7}{c}{SD+W+LD} \\

$a^{ll}_{\mu}$& $655.2(2.3)(3.9)[4.5]$ & $651.7(2.2)(3.5)[4.1]$ & $656.9(2.3)(3.2)[3.9]$ & $3.5(4.0)$ & $-1.6(3.7)$ & $-5.2(3.3)$ \\
\hline
\hline
\end{tabularx}
\end{table*}

\begin{table}[h]
\centering
\caption{Approximate absolute error budget (in units of $10^{-10}$) for uncertainties reported above for $a^{ll,\,{\mathrm {LD}}}_{\mu}$ with various choices of scale setting.  From left to right, the contributions to the error are Monte Carlo statistics and $t_{\min}$ variation in the correlator fits, continuum extrapolation and TB corrections, FV and $M_{\pi}$-mistuning corrections, scale setting, and current renormalization.}
\vspace{1mm}
\begin{tabularx}{\linewidth}{LCLLLLll}
\hline \hline 
Contribution &  Statistics, $t_{\min}$ &  $a\to0$,$\Delta_{\mathrm{TB}}$   &   $\Delta_{\textrm{FV}}$,$\Delta_{M_{\pi}}$ &   $a$  & $Z_V$ &  Total\\ 
\hline
$a^{ll,\,{\mathrm {LD}}}_{\mu} (M_\Omega)$ & 2.3 & 2.8 & 1.4 & 1.9 & 0.2 & 4.3 \\
$a^{ll,\,{\mathrm {LD}}}_{\mu} (f_\pi)$ & 2.2 & 2.7 & 1.4 & 1.4 & 0.2 & 4.0\\
$a^{ll,\,{\mathrm {LD}}}_{\mu} (\textrm{WP25 } w_0)$ & 2.3 & 2.7 & 1.4 & 0.3 & 0.2 & 3.8 \\
\hline \hline
\end{tabularx}

\end{table}

\section{Additional continuum extrapolation results}

Results for the continuum extrapolation BMA using the $M_{\Omega}$ scale for $a^{ll,\,\mathrm{LD}}_{\mu}(\mathrm{conn.})$ with the fit method are given in \cref{fig:BMALDFitOmega}.  We also give analogous results with each of these three analysis choices varied individually, {\it i.e.}, using the $f_{\pi}$ scale in \cref{fig:BMALDFitPi}, for $a^{ll,\,\mathrm{Full}}_{\mu}(\mathrm{conn.})$ in \cref{fig:BMAFullFitOmega}, and with the bounding method in \cref{fig:BMALDBoundOmega}.

\begin{figure*}
\centering
\includegraphics[height=0.45\textwidth]{BMALDllFitmOmega.pdf}\includegraphics[height=0.50\textwidth]{BMACompareLDllFitmOmega.pdf}
\vspace{-5mm}
\caption{\panel{Left plot}: Results of the BMA procedure applied to $\amuLLD$ using the fit method for correlator-reconstruction and $M_{\Omega}$ for scale setting. Squares are data points (fit results) with TB corrections added; circles are data points (fit results) without first adding TB corrections. \panel{Upper left panel}: Histogram of all continuum extrapolations used in the BMA weighted by $\pr(M|D)$, the inner light-red band includes statistical and parametric errors corresponding to the first term in \cref{eq:BMACov}, while the outer is the total error. \panel{Upper right panel}: The subset of datasets and extrapolations corresponding to correcting the local (blue unfilled) and one-link (purple filled) currents with lattice corrections from the CM correction scheme. Different extrapolations correspond to variations of the fit function and ensembles included. \panel{Lower panels}: The best fits according to the model probability, \cref{eq:modelProb}. The middle panel shows the fit results, where joint fits are indicated with mixed-filled symbols. The bottom one shows the corresponding $Q$ values \cite{FermilabLattice:2016ipl}. In both panels, the correction schemes employed for $\Delta_{\rm FV}$, $\Delta_{\rm{M}_{\pi}}$ and $\Delta_{\rm TB}$ are indicated by the symbols' color, according to the legend in the middle panel.
\panel{Right plot}: Breakdown of the results from the BMA applied to $\amuLLD$. \panel{Left panel}: From top to bottom, the first, main result (BMA) includes all datasets, schemes, and other variations. The next block of results vary the schemes for FV, $M_{\pi}$-mistuning and TB corrections. The next two correspond to different euclidean-time integration regions over which the corrections are computed. Next is a division of all the models into whether TB corrections were applied. The following three are continuum extrapolations to either both currents jointly or one of the currents individually. The next six are subsets with specific continuum fit functions, corresponding to quadratic and cubic, as well as omitting the $0.15$~fm ensemble. The unfilled circles correspond to the local current, the filled to the one-link current. The next two correspond to continuum fits with or without the sea-quark-mass--mistuning term. The final three are subsets with differing leading powers of $\alpha_s$ in the fit function where, again, unfilled symbols are the local current and the filled symbols are one-link. The inner error bars on the data points include statistical and parametric errors corresponding to the first term in \cref{eq:BMACov}, while the outer are the total errors. \panel{Right panels}: Pie charts showing the contributions to the BMA corresponding to the groupings in the left panel. The percentages are computed from \cref{eqn:ssProb} for the particular subsets. In the case of the continuum fit function subsets, which are broken up into local and one-link current variations, the left pie-chart corresponds to the local current and the right to the one-link.}\label{fig:BMALDFitOmega}
\end{figure*}

\begin{figure*}
\centering
\includegraphics[height=0.45\textwidth]{BMALDllFitfpi.pdf}\includegraphics[height=0.50\textwidth]{BMACompareLDllFitfpi.pdf}
\vspace{-5mm}
\caption{Results of the BMA procedure applied to $\amuLLD$ using the fit method correlator-reconstruction approach and $f_{\pi}$ for scale setting. See \cref{fig:BMALDFitOmega} for a complete description of the plot. The same axis scale is employed here from that plot.}\label{fig:BMALDFitPi}
\end{figure*}

\begin{figure*}
\centering
\includegraphics[height=0.45\textwidth]{BMAFullllFitmOmega.pdf}\includegraphics[height=0.50\textwidth]{BMACompareFullllFitmOmega.pdf}
\vspace{-5mm}
\caption{Results of the BMA procedure applied to $a^{ll,\,{\mathrm {Full}}}_{\mu}(\mathrm{conn.})$ using the fit method correlator-reconstruction approach and $M_{\Omega}$ for scale setting. See \cref{fig:BMALDFitOmega} for a complete description of the plot.}\label{fig:BMAFullFitOmega}
\end{figure*}

\begin{figure*}
\centering
\includegraphics[height=0.45\textwidth]{BMALDllBoundmOmega.pdf}\includegraphics[height=0.50\textwidth]{BMACompareLDllBoundmOmega.pdf}
\vspace{-5mm}
\caption{Results of the BMA procedure applied to $\amuLLD$ using the bounding method correlator-reconstruction approach and $M_{\Omega}$ for scale setting. See \cref{fig:BMALDFitOmega} for a complete description of the plot. The same axis scale is employed here from that plot.}\label{fig:BMALDBoundOmega}
\end{figure*}

%% file: main.bbl
\begin{thebibliography}{117}%
\makeatletter
\providecommand \@ifxundefined [1]{%
 \@ifx{#1\undefined}
}%
\providecommand \@ifnum [1]{%
 \ifnum #1\expandafter \@firstoftwo
 \else \expandafter \@secondoftwo
 \fi
}%
\providecommand \@ifx [1]{%
 \ifx #1\expandafter \@firstoftwo
 \else \expandafter \@secondoftwo
 \fi
}%
\providecommand \natexlab [1]{#1}%
\providecommand \enquote  [1]{``#1''}%
\providecommand \bibnamefont  [1]{#1}%
\providecommand \bibfnamefont [1]{#1}%
\providecommand \citenamefont [1]{#1}%
\providecommand \href@noop [0]{\@secondoftwo}%
\providecommand \href [0]{\begingroup \@sanitize@url \@href}%
\providecommand \@href[1]{\@@startlink{#1}\@@href}%
\providecommand \@@href[1]{\endgroup#1\@@endlink}%
\providecommand \@sanitize@url [0]{\catcode `\\12\catcode `\$12\catcode `\&12\catcode `\#12\catcode `\^12\catcode `\_12\catcode `\%12\relax}%
\providecommand \@@startlink[1]{}%
\providecommand \@@endlink[0]{}%
\providecommand \url  [0]{\begingroup\@sanitize@url \@url }%
\providecommand \@url [1]{\endgroup\@href {#1}{\urlprefix }}%
\providecommand \urlprefix  [0]{URL }%
\providecommand \Eprint [0]{\href }%
\providecommand \doibase [0]{https://doi.org/}%
\providecommand \selectlanguage [0]{\@gobble}%
\providecommand \bibinfo  [0]{\@secondoftwo}%
\providecommand \bibfield  [0]{\@secondoftwo}%
\providecommand \translation [1]{[#1]}%
\providecommand \BibitemOpen [0]{}%
\providecommand \bibitemStop [0]{}%
\providecommand \bibitemNoStop [0]{.\EOS\space}%
\providecommand \EOS [0]{\spacefactor3000\relax}%
\providecommand \BibitemShut  [1]{\csname bibitem#1\endcsname}%
\let\auto@bib@innerbib\@empty
\bibitem [{\citenamefont {Abi}\ \emph {et~al.}(2021)\citenamefont {Abi} \emph {et~al.}}]{Muong-2:2021ojo}%
  \BibitemOpen
  \bibfield  {author} {\bibinfo {author} {\bibfnamefont {B.}~\bibnamefont {Abi}} \emph {et~al.} (\bibinfo {collaboration} {Muon g-2}),\ }\href {https://doi.org/10.1103/PhysRevLett.126.141801} {\bibfield  {journal} {\bibinfo  {journal} {Phys. Rev. Lett.}\ }\textbf {\bibinfo {volume} {126}},\ \bibinfo {pages} {141801} (\bibinfo {year} {2021})},\ \Eprint {https://arxiv.org/abs/2104.03281} {arXiv:2104.03281 [hep-ex]} \BibitemShut {NoStop}%
\bibitem [{\citenamefont {Aguillard}\ \emph {et~al.}(2023)\citenamefont {Aguillard} \emph {et~al.}}]{Muong-2:2023cdq}%
  \BibitemOpen
  \bibfield  {author} {\bibinfo {author} {\bibfnamefont {D.~P.}\ \bibnamefont {Aguillard}} \emph {et~al.} (\bibinfo {collaboration} {Muon g-2}),\ }\href {https://doi.org/10.1103/PhysRevLett.131.161802} {\bibfield  {journal} {\bibinfo  {journal} {Phys. Rev. Lett.}\ }\textbf {\bibinfo {volume} {131}},\ \bibinfo {pages} {161802} (\bibinfo {year} {2023})},\ \Eprint {https://arxiv.org/abs/2308.06230} {arXiv:2308.06230 [hep-ex]} \BibitemShut {NoStop}%
\bibitem [{\citenamefont {Aoyama}\ \emph {et~al.}(2020)\citenamefont {Aoyama} \emph {et~al.}}]{Aoyama:2020ynm}%
  \BibitemOpen
  \bibfield  {author} {\bibinfo {author} {\bibfnamefont {T.}~\bibnamefont {Aoyama}} \emph {et~al.},\ }\href {https://doi.org/10.1016/j.physrep.2020.07.006} {\bibfield  {journal} {\bibinfo  {journal} {Phys. Rept.}\ }\textbf {\bibinfo {volume} {887}},\ \bibinfo {pages} {1} (\bibinfo {year} {2020})}\BibitemShut {NoStop}%
\bibitem [{\citenamefont {Bennett}\ \emph {et~al.}(2006)\citenamefont {Bennett} \emph {et~al.}}]{Muong-2:2006rrc}%
  \BibitemOpen
  \bibfield  {author} {\bibinfo {author} {\bibfnamefont {G.~W.}\ \bibnamefont {Bennett}} \emph {et~al.} (\bibinfo {collaboration} {Muon g-2}),\ }\href {https://doi.org/10.1103/PhysRevD.73.072003} {\bibfield  {journal} {\bibinfo  {journal} {Phys. Rev. D}\ }\textbf {\bibinfo {volume} {73}},\ \bibinfo {pages} {072003} (\bibinfo {year} {2006})},\ \Eprint {https://arxiv.org/abs/hep-ex/0602035} {arXiv:hep-ex/0602035} \BibitemShut {NoStop}%
\bibitem [{\citenamefont {Abe}\ \emph {et~al.}(2019)\citenamefont {Abe} \emph {et~al.}}]{Abe:2019thb}%
  \BibitemOpen
  \bibfield  {author} {\bibinfo {author} {\bibfnamefont {M.}~\bibnamefont {Abe}} \emph {et~al.},\ }\href {https://doi.org/10.1093/ptep/ptz030} {\bibfield  {journal} {\bibinfo  {journal} {PTEP}\ }\textbf {\bibinfo {volume} {2019}},\ \bibinfo {pages} {053C02} (\bibinfo {year} {2019})}\BibitemShut {NoStop}%
\bibitem [{\citenamefont {{J-PARC muon $g-2$/EDM experiment}}()}]{E34webpage}%
  \BibitemOpen
  \bibfield  {author} {\bibinfo {author} {\bibnamefont {{J-PARC muon $g-2$/EDM experiment}}},\ }\href@noop {} {}\bibinfo {howpublished} {\url{https://g-2.kek.jp/}}\BibitemShut {NoStop}%
\bibitem [{\citenamefont {Colangelo}\ \emph {et~al.}(2022{\natexlab{a}})\citenamefont {Colangelo} \emph {et~al.}}]{Colangelo:2022jxc}%
  \BibitemOpen
  \bibfield  {author} {\bibinfo {author} {\bibfnamefont {G.}~\bibnamefont {Colangelo}} \emph {et~al.},\ }\Eprint {https://arxiv.org/abs/2203.15810} {arXiv:2203.15810 [hep-ph]}  (\bibinfo {year} {2022}{\natexlab{a}})\BibitemShut {NoStop}%
\bibitem [{\citenamefont {Melnikov}\ and\ \citenamefont {Vainshtein}(2004)}]{Melnikov:2003xd}%
  \BibitemOpen
  \bibfield  {author} {\bibinfo {author} {\bibfnamefont {K.}~\bibnamefont {Melnikov}}\ and\ \bibinfo {author} {\bibfnamefont {A.}~\bibnamefont {Vainshtein}},\ }\href {https://doi.org/10.1103/PhysRevD.70.113006} {\bibfield  {journal} {\bibinfo  {journal} {Phys. Rev.}\ }\textbf {\bibinfo {volume} {D70}},\ \bibinfo {pages} {113006} (\bibinfo {year} {2004})}\BibitemShut {NoStop}%
\bibitem [{\citenamefont {Masjuan}\ and\ \citenamefont {S{\'a}nchez-Puertas}(2017)}]{Masjuan:2017tvw}%
  \BibitemOpen
  \bibfield  {author} {\bibinfo {author} {\bibfnamefont {P.}~\bibnamefont {Masjuan}}\ and\ \bibinfo {author} {\bibfnamefont {P.}~\bibnamefont {S{\'a}nchez-Puertas}},\ }\href {https://doi.org/10.1103/PhysRevD.95.054026} {\bibfield  {journal} {\bibinfo  {journal} {Phys. Rev.}\ }\textbf {\bibinfo {volume} {D95}},\ \bibinfo {pages} {054026} (\bibinfo {year} {2017})}\BibitemShut {NoStop}%
\bibitem [{\citenamefont {Colangelo}\ \emph {et~al.}(2017)\citenamefont {Colangelo}, \citenamefont {Hoferichter}, \citenamefont {Procura},\ and\ \citenamefont {Stoffer}}]{Colangelo:2017fiz}%
  \BibitemOpen
  \bibfield  {author} {\bibinfo {author} {\bibfnamefont {G.}~\bibnamefont {Colangelo}}, \bibinfo {author} {\bibfnamefont {M.}~\bibnamefont {Hoferichter}}, \bibinfo {author} {\bibfnamefont {M.}~\bibnamefont {Procura}},\ and\ \bibinfo {author} {\bibfnamefont {P.}~\bibnamefont {Stoffer}},\ }\href {https://doi.org/10.1007/JHEP04(2017)161} {\bibfield  {journal} {\bibinfo  {journal} {JHEP}\ }\textbf {\bibinfo {volume} {04}}\bibinfo  {number} { (2017)},\ \bibinfo {pages} {161}}\BibitemShut {NoStop}%
\bibitem [{\citenamefont {Hoferichter}\ \emph {et~al.}(2018)\citenamefont {Hoferichter}, \citenamefont {Hoid}, \citenamefont {Kubis}, \citenamefont {Leupold},\ and\ \citenamefont {Schneider}}]{Hoferichter:2018kwz}%
  \BibitemOpen
\bibfield  {number} {  }\bibfield  {author} {\bibinfo {author} {\bibfnamefont {M.}~\bibnamefont {Hoferichter}}, \bibinfo {author} {\bibfnamefont {B.-L.}\ \bibnamefont {Hoid}}, \bibinfo {author} {\bibfnamefont {B.}~\bibnamefont {Kubis}}, \bibinfo {author} {\bibfnamefont {S.}~\bibnamefont {Leupold}},\ and\ \bibinfo {author} {\bibfnamefont {S.~P.}\ \bibnamefont {Schneider}},\ }\href {https://doi.org/10.1007/JHEP10(2018)141} {\bibfield  {journal} {\bibinfo  {journal} {JHEP}\ }\textbf {\bibinfo {volume} {10}}\bibinfo  {number} { (2018)},\ \bibinfo {pages} {141}}\BibitemShut {NoStop}%
\bibitem [{\citenamefont {G{\'e}rardin}\ \emph {et~al.}(2019)\citenamefont {G{\'e}rardin}, \citenamefont {Meyer},\ and\ \citenamefont {Nyffeler}}]{Gerardin:2019vio}%
  \BibitemOpen
\bibfield  {number} {  }\bibfield  {author} {\bibinfo {author} {\bibfnamefont {A.}~\bibnamefont {G{\'e}rardin}}, \bibinfo {author} {\bibfnamefont {H.~B.}\ \bibnamefont {Meyer}},\ and\ \bibinfo {author} {\bibfnamefont {A.}~\bibnamefont {Nyffeler}},\ }\href {https://doi.org/10.1103/PhysRevD.100.034520} {\bibfield  {journal} {\bibinfo  {journal} {Phys. Rev.}\ }\textbf {\bibinfo {volume} {D100}},\ \bibinfo {pages} {034520} (\bibinfo {year} {2019})}\BibitemShut {NoStop}%
\bibitem [{\citenamefont {Bijnens}\ \emph {et~al.}(2019)\citenamefont {Bijnens}, \citenamefont {Hermansson-Truedsson},\ and\ \citenamefont {Rodr{\'i}guez-S{\'a}nchez}}]{Bijnens:2019ghy}%
  \BibitemOpen
  \bibfield  {author} {\bibinfo {author} {\bibfnamefont {J.}~\bibnamefont {Bijnens}}, \bibinfo {author} {\bibfnamefont {N.}~\bibnamefont {Hermansson-Truedsson}},\ and\ \bibinfo {author} {\bibfnamefont {A.}~\bibnamefont {Rodr{\'i}guez-S{\'a}nchez}},\ }\href {https://doi.org/10.1016/j.physletb.2019.134994} {\bibfield  {journal} {\bibinfo  {journal} {Phys. Lett.}\ }\textbf {\bibinfo {volume} {B798}},\ \bibinfo {pages} {134994} (\bibinfo {year} {2019})}\BibitemShut {NoStop}%
\bibitem [{\citenamefont {Colangelo}\ \emph {et~al.}(2020)\citenamefont {Colangelo}, \citenamefont {Hagelstein}, \citenamefont {Hoferichter}, \citenamefont {Laub},\ and\ \citenamefont {Stoffer}}]{Colangelo:2019uex}%
  \BibitemOpen
  \bibfield  {author} {\bibinfo {author} {\bibfnamefont {G.}~\bibnamefont {Colangelo}}, \bibinfo {author} {\bibfnamefont {F.}~\bibnamefont {Hagelstein}}, \bibinfo {author} {\bibfnamefont {M.}~\bibnamefont {Hoferichter}}, \bibinfo {author} {\bibfnamefont {L.}~\bibnamefont {Laub}},\ and\ \bibinfo {author} {\bibfnamefont {P.}~\bibnamefont {Stoffer}},\ }\href {https://doi.org/10.1007/JHEP03(2020)101} {\bibfield  {journal} {\bibinfo  {journal} {JHEP}\ }\textbf {\bibinfo {volume} {03}}\bibinfo  {number} { (2020)},\ \bibinfo {pages} {101}}\BibitemShut {NoStop}%
\bibitem [{\citenamefont {Pauk}\ and\ \citenamefont {Vanderhaeghen}(2014)}]{Pauk:2014rta}%
  \BibitemOpen
\bibfield  {number} {  }\bibfield  {author} {\bibinfo {author} {\bibfnamefont {V.}~\bibnamefont {Pauk}}\ and\ \bibinfo {author} {\bibfnamefont {M.}~\bibnamefont {Vanderhaeghen}},\ }\href {https://doi.org/10.1140/epjc/s10052-014-3008-y} {\bibfield  {journal} {\bibinfo  {journal} {Eur. Phys. J.}\ }\textbf {\bibinfo {volume} {C74}},\ \bibinfo {pages} {3008} (\bibinfo {year} {2014})},\ \Eprint {https://arxiv.org/abs/1401.0832} {arXiv:1401.0832 [hep-ph]} \BibitemShut {NoStop}%
\bibitem [{\citenamefont {Danilkin}\ and\ \citenamefont {Vanderhaeghen}(2017)}]{Danilkin:2016hnh}%
  \BibitemOpen
  \bibfield  {author} {\bibinfo {author} {\bibfnamefont {I.}~\bibnamefont {Danilkin}}\ and\ \bibinfo {author} {\bibfnamefont {M.}~\bibnamefont {Vanderhaeghen}},\ }\href {https://doi.org/10.1103/PhysRevD.95.014019} {\bibfield  {journal} {\bibinfo  {journal} {Phys. Rev.}\ }\textbf {\bibinfo {volume} {D95}},\ \bibinfo {pages} {014019} (\bibinfo {year} {2017})},\ \Eprint {https://arxiv.org/abs/1611.04646} {arXiv:1611.04646 [hep-ph]} \BibitemShut {NoStop}%
\bibitem [{\citenamefont {Jegerlehner}(2017)}]{Jegerlehner:2017gek}%
  \BibitemOpen
  \bibfield  {author} {\bibinfo {author} {\bibfnamefont {F.}~\bibnamefont {Jegerlehner}},\ }\href {https://doi.org/10.1007/978-3-319-63577-4} {\emph {\bibinfo {title} {{The Anomalous Magnetic Moment of the Muon}}}},\ Vol.\ \bibinfo {volume} {274}\ (\bibinfo  {publisher} {Springer},\ \bibinfo {address} {Cham},\ \bibinfo {year} {2017})\BibitemShut {NoStop}%
\bibitem [{\citenamefont {Knecht}\ \emph {et~al.}(2018)\citenamefont {Knecht}, \citenamefont {Narison}, \citenamefont {Rabemananjara},\ and\ \citenamefont {Rabetiarivony}}]{Knecht:2018sci}%
  \BibitemOpen
  \bibfield  {author} {\bibinfo {author} {\bibfnamefont {M.}~\bibnamefont {Knecht}}, \bibinfo {author} {\bibfnamefont {S.}~\bibnamefont {Narison}}, \bibinfo {author} {\bibfnamefont {A.}~\bibnamefont {Rabemananjara}},\ and\ \bibinfo {author} {\bibfnamefont {D.}~\bibnamefont {Rabetiarivony}},\ }\href {https://doi.org/10.1016/j.physletb.2018.10.048} {\bibfield  {journal} {\bibinfo  {journal} {Phys. Lett.}\ }\textbf {\bibinfo {volume} {B787}},\ \bibinfo {pages} {111} (\bibinfo {year} {2018})},\ \Eprint {https://arxiv.org/abs/1808.03848} {arXiv:1808.03848 [hep-ph]} \BibitemShut {NoStop}%
\bibitem [{\citenamefont {Eichmann}\ \emph {et~al.}(2020)\citenamefont {Eichmann}, \citenamefont {Fischer},\ and\ \citenamefont {Williams}}]{Eichmann:2019bqf}%
  \BibitemOpen
  \bibfield  {author} {\bibinfo {author} {\bibfnamefont {G.}~\bibnamefont {Eichmann}}, \bibinfo {author} {\bibfnamefont {C.~S.}\ \bibnamefont {Fischer}},\ and\ \bibinfo {author} {\bibfnamefont {R.}~\bibnamefont {Williams}},\ }\href {https://doi.org/10.1103/PhysRevD.101.054015} {\bibfield  {journal} {\bibinfo  {journal} {Phys. Rev.}\ }\textbf {\bibinfo {volume} {D101}},\ \bibinfo {pages} {054015} (\bibinfo {year} {2020})},\ \Eprint {https://arxiv.org/abs/1910.06795} {arXiv:1910.06795 [hep-ph]} \BibitemShut {NoStop}%
\bibitem [{\citenamefont {Roig}\ and\ \citenamefont {S{\'a}nchez-Puertas}(2020)}]{Roig:2019reh}%
  \BibitemOpen
  \bibfield  {author} {\bibinfo {author} {\bibfnamefont {P.}~\bibnamefont {Roig}}\ and\ \bibinfo {author} {\bibfnamefont {P.}~\bibnamefont {S{\'a}nchez-Puertas}},\ }\href {https://doi.org/10.1103/PhysRevD.101.074019} {\bibfield  {journal} {\bibinfo  {journal} {Phys. Rev.}\ }\textbf {\bibinfo {volume} {D101}},\ \bibinfo {pages} {074019} (\bibinfo {year} {2020})},\ \Eprint {https://arxiv.org/abs/1910.02881} {arXiv:1910.02881 [hep-ph]} \BibitemShut {NoStop}%
\bibitem [{\citenamefont {Leutgeb}\ \emph {et~al.}(2023)\citenamefont {Leutgeb}, \citenamefont {Mager},\ and\ \citenamefont {Rebhan}}]{Leutgeb:2022lqw}%
  \BibitemOpen
  \bibfield  {author} {\bibinfo {author} {\bibfnamefont {J.}~\bibnamefont {Leutgeb}}, \bibinfo {author} {\bibfnamefont {J.}~\bibnamefont {Mager}},\ and\ \bibinfo {author} {\bibfnamefont {A.}~\bibnamefont {Rebhan}},\ }\href {https://doi.org/10.1103/PhysRevD.107.054021} {\bibfield  {journal} {\bibinfo  {journal} {Phys. Rev. D}\ }\textbf {\bibinfo {volume} {107}},\ \bibinfo {pages} {054021} (\bibinfo {year} {2023})},\ \Eprint {https://arxiv.org/abs/2211.16562} {arXiv:2211.16562 [hep-ph]} \BibitemShut {NoStop}%
\bibitem [{\citenamefont {Blum}\ \emph {et~al.}(2020)\citenamefont {Blum}, \citenamefont {Christ}, \citenamefont {Hayakawa}, \citenamefont {Izubuchi}, \citenamefont {Jin}, \citenamefont {Jung},\ and\ \citenamefont {Lehner}}]{Blum:2019ugy}%
  \BibitemOpen
  \bibfield  {author} {\bibinfo {author} {\bibfnamefont {T.}~\bibnamefont {Blum}}, \bibinfo {author} {\bibfnamefont {N.}~\bibnamefont {Christ}}, \bibinfo {author} {\bibfnamefont {M.}~\bibnamefont {Hayakawa}}, \bibinfo {author} {\bibfnamefont {T.}~\bibnamefont {Izubuchi}}, \bibinfo {author} {\bibfnamefont {L.}~\bibnamefont {Jin}}, \bibinfo {author} {\bibfnamefont {C.}~\bibnamefont {Jung}},\ and\ \bibinfo {author} {\bibfnamefont {C.}~\bibnamefont {Lehner}} (\bibinfo {collaboration} {RBC}),\ }\href {https://doi.org/10.1103/PhysRevLett.124.132002} {\bibfield  {journal} {\bibinfo  {journal} {Phys. Rev. Lett.}\ }\textbf {\bibinfo {volume} {124}},\ \bibinfo {pages} {132002} (\bibinfo {year} {2020})}\BibitemShut {NoStop}%
\bibitem [{\citenamefont {Chao}\ \emph {et~al.}(2021)\citenamefont {Chao}, \citenamefont {Hudspith}, \citenamefont {G\'erardin}, \citenamefont {Green}, \citenamefont {Meyer},\ and\ \citenamefont {Ottnad}}]{Chao:2021tvp}%
  \BibitemOpen
  \bibfield  {author} {\bibinfo {author} {\bibfnamefont {E.-H.}\ \bibnamefont {Chao}}, \bibinfo {author} {\bibfnamefont {R.~J.}\ \bibnamefont {Hudspith}}, \bibinfo {author} {\bibfnamefont {A.}~\bibnamefont {G\'erardin}}, \bibinfo {author} {\bibfnamefont {J.~R.}\ \bibnamefont {Green}}, \bibinfo {author} {\bibfnamefont {H.~B.}\ \bibnamefont {Meyer}},\ and\ \bibinfo {author} {\bibfnamefont {K.}~\bibnamefont {Ottnad}},\ }\href {https://doi.org/10.1140/epjc/s10052-021-09455-4} {\bibfield  {journal} {\bibinfo  {journal} {Eur. Phys. J. C}\ }\textbf {\bibinfo {volume} {81}},\ \bibinfo {pages} {651} (\bibinfo {year} {2021})},\ \Eprint {https://arxiv.org/abs/2104.02632} {arXiv:2104.02632 [hep-lat]} \BibitemShut {NoStop}%
\bibitem [{\citenamefont {Bijnens}\ \emph {et~al.}(2020)\citenamefont {Bijnens}, \citenamefont {Hermansson-Truedsson}, \citenamefont {Laub},\ and\ \citenamefont {Rodr\'\i{}guez-S\'anchez}}]{Bijnens:2020xnl}%
  \BibitemOpen
  \bibfield  {author} {\bibinfo {author} {\bibfnamefont {J.}~\bibnamefont {Bijnens}}, \bibinfo {author} {\bibfnamefont {N.}~\bibnamefont {Hermansson-Truedsson}}, \bibinfo {author} {\bibfnamefont {L.}~\bibnamefont {Laub}},\ and\ \bibinfo {author} {\bibfnamefont {A.}~\bibnamefont {Rodr\'\i{}guez-S\'anchez}},\ }\href {https://doi.org/10.1007/JHEP10(2020)203} {\bibfield  {journal} {\bibinfo  {journal} {JHEP}\ }\textbf {\bibinfo {volume} {10}},\ \bibinfo {pages} {203}},\ \Eprint {https://arxiv.org/abs/2008.13487} {arXiv:2008.13487 [hep-ph]} \BibitemShut {NoStop}%
\bibitem [{\citenamefont {L\"udtke}\ and\ \citenamefont {Procura}(2020)}]{Ludtke:2020moa}%
  \BibitemOpen
  \bibfield  {author} {\bibinfo {author} {\bibfnamefont {J.}~\bibnamefont {L\"udtke}}\ and\ \bibinfo {author} {\bibfnamefont {M.}~\bibnamefont {Procura}},\ }\href {https://doi.org/10.1140/epjc/s10052-020-08611-6} {\bibfield  {journal} {\bibinfo  {journal} {Eur. Phys. J. C}\ }\textbf {\bibinfo {volume} {80}},\ \bibinfo {pages} {1108} (\bibinfo {year} {2020})},\ \Eprint {https://arxiv.org/abs/2006.00007} {arXiv:2006.00007 [hep-ph]} \BibitemShut {NoStop}%
\bibitem [{\citenamefont {Bijnens}\ \emph {et~al.}(2021)\citenamefont {Bijnens}, \citenamefont {Hermansson-Truedsson}, \citenamefont {Laub},\ and\ \citenamefont {Rodr\'\i{}guez-S\'anchez}}]{Bijnens:2021jqo}%
  \BibitemOpen
  \bibfield  {author} {\bibinfo {author} {\bibfnamefont {J.}~\bibnamefont {Bijnens}}, \bibinfo {author} {\bibfnamefont {N.}~\bibnamefont {Hermansson-Truedsson}}, \bibinfo {author} {\bibfnamefont {L.}~\bibnamefont {Laub}},\ and\ \bibinfo {author} {\bibfnamefont {A.}~\bibnamefont {Rodr\'\i{}guez-S\'anchez}},\ }\href {https://doi.org/10.1007/JHEP04(2021)240} {\bibfield  {journal} {\bibinfo  {journal} {JHEP}\ }\textbf {\bibinfo {volume} {04}},\ \bibinfo {pages} {240}},\ \Eprint {https://arxiv.org/abs/2101.09169} {arXiv:2101.09169 [hep-ph]} \BibitemShut {NoStop}%
\bibitem [{\citenamefont {Hoferichter}\ and\ \citenamefont {Stoffer}(2020)}]{Hoferichter:2020lap}%
  \BibitemOpen
  \bibfield  {author} {\bibinfo {author} {\bibfnamefont {M.}~\bibnamefont {Hoferichter}}\ and\ \bibinfo {author} {\bibfnamefont {P.}~\bibnamefont {Stoffer}},\ }\href {https://doi.org/10.1007/JHEP05(2020)159} {\bibfield  {journal} {\bibinfo  {journal} {JHEP}\ }\textbf {\bibinfo {volume} {05}},\ \bibinfo {pages} {159}},\ \Eprint {https://arxiv.org/abs/2004.06127} {arXiv:2004.06127 [hep-ph]} \BibitemShut {NoStop}%
\bibitem [{\citenamefont {Leutgeb}\ and\ \citenamefont {Rebhan}(2021)}]{Leutgeb:2021mpu}%
  \BibitemOpen
  \bibfield  {author} {\bibinfo {author} {\bibfnamefont {J.}~\bibnamefont {Leutgeb}}\ and\ \bibinfo {author} {\bibfnamefont {A.}~\bibnamefont {Rebhan}},\ }\href {https://doi.org/10.1103/PhysRevD.104.094017} {\bibfield  {journal} {\bibinfo  {journal} {Phys. Rev. D}\ }\textbf {\bibinfo {volume} {104}},\ \bibinfo {pages} {094017} (\bibinfo {year} {2021})},\ \Eprint {https://arxiv.org/abs/2108.12345} {arXiv:2108.12345 [hep-ph]} \BibitemShut {NoStop}%
\bibitem [{\citenamefont {Zanke}\ \emph {et~al.}(2021)\citenamefont {Zanke}, \citenamefont {Hoferichter},\ and\ \citenamefont {Kubis}}]{Zanke:2021wiq}%
  \BibitemOpen
  \bibfield  {author} {\bibinfo {author} {\bibfnamefont {M.}~\bibnamefont {Zanke}}, \bibinfo {author} {\bibfnamefont {M.}~\bibnamefont {Hoferichter}},\ and\ \bibinfo {author} {\bibfnamefont {B.}~\bibnamefont {Kubis}},\ }\href {https://doi.org/10.1007/JHEP07(2021)106} {\bibfield  {journal} {\bibinfo  {journal} {JHEP}\ }\textbf {\bibinfo {volume} {07}},\ \bibinfo {pages} {106}},\ \Eprint {https://arxiv.org/abs/2103.09829} {arXiv:2103.09829 [hep-ph]} \BibitemShut {NoStop}%
\bibitem [{\citenamefont {Danilkin}\ \emph {et~al.}(2021)\citenamefont {Danilkin}, \citenamefont {Hoferichter},\ and\ \citenamefont {Stoffer}}]{Danilkin:2021icn}%
  \BibitemOpen
  \bibfield  {author} {\bibinfo {author} {\bibfnamefont {I.}~\bibnamefont {Danilkin}}, \bibinfo {author} {\bibfnamefont {M.}~\bibnamefont {Hoferichter}},\ and\ \bibinfo {author} {\bibfnamefont {P.}~\bibnamefont {Stoffer}},\ }\href {https://doi.org/10.1016/j.physletb.2021.136502} {\bibfield  {journal} {\bibinfo  {journal} {Phys. Lett. B}\ }\textbf {\bibinfo {volume} {820}},\ \bibinfo {pages} {136502} (\bibinfo {year} {2021})},\ \Eprint {https://arxiv.org/abs/2105.01666} {arXiv:2105.01666 [hep-ph]} \BibitemShut {NoStop}%
\bibitem [{\citenamefont {Colangelo}\ \emph {et~al.}(2021)\citenamefont {Colangelo}, \citenamefont {Hagelstein}, \citenamefont {Hoferichter}, \citenamefont {Laub},\ and\ \citenamefont {Stoffer}}]{Colangelo:2021nkr}%
  \BibitemOpen
  \bibfield  {author} {\bibinfo {author} {\bibfnamefont {G.}~\bibnamefont {Colangelo}}, \bibinfo {author} {\bibfnamefont {F.}~\bibnamefont {Hagelstein}}, \bibinfo {author} {\bibfnamefont {M.}~\bibnamefont {Hoferichter}}, \bibinfo {author} {\bibfnamefont {L.}~\bibnamefont {Laub}},\ and\ \bibinfo {author} {\bibfnamefont {P.}~\bibnamefont {Stoffer}},\ }\href {https://doi.org/10.1140/epjc/s10052-021-09513-x} {\bibfield  {journal} {\bibinfo  {journal} {Eur. Phys. J. C}\ }\textbf {\bibinfo {volume} {81}},\ \bibinfo {pages} {702} (\bibinfo {year} {2021})},\ \Eprint {https://arxiv.org/abs/2106.13222} {arXiv:2106.13222 [hep-ph]} \BibitemShut {NoStop}%
\bibitem [{\citenamefont {Cappiello}\ \emph {et~al.}(2022)\citenamefont {Cappiello}, \citenamefont {Cat\`a},\ and\ \citenamefont {D'Ambrosio}}]{Cappiello:2021vzi}%
  \BibitemOpen
  \bibfield  {author} {\bibinfo {author} {\bibfnamefont {L.}~\bibnamefont {Cappiello}}, \bibinfo {author} {\bibfnamefont {O.}~\bibnamefont {Cat\`a}},\ and\ \bibinfo {author} {\bibfnamefont {G.}~\bibnamefont {D'Ambrosio}},\ }\href {https://doi.org/10.1103/PhysRevD.105.056020} {\bibfield  {journal} {\bibinfo  {journal} {Phys. Rev. D}\ }\textbf {\bibinfo {volume} {105}},\ \bibinfo {pages} {056020} (\bibinfo {year} {2022})},\ \Eprint {https://arxiv.org/abs/2110.05962} {arXiv:2110.05962 [hep-ph]} \BibitemShut {NoStop}%
\bibitem [{\citenamefont {Bijnens}\ \emph {et~al.}(2023)\citenamefont {Bijnens}, \citenamefont {Hermansson-Truedsson},\ and\ \citenamefont {Rodr\'\i{}guez-S\'anchez}}]{Bijnens:2022itw}%
  \BibitemOpen
  \bibfield  {author} {\bibinfo {author} {\bibfnamefont {J.}~\bibnamefont {Bijnens}}, \bibinfo {author} {\bibfnamefont {N.}~\bibnamefont {Hermansson-Truedsson}},\ and\ \bibinfo {author} {\bibfnamefont {A.}~\bibnamefont {Rodr\'\i{}guez-S\'anchez}},\ }\href {https://doi.org/10.1007/JHEP02(2023)167} {\bibfield  {journal} {\bibinfo  {journal} {JHEP}\ }\textbf {\bibinfo {volume} {02}},\ \bibinfo {pages} {167}},\ \Eprint {https://arxiv.org/abs/2211.17183} {arXiv:2211.17183 [hep-ph]} \BibitemShut {NoStop}%
\bibitem [{\citenamefont {Hoferichter}\ \emph {et~al.}(2023)\citenamefont {Hoferichter}, \citenamefont {Kubis},\ and\ \citenamefont {Zanke}}]{Hoferichter:2023tgp}%
  \BibitemOpen
  \bibfield  {author} {\bibinfo {author} {\bibfnamefont {M.}~\bibnamefont {Hoferichter}}, \bibinfo {author} {\bibfnamefont {B.}~\bibnamefont {Kubis}},\ and\ \bibinfo {author} {\bibfnamefont {M.}~\bibnamefont {Zanke}},\ }\href {https://doi.org/10.1007/JHEP08(2023)209} {\bibfield  {journal} {\bibinfo  {journal} {JHEP}\ }\textbf {\bibinfo {volume} {08}},\ \bibinfo {pages} {209}},\ \Eprint {https://arxiv.org/abs/2307.14413} {arXiv:2307.14413 [hep-ph]} \BibitemShut {NoStop}%
\bibitem [{\citenamefont {L\"udtke}\ \emph {et~al.}(2023)\citenamefont {L\"udtke}, \citenamefont {Procura},\ and\ \citenamefont {Stoffer}}]{Ludtke:2023hvz}%
  \BibitemOpen
  \bibfield  {author} {\bibinfo {author} {\bibfnamefont {J.}~\bibnamefont {L\"udtke}}, \bibinfo {author} {\bibfnamefont {M.}~\bibnamefont {Procura}},\ and\ \bibinfo {author} {\bibfnamefont {P.}~\bibnamefont {Stoffer}},\ }\href {https://doi.org/10.1007/JHEP04(2023)125} {\bibfield  {journal} {\bibinfo  {journal} {JHEP}\ }\textbf {\bibinfo {volume} {04}},\ \bibinfo {pages} {125}},\ \Eprint {https://arxiv.org/abs/2302.12264} {arXiv:2302.12264 [hep-ph]} \BibitemShut {NoStop}%
\bibitem [{\citenamefont {Colangelo}\ \emph {et~al.}(2023)\citenamefont {Colangelo}, \citenamefont {Giannuzzi},\ and\ \citenamefont {Nicotri}}]{Colangelo:2023een}%
  \BibitemOpen
  \bibfield  {author} {\bibinfo {author} {\bibfnamefont {P.}~\bibnamefont {Colangelo}}, \bibinfo {author} {\bibfnamefont {F.}~\bibnamefont {Giannuzzi}},\ and\ \bibinfo {author} {\bibfnamefont {S.}~\bibnamefont {Nicotri}},\ }\href {https://doi.org/10.1016/j.physletb.2023.137878} {\bibfield  {journal} {\bibinfo  {journal} {Phys. Lett. B}\ }\textbf {\bibinfo {volume} {840}},\ \bibinfo {pages} {137878} (\bibinfo {year} {2023})},\ \Eprint {https://arxiv.org/abs/2301.06456} {arXiv:2301.06456 [hep-ph]} \BibitemShut {NoStop}%
\bibitem [{\citenamefont {Hoferichter}\ \emph {et~al.}(2024{\natexlab{a}})\citenamefont {Hoferichter}, \citenamefont {Stoffer},\ and\ \citenamefont {Zillinger}}]{Hoferichter:2024fsj}%
  \BibitemOpen
  \bibfield  {author} {\bibinfo {author} {\bibfnamefont {M.}~\bibnamefont {Hoferichter}}, \bibinfo {author} {\bibfnamefont {P.}~\bibnamefont {Stoffer}},\ and\ \bibinfo {author} {\bibfnamefont {M.}~\bibnamefont {Zillinger}},\ }\href {https://doi.org/10.1007/JHEP04(2024)092} {\bibfield  {journal} {\bibinfo  {journal} {JHEP}\ }\textbf {\bibinfo {volume} {04}},\ \bibinfo {pages} {092}},\ \Eprint {https://arxiv.org/abs/2402.14060} {arXiv:2402.14060 [hep-ph]} \BibitemShut {NoStop}%
\bibitem [{\citenamefont {Hoferichter}\ \emph {et~al.}(2024{\natexlab{b}})\citenamefont {Hoferichter}, \citenamefont {Stoffer},\ and\ \citenamefont {Zillinger}}]{Hoferichter:2024vbu}%
  \BibitemOpen
  \bibfield  {author} {\bibinfo {author} {\bibfnamefont {M.}~\bibnamefont {Hoferichter}}, \bibinfo {author} {\bibfnamefont {P.}~\bibnamefont {Stoffer}},\ and\ \bibinfo {author} {\bibfnamefont {M.}~\bibnamefont {Zillinger}},\ }\Eprint {https://arxiv.org/abs/2412.00190} {arXiv:2412.00190 [hep-ph]}  (\bibinfo {year} {2024}{\natexlab{b}})\BibitemShut {NoStop}%
\bibitem [{\citenamefont {Hoferichter}\ \emph {et~al.}(2024{\natexlab{c}})\citenamefont {Hoferichter}, \citenamefont {Stoffer},\ and\ \citenamefont {Zillinger}}]{Hoferichter:2024bae}%
  \BibitemOpen
  \bibfield  {author} {\bibinfo {author} {\bibfnamefont {M.}~\bibnamefont {Hoferichter}}, \bibinfo {author} {\bibfnamefont {P.}~\bibnamefont {Stoffer}},\ and\ \bibinfo {author} {\bibfnamefont {M.}~\bibnamefont {Zillinger}},\ }\Eprint {https://arxiv.org/abs/2412.00178} {arXiv:2412.00178 [hep-ph]}  (\bibinfo {year} {2024}{\natexlab{c}})\BibitemShut {NoStop}%
\bibitem [{\citenamefont {Bijnens}\ \emph {et~al.}(2024)\citenamefont {Bijnens}, \citenamefont {Hermansson-Truedsson},\ and\ \citenamefont {Rodr\'\i{}guez-S\'anchez}}]{Bijnens:2024jgh}%
  \BibitemOpen
  \bibfield  {author} {\bibinfo {author} {\bibfnamefont {J.}~\bibnamefont {Bijnens}}, \bibinfo {author} {\bibfnamefont {N.}~\bibnamefont {Hermansson-Truedsson}},\ and\ \bibinfo {author} {\bibfnamefont {A.}~\bibnamefont {Rodr\'\i{}guez-S\'anchez}},\ }\Eprint {https://arxiv.org/abs/2411.09578} {arXiv:2411.09578 [hep-ph]}  (\bibinfo {year} {2024})\BibitemShut {NoStop}%
\bibitem [{\citenamefont {Holz}\ \emph {et~al.}(2022)\citenamefont {Holz}, \citenamefont {Hanhart}, \citenamefont {Hoferichter},\ and\ \citenamefont {Kubis}}]{Holz:2022hwz}%
  \BibitemOpen
  \bibfield  {author} {\bibinfo {author} {\bibfnamefont {S.}~\bibnamefont {Holz}}, \bibinfo {author} {\bibfnamefont {C.}~\bibnamefont {Hanhart}}, \bibinfo {author} {\bibfnamefont {M.}~\bibnamefont {Hoferichter}},\ and\ \bibinfo {author} {\bibfnamefont {B.}~\bibnamefont {Kubis}},\ }\href {https://doi.org/10.1140/epjc/s10052-022-10247-7} {\bibfield  {journal} {\bibinfo  {journal} {Eur. Phys. J. C}\ }\textbf {\bibinfo {volume} {82}},\ \bibinfo {pages} {434} (\bibinfo {year} {2022})},\ \bibinfo {note} {[Addendum: Eur.Phys.J.C 82, 1159 (2022)]},\ \Eprint {https://arxiv.org/abs/2202.05846} {arXiv:2202.05846 [hep-ph]} \BibitemShut {NoStop}%
\bibitem [{\citenamefont {Holz}\ \emph {et~al.}(2024)\citenamefont {Holz}, \citenamefont {Hoferichter}, \citenamefont {Hoid},\ and\ \citenamefont {Kubis}}]{Holz:2024lom}%
  \BibitemOpen
  \bibfield  {author} {\bibinfo {author} {\bibfnamefont {S.}~\bibnamefont {Holz}}, \bibinfo {author} {\bibfnamefont {M.}~\bibnamefont {Hoferichter}}, \bibinfo {author} {\bibfnamefont {B.-L.}\ \bibnamefont {Hoid}},\ and\ \bibinfo {author} {\bibfnamefont {B.}~\bibnamefont {Kubis}},\ }\Eprint {https://arxiv.org/abs/2411.08098} {arXiv:2411.08098 [hep-ph]}  (\bibinfo {year} {2024})\BibitemShut {NoStop}%
\bibitem [{\citenamefont {Leutgeb}\ \emph {et~al.}(2024)\citenamefont {Leutgeb}, \citenamefont {Mager},\ and\ \citenamefont {Rebhan}}]{Leutgeb:2024rfs}%
  \BibitemOpen
  \bibfield  {author} {\bibinfo {author} {\bibfnamefont {J.}~\bibnamefont {Leutgeb}}, \bibinfo {author} {\bibfnamefont {J.}~\bibnamefont {Mager}},\ and\ \bibinfo {author} {\bibfnamefont {A.}~\bibnamefont {Rebhan}},\ }\Eprint {https://arxiv.org/abs/2411.10432} {arXiv:2411.10432 [hep-ph]}  (\bibinfo {year} {2024})\BibitemShut {NoStop}%
\bibitem [{\citenamefont {Colangelo}\ \emph {et~al.}(2024)\citenamefont {Colangelo}, \citenamefont {Giannuzzi},\ and\ \citenamefont {Nicotri}}]{Colangelo:2024xfh}%
  \BibitemOpen
  \bibfield  {author} {\bibinfo {author} {\bibfnamefont {P.}~\bibnamefont {Colangelo}}, \bibinfo {author} {\bibfnamefont {F.}~\bibnamefont {Giannuzzi}},\ and\ \bibinfo {author} {\bibfnamefont {S.}~\bibnamefont {Nicotri}},\ }\href {https://doi.org/10.1103/PhysRevD.109.094036} {\bibfield  {journal} {\bibinfo  {journal} {Phys. Rev. D}\ }\textbf {\bibinfo {volume} {109}},\ \bibinfo {pages} {094036} (\bibinfo {year} {2024})},\ \Eprint {https://arxiv.org/abs/2402.07579} {arXiv:2402.07579 [hep-ph]} \BibitemShut {NoStop}%
\bibitem [{\citenamefont {Estrada}\ \emph {et~al.}(2024{\natexlab{a}})\citenamefont {Estrada}, \citenamefont {M\'arquez}, \citenamefont {Portillo-S\'anchez},\ and\ \citenamefont {Roig}}]{Estrada:2024rfi}%
  \BibitemOpen
  \bibfield  {author} {\bibinfo {author} {\bibfnamefont {E.~J.}\ \bibnamefont {Estrada}}, \bibinfo {author} {\bibfnamefont {J.~M.}\ \bibnamefont {M\'arquez}}, \bibinfo {author} {\bibfnamefont {D.}~\bibnamefont {Portillo-S\'anchez}},\ and\ \bibinfo {author} {\bibfnamefont {P.}~\bibnamefont {Roig}},\ }\Eprint {https://arxiv.org/abs/2411.07115} {arXiv:2411.07115 [hep-ph]}  (\bibinfo {year} {2024}{\natexlab{a}})\BibitemShut {NoStop}%
\bibitem [{\citenamefont {Miramontes}\ \emph {et~al.}(2024)\citenamefont {Miramontes}, \citenamefont {Raya}, \citenamefont {Bashir}, \citenamefont {Roig},\ and\ \citenamefont {Paredes-Torres}}]{Miramontes:2024fgo}%
  \BibitemOpen
  \bibfield  {author} {\bibinfo {author} {\bibfnamefont {A.~S.}\ \bibnamefont {Miramontes}}, \bibinfo {author} {\bibfnamefont {K.}~\bibnamefont {Raya}}, \bibinfo {author} {\bibfnamefont {A.}~\bibnamefont {Bashir}}, \bibinfo {author} {\bibfnamefont {P.}~\bibnamefont {Roig}},\ and\ \bibinfo {author} {\bibfnamefont {G.}~\bibnamefont {Paredes-Torres}},\ }\Eprint {https://arxiv.org/abs/2411.02218} {arXiv:2411.02218 [hep-ph]}  (\bibinfo {year} {2024})\BibitemShut {NoStop}%
\bibitem [{\citenamefont {Estrada}\ \emph {et~al.}(2024{\natexlab{b}})\citenamefont {Estrada}, \citenamefont {Gonz\`alez-Sol\'\i{}s}, \citenamefont {Guevara},\ and\ \citenamefont {Roig}}]{Estrada:2024cfy}%
  \BibitemOpen
  \bibfield  {author} {\bibinfo {author} {\bibfnamefont {E.~J.}\ \bibnamefont {Estrada}}, \bibinfo {author} {\bibfnamefont {S.}~\bibnamefont {Gonz\`alez-Sol\'\i{}s}}, \bibinfo {author} {\bibfnamefont {A.}~\bibnamefont {Guevara}},\ and\ \bibinfo {author} {\bibfnamefont {P.}~\bibnamefont {Roig}},\ }\Eprint {https://arxiv.org/abs/2409.10503} {arXiv:2409.10503 [hep-ph]}  (\bibinfo {year} {2024}{\natexlab{b}})\BibitemShut {NoStop}%
\bibitem [{\citenamefont {Asmussen}\ \emph {et~al.}(2023)\citenamefont {Asmussen}, \citenamefont {Chao}, \citenamefont {G\'erardin}, \citenamefont {Green}, \citenamefont {Hudspith}, \citenamefont {Meyer},\ and\ \citenamefont {Nyffeler}}]{Asmussen:2022oql}%
  \BibitemOpen
  \bibfield  {author} {\bibinfo {author} {\bibfnamefont {N.}~\bibnamefont {Asmussen}}, \bibinfo {author} {\bibfnamefont {E.-H.}\ \bibnamefont {Chao}}, \bibinfo {author} {\bibfnamefont {A.}~\bibnamefont {G\'erardin}}, \bibinfo {author} {\bibfnamefont {J.~R.}\ \bibnamefont {Green}}, \bibinfo {author} {\bibfnamefont {R.~J.}\ \bibnamefont {Hudspith}}, \bibinfo {author} {\bibfnamefont {H.~B.}\ \bibnamefont {Meyer}},\ and\ \bibinfo {author} {\bibfnamefont {A.}~\bibnamefont {Nyffeler}},\ }\href {https://doi.org/10.1007/JHEP04(2023)040} {\bibfield  {journal} {\bibinfo  {journal} {JHEP}\ }\textbf {\bibinfo {volume} {04}},\ \bibinfo {pages} {040}},\ \Eprint {https://arxiv.org/abs/2210.12263} {arXiv:2210.12263 [hep-lat]} \BibitemShut {NoStop}%
\bibitem [{\citenamefont {Chao}\ \emph {et~al.}(2022)\citenamefont {Chao}, \citenamefont {Hudspith}, \citenamefont {G\'erardin}, \citenamefont {Green},\ and\ \citenamefont {Meyer}}]{Chao:2022xzg}%
  \BibitemOpen
  \bibfield  {author} {\bibinfo {author} {\bibfnamefont {E.-H.}\ \bibnamefont {Chao}}, \bibinfo {author} {\bibfnamefont {R.~J.}\ \bibnamefont {Hudspith}}, \bibinfo {author} {\bibfnamefont {A.}~\bibnamefont {G\'erardin}}, \bibinfo {author} {\bibfnamefont {J.~R.}\ \bibnamefont {Green}},\ and\ \bibinfo {author} {\bibfnamefont {H.~B.}\ \bibnamefont {Meyer}},\ }\href {https://doi.org/10.1140/epjc/s10052-022-10589-2} {\bibfield  {journal} {\bibinfo  {journal} {Eur. Phys. J. C}\ }\textbf {\bibinfo {volume} {82}},\ \bibinfo {pages} {664} (\bibinfo {year} {2022})},\ \Eprint {https://arxiv.org/abs/2204.08844} {arXiv:2204.08844 [hep-lat]} \BibitemShut {NoStop}%
\bibitem [{\citenamefont {Blum}\ \emph {et~al.}(2023{\natexlab{a}})\citenamefont {Blum}, \citenamefont {Christ}, \citenamefont {Hayakawa}, \citenamefont {Izubuchi}, \citenamefont {Jin}, \citenamefont {Jung}, \citenamefont {Lehner},\ and\ \citenamefont {Tu}}]{Blum:2023vlm}%
  \BibitemOpen
  \bibfield  {author} {\bibinfo {author} {\bibfnamefont {T.}~\bibnamefont {Blum}}, \bibinfo {author} {\bibfnamefont {N.}~\bibnamefont {Christ}}, \bibinfo {author} {\bibfnamefont {M.}~\bibnamefont {Hayakawa}}, \bibinfo {author} {\bibfnamefont {T.}~\bibnamefont {Izubuchi}}, \bibinfo {author} {\bibfnamefont {L.}~\bibnamefont {Jin}}, \bibinfo {author} {\bibfnamefont {C.}~\bibnamefont {Jung}}, \bibinfo {author} {\bibfnamefont {C.}~\bibnamefont {Lehner}},\ and\ \bibinfo {author} {\bibfnamefont {C.}~\bibnamefont {Tu}},\ }\Eprint {https://arxiv.org/abs/2304.04423} {arXiv:2304.04423 [hep-lat]}  (\bibinfo {year} {2023}{\natexlab{a}})\BibitemShut {NoStop}%
\bibitem [{\citenamefont {G\'erardin}\ \emph {et~al.}(2023)\citenamefont {G\'erardin}, \citenamefont {Verplanke}, \citenamefont {Wang}, \citenamefont {Fodor}, \citenamefont {Guenther}, \citenamefont {Lellouch}, \citenamefont {Szabo},\ and\ \citenamefont {Varnhorst}}]{Gerardin:2023naa}%
  \BibitemOpen
  \bibfield  {author} {\bibinfo {author} {\bibfnamefont {A.}~\bibnamefont {G\'erardin}}, \bibinfo {author} {\bibfnamefont {W.~E.~A.}\ \bibnamefont {Verplanke}}, \bibinfo {author} {\bibfnamefont {G.}~\bibnamefont {Wang}}, \bibinfo {author} {\bibfnamefont {Z.}~\bibnamefont {Fodor}}, \bibinfo {author} {\bibfnamefont {J.~N.}\ \bibnamefont {Guenther}}, \bibinfo {author} {\bibfnamefont {L.}~\bibnamefont {Lellouch}}, \bibinfo {author} {\bibfnamefont {K.~K.}\ \bibnamefont {Szabo}},\ and\ \bibinfo {author} {\bibfnamefont {L.}~\bibnamefont {Varnhorst}},\ }\Eprint {https://arxiv.org/abs/2305.04570} {arXiv:2305.04570 [hep-lat]}  (\bibinfo {year} {2023})\BibitemShut {NoStop}%
\bibitem [{\citenamefont {Alexandrou}\ \emph {et~al.}(2023{\natexlab{a}})\citenamefont {Alexandrou} \emph {et~al.}}]{ExtendedTwistedMass:2023hin}%
  \BibitemOpen
  \bibfield  {author} {\bibinfo {author} {\bibfnamefont {C.}~\bibnamefont {Alexandrou}} \emph {et~al.} (\bibinfo {collaboration} {Extended Twisted Mass}),\ }\href {https://doi.org/10.1103/PhysRevD.108.094514} {\bibfield  {journal} {\bibinfo  {journal} {Phys. Rev. D}\ }\textbf {\bibinfo {volume} {108}},\ \bibinfo {pages} {094514} (\bibinfo {year} {2023}{\natexlab{a}})},\ \Eprint {https://arxiv.org/abs/2308.12458} {arXiv:2308.12458 [hep-lat]} \BibitemShut {NoStop}%
\bibitem [{\citenamefont {Lin}\ \emph {et~al.}(2024)\citenamefont {Lin}, \citenamefont {Bruno}, \citenamefont {Feng}, \citenamefont {Jin}, \citenamefont {Lehner}, \citenamefont {Liu},\ and\ \citenamefont {Luo}}]{Lin:2024khg}%
  \BibitemOpen
  \bibfield  {author} {\bibinfo {author} {\bibfnamefont {T.}~\bibnamefont {Lin}}, \bibinfo {author} {\bibfnamefont {M.}~\bibnamefont {Bruno}}, \bibinfo {author} {\bibfnamefont {X.}~\bibnamefont {Feng}}, \bibinfo {author} {\bibfnamefont {L.-C.}\ \bibnamefont {Jin}}, \bibinfo {author} {\bibfnamefont {C.}~\bibnamefont {Lehner}}, \bibinfo {author} {\bibfnamefont {C.}~\bibnamefont {Liu}},\ and\ \bibinfo {author} {\bibfnamefont {Q.-Y.}\ \bibnamefont {Luo}},\ }\Eprint {https://arxiv.org/abs/2411.06349} {arXiv:2411.06349 [hep-lat]}  (\bibinfo {year} {2024})\BibitemShut {NoStop}%
\bibitem [{\citenamefont {Fodor}\ \emph {et~al.}(2024)\citenamefont {Fodor}, \citenamefont {Gerardin}, \citenamefont {Lellouch}, \citenamefont {Szabo}, \citenamefont {Toth},\ and\ \citenamefont {Zimmermann}}]{Fodor:2024jyn}%
  \BibitemOpen
  \bibfield  {author} {\bibinfo {author} {\bibfnamefont {Z.}~\bibnamefont {Fodor}}, \bibinfo {author} {\bibfnamefont {A.}~\bibnamefont {Gerardin}}, \bibinfo {author} {\bibfnamefont {L.}~\bibnamefont {Lellouch}}, \bibinfo {author} {\bibfnamefont {K.~K.}\ \bibnamefont {Szabo}}, \bibinfo {author} {\bibfnamefont {B.~C.}\ \bibnamefont {Toth}},\ and\ \bibinfo {author} {\bibfnamefont {C.}~\bibnamefont {Zimmermann}},\ }\Eprint {https://arxiv.org/abs/2411.11719} {arXiv:2411.11719 [hep-lat]}  (\bibinfo {year} {2024})\BibitemShut {NoStop}%
\bibitem [{\citenamefont {Borsanyi}\ \emph {et~al.}(2021)\citenamefont {Borsanyi} \emph {et~al.}}]{Borsanyi:2020mff}%
  \BibitemOpen
  \bibfield  {author} {\bibinfo {author} {\bibfnamefont {S.}~\bibnamefont {Borsanyi}} \emph {et~al.} (\bibinfo {collaboration} {BMW}),\ }\href {https://doi.org/10.1038/s41586-021-03418-1} {\bibfield  {journal} {\bibinfo  {journal} {Nature}\ }\textbf {\bibinfo {volume} {593}},\ \bibinfo {pages} {51} (\bibinfo {year} {2021})}\BibitemShut {NoStop}%
\bibitem [{\citenamefont {Aubin}\ \emph {et~al.}(2022)\citenamefont {Aubin}, \citenamefont {Blum}, \citenamefont {Golterman},\ and\ \citenamefont {Peris}}]{Aubin:2022hgm}%
  \BibitemOpen
  \bibfield  {author} {\bibinfo {author} {\bibfnamefont {C.}~\bibnamefont {Aubin}}, \bibinfo {author} {\bibfnamefont {T.}~\bibnamefont {Blum}}, \bibinfo {author} {\bibfnamefont {M.}~\bibnamefont {Golterman}},\ and\ \bibinfo {author} {\bibfnamefont {S.}~\bibnamefont {Peris}},\ }\href {https://doi.org/10.1103/PhysRevD.106.054503} {\bibfield  {journal} {\bibinfo  {journal} {Phys. Rev. D}\ }\textbf {\bibinfo {volume} {106}},\ \bibinfo {pages} {054503} (\bibinfo {year} {2022})}\BibitemShut {NoStop}%
\bibitem [{\citenamefont {Alexandrou}\ \emph {et~al.}(2023{\natexlab{b}})\citenamefont {Alexandrou} \emph {et~al.}}]{Alexandrou:2022amy}%
  \BibitemOpen
  \bibfield  {author} {\bibinfo {author} {\bibfnamefont {C.}~\bibnamefont {Alexandrou}} \emph {et~al.} (\bibinfo {collaboration} {Extended Twisted Mass}),\ }\href {https://doi.org/10.1103/PhysRevD.107.074506} {\bibfield  {journal} {\bibinfo  {journal} {Phys. Rev. D}\ }\textbf {\bibinfo {volume} {107}},\ \bibinfo {pages} {074506} (\bibinfo {year} {2023}{\natexlab{b}})},\ \Eprint {https://arxiv.org/abs/2206.15084} {arXiv:2206.15084 [hep-lat]} \BibitemShut {NoStop}%
\bibitem [{\citenamefont {C\`e}\ \emph {et~al.}(2022)\citenamefont {C\`e} \emph {et~al.}}]{Ce:2022kxy}%
  \BibitemOpen
  \bibfield  {author} {\bibinfo {author} {\bibfnamefont {M.}~\bibnamefont {C\`e}} \emph {et~al.},\ }\href {https://doi.org/10.1103/PhysRevD.106.114502} {\bibfield  {journal} {\bibinfo  {journal} {Phys. Rev. D}\ }\textbf {\bibinfo {volume} {106}},\ \bibinfo {pages} {114502} (\bibinfo {year} {2022})}\BibitemShut {NoStop}%
\bibitem [{\citenamefont {Bazavov}\ \emph {et~al.}(2023)\citenamefont {Bazavov} \emph {et~al.}}]{FermilabLatticeHPQCD:2023jof}%
  \BibitemOpen
  \bibfield  {author} {\bibinfo {author} {\bibfnamefont {A.}~\bibnamefont {Bazavov}} \emph {et~al.} (\bibinfo {collaboration} {Fermilab Lattice, HPQCD, MILC}),\ }\href {https://doi.org/10.1103/PhysRevD.107.114514} {\bibfield  {journal} {\bibinfo  {journal} {Phys. Rev. D}\ }\textbf {\bibinfo {volume} {107}},\ \bibinfo {pages} {114514} (\bibinfo {year} {2023})},\ \Eprint {https://arxiv.org/abs/2301.08274} {arXiv:2301.08274 [hep-lat]} \BibitemShut {NoStop}%
\bibitem [{\citenamefont {Blum}\ \emph {et~al.}(2023{\natexlab{b}})\citenamefont {Blum} \emph {et~al.}}]{Blum:2023qou}%
  \BibitemOpen
  \bibfield  {author} {\bibinfo {author} {\bibfnamefont {T.}~\bibnamefont {Blum}} \emph {et~al.} (\bibinfo {collaboration} {RBC, UKQCD}),\ }\href {https://doi.org/10.1103/PhysRevD.108.054507} {\bibfield  {journal} {\bibinfo  {journal} {Phys. Rev. D}\ }\textbf {\bibinfo {volume} {108}},\ \bibinfo {pages} {054507} (\bibinfo {year} {2023}{\natexlab{b}})},\ \Eprint {https://arxiv.org/abs/2301.08696} {arXiv:2301.08696 [hep-lat]} \BibitemShut {NoStop}%
\bibitem [{\citenamefont {Boccaletti}\ \emph {et~al.}(2024)\citenamefont {Boccaletti} \emph {et~al.}}]{Boccaletti:2024guq}%
  \BibitemOpen
  \bibfield  {author} {\bibinfo {author} {\bibfnamefont {A.}~\bibnamefont {Boccaletti}} \emph {et~al.},\ }\href@noop {} {\  (\bibinfo {year} {2024})},\ \Eprint {https://arxiv.org/abs/2407.10913} {arXiv:2407.10913 [hep-lat]} \BibitemShut {NoStop}%
\bibitem [{\citenamefont {Kuberski}\ \emph {et~al.}(2024)\citenamefont {Kuberski}, \citenamefont {C\`e}, \citenamefont {von Hippel}, \citenamefont {Meyer}, \citenamefont {Ottnad}, \citenamefont {Risch},\ and\ \citenamefont {Wittig}}]{Kuberski:2024bcj}%
  \BibitemOpen
  \bibfield  {author} {\bibinfo {author} {\bibfnamefont {S.}~\bibnamefont {Kuberski}}, \bibinfo {author} {\bibfnamefont {M.}~\bibnamefont {C\`e}}, \bibinfo {author} {\bibfnamefont {G.}~\bibnamefont {von Hippel}}, \bibinfo {author} {\bibfnamefont {H.~B.}\ \bibnamefont {Meyer}}, \bibinfo {author} {\bibfnamefont {K.}~\bibnamefont {Ottnad}}, \bibinfo {author} {\bibfnamefont {A.}~\bibnamefont {Risch}},\ and\ \bibinfo {author} {\bibfnamefont {H.}~\bibnamefont {Wittig}},\ }\href {https://doi.org/10.1007/JHEP03(2024)172} {\bibfield  {journal} {\bibinfo  {journal} {JHEP}\ }\textbf {\bibinfo {volume} {03}},\ \bibinfo {pages} {172}},\ \Eprint {https://arxiv.org/abs/2401.11895} {arXiv:2401.11895 [hep-lat]} \BibitemShut {NoStop}%
\bibitem [{\citenamefont {Blum}\ \emph {et~al.}(2024)\citenamefont {Blum} \emph {et~al.}}]{RBC:2024fic}%
  \BibitemOpen
  \bibfield  {author} {\bibinfo {author} {\bibfnamefont {T.}~\bibnamefont {Blum}} \emph {et~al.} (\bibinfo {collaboration} {RBC, UKQCD}),\ }\Eprint {https://arxiv.org/abs/2410.20590} {arXiv:2410.20590 [hep-lat]}  (\bibinfo {year} {2024})\BibitemShut {NoStop}%
\bibitem [{\citenamefont {Spiegel}\ and\ \citenamefont {Lehner}(2024)}]{Spiegel:2024dec}%
  \BibitemOpen
  \bibfield  {author} {\bibinfo {author} {\bibfnamefont {S.}~\bibnamefont {Spiegel}}\ and\ \bibinfo {author} {\bibfnamefont {C.}~\bibnamefont {Lehner}},\ }\Eprint {https://arxiv.org/abs/2410.17053} {arXiv:2410.17053 [hep-lat]}  (\bibinfo {year} {2024})\BibitemShut {NoStop}%
\bibitem [{\citenamefont {Djukanovic}\ \emph {et~al.}(2025)\citenamefont {Djukanovic}, \citenamefont {von Hippel}, \citenamefont {Kuberski}, \citenamefont {Meyer}, \citenamefont {Miller}, \citenamefont {Ottnad}, \citenamefont {Parrino}, \citenamefont {Risch},\ and\ \citenamefont {Wittig}}]{Djukanovic:2024cmq}%
  \BibitemOpen
  \bibfield  {author} {\bibinfo {author} {\bibfnamefont {D.}~\bibnamefont {Djukanovic}}, \bibinfo {author} {\bibfnamefont {G.}~\bibnamefont {von Hippel}}, \bibinfo {author} {\bibfnamefont {S.}~\bibnamefont {Kuberski}}, \bibinfo {author} {\bibfnamefont {H.~B.}\ \bibnamefont {Meyer}}, \bibinfo {author} {\bibfnamefont {N.}~\bibnamefont {Miller}}, \bibinfo {author} {\bibfnamefont {K.}~\bibnamefont {Ottnad}}, \bibinfo {author} {\bibfnamefont {J.}~\bibnamefont {Parrino}}, \bibinfo {author} {\bibfnamefont {A.}~\bibnamefont {Risch}},\ and\ \bibinfo {author} {\bibfnamefont {H.}~\bibnamefont {Wittig}},\ }\href {https://doi.org/10.1007/JHEP04(2025)098} {\bibfield  {journal} {\bibinfo  {journal} {JHEP}\ }\textbf {\bibinfo {volume} {04}},\ \bibinfo {pages} {098}},\ \Eprint {https://arxiv.org/abs/2411.07969} {arXiv:2411.07969 [hep-lat]} \BibitemShut {NoStop}%
\bibitem [{\citenamefont {Alexandrou}\ \emph {et~al.}(2024)\citenamefont {Alexandrou} \emph {et~al.}}]{ExtendedTwistedMassCollaborationETMC:2024xdf}%
  \BibitemOpen
  \bibfield  {author} {\bibinfo {author} {\bibfnamefont {C.}~\bibnamefont {Alexandrou}} \emph {et~al.} (\bibinfo {collaboration} {Extended Twisted Mass Collaboration (ETMC)}),\ }\Eprint {https://arxiv.org/abs/2411.08852} {arXiv:2411.08852 [hep-lat]}  (\bibinfo {year} {2024})\BibitemShut {NoStop}%
\bibitem [{\citenamefont {Bazavov}\ \emph {et~al.}(2024{\natexlab{a}})\citenamefont {Bazavov} \emph {et~al.}}]{FermilabLattice:2024yho}%
  \BibitemOpen
  \bibfield  {author} {\bibinfo {author} {\bibfnamefont {A.}~\bibnamefont {Bazavov}} \emph {et~al.} (\bibinfo {collaboration} {Fermilab Lattice, HPQCD, MILC}),\ }\Eprint {https://arxiv.org/abs/2411.09656} {arXiv:2411.09656 [hep-lat]}  (\bibinfo {year} {2024}{\natexlab{a}})\BibitemShut {NoStop}%
\bibitem [{\citenamefont {Benton}\ \emph {et~al.}(2024{\natexlab{a}})\citenamefont {Benton}, \citenamefont {Boito}, \citenamefont {Golterman}, \citenamefont {Keshavarzi}, \citenamefont {Maltman},\ and\ \citenamefont {Peris}}]{Benton:2024kwp}%
  \BibitemOpen
  \bibfield  {author} {\bibinfo {author} {\bibfnamefont {G.}~\bibnamefont {Benton}}, \bibinfo {author} {\bibfnamefont {D.}~\bibnamefont {Boito}}, \bibinfo {author} {\bibfnamefont {M.}~\bibnamefont {Golterman}}, \bibinfo {author} {\bibfnamefont {A.}~\bibnamefont {Keshavarzi}}, \bibinfo {author} {\bibfnamefont {K.}~\bibnamefont {Maltman}},\ and\ \bibinfo {author} {\bibfnamefont {S.}~\bibnamefont {Peris}},\ }\Eprint {https://arxiv.org/abs/2411.06637} {arXiv:2411.06637 [hep-ph]}  (\bibinfo {year} {2024}{\natexlab{a}})\BibitemShut {NoStop}%
\bibitem [{\citenamefont {Blum}\ \emph {et~al.}(2018)\citenamefont {Blum} \emph {et~al.}}]{RBC:2018dos}%
  \BibitemOpen
  \bibfield  {author} {\bibinfo {author} {\bibfnamefont {T.}~\bibnamefont {Blum}} \emph {et~al.} (\bibinfo {collaboration} {RBC, UKQCD}),\ }\href {https://doi.org/10.1103/PhysRevLett.121.022003} {\bibfield  {journal} {\bibinfo  {journal} {Phys. Rev. Lett.}\ }\textbf {\bibinfo {volume} {121}},\ \bibinfo {pages} {022003} (\bibinfo {year} {2018})}\BibitemShut {NoStop}%
\bibitem [{\citenamefont {Colangelo}\ \emph {et~al.}(2022{\natexlab{b}})\citenamefont {Colangelo}, \citenamefont {El-Khadra}, \citenamefont {Hoferichter}, \citenamefont {Keshavarzi}, \citenamefont {Lehner}, \citenamefont {Stoffer},\ and\ \citenamefont {Teubner}}]{Colangelo:2022vok}%
  \BibitemOpen
  \bibfield  {author} {\bibinfo {author} {\bibfnamefont {G.}~\bibnamefont {Colangelo}}, \bibinfo {author} {\bibfnamefont {A.~X.}\ \bibnamefont {El-Khadra}}, \bibinfo {author} {\bibfnamefont {M.}~\bibnamefont {Hoferichter}}, \bibinfo {author} {\bibfnamefont {A.}~\bibnamefont {Keshavarzi}}, \bibinfo {author} {\bibfnamefont {C.}~\bibnamefont {Lehner}}, \bibinfo {author} {\bibfnamefont {P.}~\bibnamefont {Stoffer}},\ and\ \bibinfo {author} {\bibfnamefont {T.}~\bibnamefont {Teubner}},\ }\href {https://doi.org/10.1016/j.physletb.2022.137313} {\bibfield  {journal} {\bibinfo  {journal} {Phys. Lett. B}\ }\textbf {\bibinfo {volume} {833}},\ \bibinfo {pages} {137313} (\bibinfo {year} {2022}{\natexlab{b}})}\BibitemShut {NoStop}%
\bibitem [{\citenamefont {Benton}\ \emph {et~al.}(2023)\citenamefont {Benton}, \citenamefont {Boito}, \citenamefont {Golterman}, \citenamefont {Keshavarzi}, \citenamefont {Maltman},\ and\ \citenamefont {Peris}}]{Benton:2023dci}%
  \BibitemOpen
  \bibfield  {author} {\bibinfo {author} {\bibfnamefont {G.}~\bibnamefont {Benton}}, \bibinfo {author} {\bibfnamefont {D.}~\bibnamefont {Boito}}, \bibinfo {author} {\bibfnamefont {M.}~\bibnamefont {Golterman}}, \bibinfo {author} {\bibfnamefont {A.}~\bibnamefont {Keshavarzi}}, \bibinfo {author} {\bibfnamefont {K.}~\bibnamefont {Maltman}},\ and\ \bibinfo {author} {\bibfnamefont {S.}~\bibnamefont {Peris}},\ }\href {https://doi.org/10.1103/PhysRevLett.131.251803} {\bibfield  {journal} {\bibinfo  {journal} {Phys. Rev. Lett.}\ }\textbf {\bibinfo {volume} {131}},\ \bibinfo {pages} {251803} (\bibinfo {year} {2023})},\ \Eprint {https://arxiv.org/abs/2306.16808} {arXiv:2306.16808 [hep-ph]} \BibitemShut {NoStop}%
\bibitem [{\citenamefont {Davier}\ \emph {et~al.}(2024)\citenamefont {Davier}, \citenamefont {Fodor}, \citenamefont {Gerardin}, \citenamefont {Lellouch}, \citenamefont {Malaescu}, \citenamefont {Stokes}, \citenamefont {Szabo}, \citenamefont {Toth}, \citenamefont {Varnhorst},\ and\ \citenamefont {Zhang}}]{Davier:2023cyp}%
  \BibitemOpen
  \bibfield  {author} {\bibinfo {author} {\bibfnamefont {M.}~\bibnamefont {Davier}}, \bibinfo {author} {\bibfnamefont {Z.}~\bibnamefont {Fodor}}, \bibinfo {author} {\bibfnamefont {A.}~\bibnamefont {Gerardin}}, \bibinfo {author} {\bibfnamefont {L.}~\bibnamefont {Lellouch}}, \bibinfo {author} {\bibfnamefont {B.}~\bibnamefont {Malaescu}}, \bibinfo {author} {\bibfnamefont {F.~M.}\ \bibnamefont {Stokes}}, \bibinfo {author} {\bibfnamefont {K.~K.}\ \bibnamefont {Szabo}}, \bibinfo {author} {\bibfnamefont {B.~C.}\ \bibnamefont {Toth}}, \bibinfo {author} {\bibfnamefont {L.}~\bibnamefont {Varnhorst}},\ and\ \bibinfo {author} {\bibfnamefont {Z.}~\bibnamefont {Zhang}},\ }\href {https://doi.org/10.1103/PhysRevD.109.076019} {\bibfield  {journal} {\bibinfo  {journal} {Phys. Rev. D}\ }\textbf {\bibinfo {volume} {109}},\ \bibinfo {pages} {076019} (\bibinfo {year} {2024})},\ \Eprint {https://arxiv.org/abs/2308.04221} {arXiv:2308.04221 [hep-ph]} \BibitemShut {NoStop}%
\bibitem [{\citenamefont {Benton}\ \emph {et~al.}(2024{\natexlab{b}})\citenamefont {Benton}, \citenamefont {Boito}, \citenamefont {Golterman}, \citenamefont {Keshavarzi}, \citenamefont {Maltman},\ and\ \citenamefont {Peris}}]{Benton:2023fcv}%
  \BibitemOpen
  \bibfield  {author} {\bibinfo {author} {\bibfnamefont {G.}~\bibnamefont {Benton}}, \bibinfo {author} {\bibfnamefont {D.}~\bibnamefont {Boito}}, \bibinfo {author} {\bibfnamefont {M.}~\bibnamefont {Golterman}}, \bibinfo {author} {\bibfnamefont {A.}~\bibnamefont {Keshavarzi}}, \bibinfo {author} {\bibfnamefont {K.}~\bibnamefont {Maltman}},\ and\ \bibinfo {author} {\bibfnamefont {S.}~\bibnamefont {Peris}},\ }\href {https://doi.org/10.1103/PhysRevD.109.036010} {\bibfield  {journal} {\bibinfo  {journal} {Phys. Rev. D}\ }\textbf {\bibinfo {volume} {109}},\ \bibinfo {pages} {036010} (\bibinfo {year} {2024}{\natexlab{b}})},\ \Eprint {https://arxiv.org/abs/2311.09523} {arXiv:2311.09523 [hep-ph]} \BibitemShut {NoStop}%
\bibitem [{\citenamefont {Della~Morte}\ \emph {et~al.}(2017)\citenamefont {Della~Morte}, \citenamefont {Francis}, \citenamefont {G\"ulpers}, \citenamefont {Herdo\'\i{}za}, \citenamefont {von Hippel}, \citenamefont {Horch}, \citenamefont {J\"ager}, \citenamefont {Meyer}, \citenamefont {Nyffeler},\ and\ \citenamefont {Wittig}}]{DellaMorte:2017dyu}%
  \BibitemOpen
  \bibfield  {author} {\bibinfo {author} {\bibfnamefont {M.}~\bibnamefont {Della~Morte}}, \bibinfo {author} {\bibfnamefont {A.}~\bibnamefont {Francis}}, \bibinfo {author} {\bibfnamefont {V.}~\bibnamefont {G\"ulpers}}, \bibinfo {author} {\bibfnamefont {G.}~\bibnamefont {Herdo\'\i{}za}}, \bibinfo {author} {\bibfnamefont {G.}~\bibnamefont {von Hippel}}, \bibinfo {author} {\bibfnamefont {H.}~\bibnamefont {Horch}}, \bibinfo {author} {\bibfnamefont {B.}~\bibnamefont {J\"ager}}, \bibinfo {author} {\bibfnamefont {H.~B.}\ \bibnamefont {Meyer}}, \bibinfo {author} {\bibfnamefont {A.}~\bibnamefont {Nyffeler}},\ and\ \bibinfo {author} {\bibfnamefont {H.}~\bibnamefont {Wittig}},\ }\href {https://doi.org/10.1007/JHEP10(2017)020} {\bibfield  {journal} {\bibinfo  {journal} {JHEP}\ }\textbf {\bibinfo {volume} {10}}\bibinfo  {number} { (2017)},\ \bibinfo {pages} {020}}\BibitemShut {NoStop}%
\bibitem [{\citenamefont {Bazavov}\ \emph {et~al.}(2024{\natexlab{b}})\citenamefont {Bazavov} \emph {et~al.}}]{Bazavov:2024dov}%
  \BibitemOpen
\bibfield  {number} {  }\bibfield  {author} {\bibinfo {author} {\bibfnamefont {A.}~\bibnamefont {Bazavov}} \emph {et~al.},\ }\href {https://doi.org/10.22323/1.453.0292} {\bibfield  {journal} {\bibinfo  {journal} {PoS}\ }\textbf {\bibinfo {volume} {LATTICE2023}},\ \bibinfo {pages} {292} (\bibinfo {year} {2024}{\natexlab{b}})},\ \Eprint {https://arxiv.org/abs/2401.06522} {arXiv:2401.06522 [hep-lat]} \BibitemShut {NoStop}%
\bibitem [{\citenamefont {Bazavov}\ \emph {et~al.}(2025{\natexlab{a}})\citenamefont {Bazavov} \emph {et~al.}}]{Grebe:2025}%
  \BibitemOpen
  \bibfield  {author} {\bibinfo {author} {\bibfnamefont {A.}~\bibnamefont {Bazavov}} \emph {et~al.} (\bibinfo {collaboration} {Fermilab Lattice, MILC})} (\bibinfo {year} {2025}{\natexlab{a}}),\ \bibinfo {note} {in preparation.}\BibitemShut {Stop}%
\bibitem [{\citenamefont {Blum}(2003)}]{Blum:2002ii}%
  \BibitemOpen
  \bibfield  {author} {\bibinfo {author} {\bibfnamefont {T.}~\bibnamefont {Blum}},\ }\href {https://doi.org/10.1103/PhysRevLett.91.052001} {\bibfield  {journal} {\bibinfo  {journal} {Phys. Rev. Lett.}\ }\textbf {\bibinfo {volume} {91}},\ \bibinfo {pages} {052001} (\bibinfo {year} {2003})},\ \Eprint {https://arxiv.org/abs/hep-lat/0212018} {arXiv:hep-lat/0212018} \BibitemShut {NoStop}%
\bibitem [{\citenamefont {Bernecker}\ and\ \citenamefont {Meyer}(2011)}]{Bernecker:2011gh}%
  \BibitemOpen
  \bibfield  {author} {\bibinfo {author} {\bibfnamefont {D.}~\bibnamefont {Bernecker}}\ and\ \bibinfo {author} {\bibfnamefont {H.~B.}\ \bibnamefont {Meyer}},\ }\href {https://doi.org/10.1140/epja/i2011-11148-6} {\bibfield  {journal} {\bibinfo  {journal} {Eur. Phys. J. A}\ }\textbf {\bibinfo {volume} {47}},\ \bibinfo {pages} {148} (\bibinfo {year} {2011})}\BibitemShut {NoStop}%
\bibitem [{\citenamefont {Bazavov}\ \emph {et~al.}(2025{\natexlab{b}})\citenamefont {Bazavov} \emph {et~al.}}]{QEDpaper}%
  \BibitemOpen
  \bibfield  {author} {\bibinfo {author} {\bibfnamefont {A.}~\bibnamefont {Bazavov}} \emph {et~al.} (\bibinfo {collaboration} {Fermilab Lattice, HPQCD, MILC})} (\bibinfo {year} {2025}{\natexlab{b}}),\ \bibinfo {note} {in preparation.}\BibitemShut {Stop}%
\bibitem [{sup()}]{supplement}%
  \BibitemOpen
  \href@noop {} {}\bibinfo {note} {See Supplemental Material at [URL will be inserted by publisher], which includes Refs. [81-84], for additional details on the noise-reduction strategy, lattice corrections, continuum extrapolation, and Bayesian model averaging procedures.}\BibitemShut {Stop}%
\bibitem [{\citenamefont {Borsanyi}\ \emph {et~al.}(2018)\citenamefont {Borsanyi} \emph {et~al.}}]{Budapest-Marseille-Wuppertal:2017okr}%
  \BibitemOpen
  \bibfield  {author} {\bibinfo {author} {\bibfnamefont {S.}~\bibnamefont {Borsanyi}} \emph {et~al.} (\bibinfo {collaboration} {Budapest-Marseille-Wuppertal}),\ }\href {https://doi.org/10.1103/PhysRevLett.121.022002} {\bibfield  {journal} {\bibinfo  {journal} {Phys. Rev. Lett.}\ }\textbf {\bibinfo {volume} {121}},\ \bibinfo {pages} {022002} (\bibinfo {year} {2018})},\ \Eprint {https://arxiv.org/abs/1711.04980} {arXiv:1711.04980 [hep-lat]} \BibitemShut {NoStop}%
\bibitem [{\citenamefont {Lepage}\ \emph {et~al.}(2002)\citenamefont {Lepage} \emph {et~al.}}]{Lepage:2001ym}%
  \BibitemOpen
  \bibfield  {author} {\bibinfo {author} {\bibfnamefont {G.~P.}\ \bibnamefont {Lepage}} \emph {et~al.},\ }\href {https://doi.org/10.1016/S0920-5632(01)01638-3} {\bibfield  {journal} {\bibinfo  {journal} {Nucl. Phys. B Proc. Suppl.}\ }\textbf {\bibinfo {volume} {106}},\ \bibinfo {pages} {12} (\bibinfo {year} {2002})}\BibitemShut {NoStop}%
\bibitem [{\citenamefont {Bouchard}\ \emph {et~al.}(2014)\citenamefont {Bouchard}, \citenamefont {Lepage}, \citenamefont {Monahan}, \citenamefont {Na},\ and\ \citenamefont {Shigemitsu}}]{Bouchard:2014ypa}%
  \BibitemOpen
  \bibfield  {author} {\bibinfo {author} {\bibfnamefont {C.~M.}\ \bibnamefont {Bouchard}}, \bibinfo {author} {\bibfnamefont {G.~P.}\ \bibnamefont {Lepage}}, \bibinfo {author} {\bibfnamefont {C.}~\bibnamefont {Monahan}}, \bibinfo {author} {\bibfnamefont {H.}~\bibnamefont {Na}},\ and\ \bibinfo {author} {\bibfnamefont {J.}~\bibnamefont {Shigemitsu}},\ }\href {https://doi.org/10.1103/PhysRevD.90.054506} {\bibfield  {journal} {\bibinfo  {journal} {Phys. Rev. D}\ }\textbf {\bibinfo {volume} {90}},\ \bibinfo {pages} {054506} (\bibinfo {year} {2014})},\ \Eprint {https://arxiv.org/abs/1406.2279} {arXiv:1406.2279 [hep-lat]} \BibitemShut {NoStop}%
\bibitem [{\citenamefont {Bazavov}\ \emph {et~al.}(2016)\citenamefont {Bazavov} \emph {et~al.}}]{FermilabLattice:2016ipl}%
  \BibitemOpen
  \bibfield  {author} {\bibinfo {author} {\bibfnamefont {A.}~\bibnamefont {Bazavov}} \emph {et~al.} (\bibinfo {collaboration} {Fermilab Lattice, MILC}),\ }\href {https://doi.org/10.1103/PhysRevD.93.113016} {\bibfield  {journal} {\bibinfo  {journal} {Phys. Rev. D}\ }\textbf {\bibinfo {volume} {93}},\ \bibinfo {pages} {113016} (\bibinfo {year} {2016})}\BibitemShut {NoStop}%
\bibitem [{\citenamefont {Aoki}\ \emph {et~al.}(2024)\citenamefont {Aoki} \emph {et~al.}}]{FlavourLatticeAveragingGroupFLAG:2024oxs}%
  \BibitemOpen
  \bibfield  {author} {\bibinfo {author} {\bibfnamefont {Y.}~\bibnamefont {Aoki}} \emph {et~al.} (\bibinfo {collaboration} {Flavour Lattice Averaging Group (FLAG)}),\ }\Eprint {https://arxiv.org/abs/2411.04268} {arXiv:2411.04268 [hep-lat]}  (\bibinfo {year} {2024})\BibitemShut {NoStop}%
\bibitem [{\citenamefont {Bazavov}\ \emph {et~al.}(2014)\citenamefont {Bazavov} \emph {et~al.}}]{FermilabLattice:2014tsy}%
  \BibitemOpen
  \bibfield  {author} {\bibinfo {author} {\bibfnamefont {A.}~\bibnamefont {Bazavov}} \emph {et~al.} (\bibinfo {collaboration} {Fermilab Lattice, MILC}),\ }\href {https://doi.org/10.1103/PhysRevD.90.074509} {\bibfield  {journal} {\bibinfo  {journal} {Phys. Rev. D}\ }\textbf {\bibinfo {volume} {90}},\ \bibinfo {pages} {074509} (\bibinfo {year} {2014})}\BibitemShut {NoStop}%
\bibitem [{\citenamefont {Chakraborty}\ \emph {et~al.}(2018)\citenamefont {Chakraborty} \emph {et~al.}}]{FermilabLattice:2017wgj}%
  \BibitemOpen
  \bibfield  {author} {\bibinfo {author} {\bibfnamefont {B.}~\bibnamefont {Chakraborty}} \emph {et~al.} (\bibinfo {collaboration} {Fermilab Lattice, HPQCD, MILC}),\ }\href {https://doi.org/10.1103/PhysRevLett.120.152001} {\bibfield  {journal} {\bibinfo  {journal} {Phys. Rev. Lett.}\ }\textbf {\bibinfo {volume} {120}},\ \bibinfo {pages} {152001} (\bibinfo {year} {2018})}\BibitemShut {NoStop}%
\bibitem [{\citenamefont {Bazavov}\ \emph {et~al.}(2025{\natexlab{c}})\citenamefont {Bazavov}, \citenamefont {Bernard} \emph {et~al.}}]{Bernard:2025}%
  \BibitemOpen
  \bibfield  {author} {\bibinfo {author} {\bibfnamefont {A.}~\bibnamefont {Bazavov}}, \bibinfo {author} {\bibfnamefont {C.}~\bibnamefont {Bernard}}, \emph {et~al.} (\bibinfo {collaboration} {Fermilab Lattice, MILC})} (\bibinfo {year} {2025}{\natexlab{c}}),\ \bibinfo {note} {in preparation.}\BibitemShut {Stop}%
\bibitem [{\citenamefont {Workman}\ \emph {et~al.}(2022)\citenamefont {Workman} \emph {et~al.}}]{ParticleDataGroup:2022pth}%
  \BibitemOpen
  \bibfield  {author} {\bibinfo {author} {\bibfnamefont {R.~L.}\ \bibnamefont {Workman}} \emph {et~al.} (\bibinfo {collaboration} {Particle Data Group}),\ }\href {https://doi.org/10.1093/ptep/ptac097} {\bibfield  {journal} {\bibinfo  {journal} {PTEP}\ }\textbf {\bibinfo {volume} {2022}},\ \bibinfo {pages} {083C01} (\bibinfo {year} {2022})}\BibitemShut {NoStop}%
\bibitem [{\citenamefont {Bazavov}\ \emph {et~al.}(2010)\citenamefont {Bazavov} \emph {et~al.}}]{MILC:2010pul}%
  \BibitemOpen
  \bibfield  {author} {\bibinfo {author} {\bibfnamefont {A.}~\bibnamefont {Bazavov}} \emph {et~al.} (\bibinfo {collaboration} {MILC}),\ }\href {https://doi.org/10.1103/PhysRevD.82.074501} {\bibfield  {journal} {\bibinfo  {journal} {Phys. Rev. D}\ }\textbf {\bibinfo {volume} {82}},\ \bibinfo {pages} {074501} (\bibinfo {year} {2010})}\BibitemShut {NoStop}%
\bibitem [{\citenamefont {Bazavov}\ \emph {et~al.}(2013)\citenamefont {Bazavov} \emph {et~al.}}]{MILC:2012znn}%
  \BibitemOpen
  \bibfield  {author} {\bibinfo {author} {\bibfnamefont {A.}~\bibnamefont {Bazavov}} \emph {et~al.} (\bibinfo {collaboration} {MILC}),\ }\href {https://doi.org/10.1103/PhysRevD.87.054505} {\bibfield  {journal} {\bibinfo  {journal} {Phys. Rev. D}\ }\textbf {\bibinfo {volume} {87}},\ \bibinfo {pages} {054505} (\bibinfo {year} {2013})}\BibitemShut {NoStop}%
\bibitem [{\citenamefont {Bazavov}\ \emph {et~al.}(2018)\citenamefont {Bazavov} \emph {et~al.}}]{Bazavov:2017lyh}%
  \BibitemOpen
  \bibfield  {author} {\bibinfo {author} {\bibfnamefont {A.}~\bibnamefont {Bazavov}} \emph {et~al.} (\bibinfo {collaboration} {Fermilab Lattice, MILC}),\ }\href {https://doi.org/10.1103/PhysRevD.98.074512} {\bibfield  {journal} {\bibinfo  {journal} {Phys. Rev. D}\ }\textbf {\bibinfo {volume} {98}},\ \bibinfo {pages} {074512} (\bibinfo {year} {2018})}\BibitemShut {NoStop}%
\bibitem [{\citenamefont {MILC}(2021)}]{MILCConfigsGitHub}%
  \BibitemOpen
  \bibfield  {author} {\bibinfo {author} {\bibnamefont {MILC}},\ }\href@noop {} {}\bibinfo {howpublished} {\href{https://github.com/milc-qcd/sharing/wiki/LatticeSharing}{\texttt{HISQ Gauge Configurations}}} (\bibinfo {year} {2021})\BibitemShut {NoStop}%
\bibitem [{\citenamefont {Miller}\ \emph {et~al.}(2021)\citenamefont {Miller} \emph {et~al.}}]{Miller:2020evg}%
  \BibitemOpen
  \bibfield  {author} {\bibinfo {author} {\bibfnamefont {N.}~\bibnamefont {Miller}} \emph {et~al.},\ }\href {https://doi.org/10.1103/PhysRevD.103.054511} {\bibfield  {journal} {\bibinfo  {journal} {Phys. Rev. D}\ }\textbf {\bibinfo {volume} {103}},\ \bibinfo {pages} {054511} (\bibinfo {year} {2021})},\ \Eprint {https://arxiv.org/abs/2011.12166} {arXiv:2011.12166 [hep-lat]} \BibitemShut {NoStop}%
\bibitem [{\citenamefont {Hatton}\ \emph {et~al.}(2019)\citenamefont {Hatton}, \citenamefont {Davies}, \citenamefont {Lepage},\ and\ \citenamefont {Lytle}}]{Hatton:2019gha}%
  \BibitemOpen
  \bibfield  {author} {\bibinfo {author} {\bibfnamefont {D.}~\bibnamefont {Hatton}}, \bibinfo {author} {\bibfnamefont {C.~T.~H.}\ \bibnamefont {Davies}}, \bibinfo {author} {\bibfnamefont {G.~P.}\ \bibnamefont {Lepage}},\ and\ \bibinfo {author} {\bibfnamefont {A.~T.}\ \bibnamefont {Lytle}} (\bibinfo {collaboration} {HPQCD}),\ }\href {https://doi.org/10.1103/PhysRevD.100.114513} {\bibfield  {journal} {\bibinfo  {journal} {Phys. Rev. D}\ }\textbf {\bibinfo {volume} {100}},\ \bibinfo {pages} {114513} (\bibinfo {year} {2019})},\ \Eprint {https://arxiv.org/abs/1909.00756} {arXiv:1909.00756 [hep-lat]} \BibitemShut {NoStop}%
\bibitem [{\citenamefont {Hatton}\ \emph {et~al.}(2020)\citenamefont {Hatton}, \citenamefont {Davies}, \citenamefont {Galloway}, \citenamefont {Koponen}, \citenamefont {Lepage},\ and\ \citenamefont {Lytle}}]{Hatton:2020qhk}%
  \BibitemOpen
  \bibfield  {author} {\bibinfo {author} {\bibfnamefont {D.}~\bibnamefont {Hatton}}, \bibinfo {author} {\bibfnamefont {C.~T.~H.}\ \bibnamefont {Davies}}, \bibinfo {author} {\bibfnamefont {B.}~\bibnamefont {Galloway}}, \bibinfo {author} {\bibfnamefont {J.}~\bibnamefont {Koponen}}, \bibinfo {author} {\bibfnamefont {G.~P.}\ \bibnamefont {Lepage}},\ and\ \bibinfo {author} {\bibfnamefont {A.~T.}\ \bibnamefont {Lytle}} (\bibinfo {collaboration} {HPQCD}),\ }\href {https://doi.org/10.1103/PhysRevD.102.054511} {\bibfield  {journal} {\bibinfo  {journal} {Phys. Rev. D}\ }\textbf {\bibinfo {volume} {102}},\ \bibinfo {pages} {054511} (\bibinfo {year} {2020})}\BibitemShut {NoStop}%
\bibitem [{\citenamefont {DeGrand}\ and\ \citenamefont {Schaefer}(2004)}]{DeGrand:2004qw}%
  \BibitemOpen
  \bibfield  {author} {\bibinfo {author} {\bibfnamefont {T.~A.}\ \bibnamefont {DeGrand}}\ and\ \bibinfo {author} {\bibfnamefont {S.}~\bibnamefont {Schaefer}},\ }\href {https://doi.org/10.1016/j.cpc.2004.02.006} {\bibfield  {journal} {\bibinfo  {journal} {Comput. Phys. Commun.}\ }\textbf {\bibinfo {volume} {159}},\ \bibinfo {pages} {185} (\bibinfo {year} {2004})}\BibitemShut {NoStop}%
\bibitem [{\citenamefont {Giusti}\ \emph {et~al.}(2004)\citenamefont {Giusti}, \citenamefont {Hernandez}, \citenamefont {Laine}, \citenamefont {Weisz},\ and\ \citenamefont {Wittig}}]{Giusti:2004yp}%
  \BibitemOpen
  \bibfield  {author} {\bibinfo {author} {\bibfnamefont {L.}~\bibnamefont {Giusti}}, \bibinfo {author} {\bibfnamefont {P.}~\bibnamefont {Hernandez}}, \bibinfo {author} {\bibfnamefont {M.}~\bibnamefont {Laine}}, \bibinfo {author} {\bibfnamefont {P.}~\bibnamefont {Weisz}},\ and\ \bibinfo {author} {\bibfnamefont {H.}~\bibnamefont {Wittig}},\ }\href {https://doi.org/10.1088/1126-6708/2004/04/013} {\bibfield  {journal} {\bibinfo  {journal} {JHEP}\ }\textbf {\bibinfo {volume} {04}}\bibinfo  {number} { (2004)},\ \bibinfo {pages} {013}}\BibitemShut {NoStop}%
\bibitem [{\citenamefont {Blum}\ \emph {et~al.}(2013)\citenamefont {Blum}, \citenamefont {Izubuchi},\ and\ \citenamefont {Shintani}}]{Blum:2012uh}%
  \BibitemOpen
\bibfield  {number} {  }\bibfield  {author} {\bibinfo {author} {\bibfnamefont {T.}~\bibnamefont {Blum}}, \bibinfo {author} {\bibfnamefont {T.}~\bibnamefont {Izubuchi}},\ and\ \bibinfo {author} {\bibfnamefont {E.}~\bibnamefont {Shintani}},\ }\href {https://doi.org/10.1103/PhysRevD.88.094503} {\bibfield  {journal} {\bibinfo  {journal} {Phys. Rev. D}\ }\textbf {\bibinfo {volume} {88}},\ \bibinfo {pages} {094503} (\bibinfo {year} {2013})},\ \Eprint {https://arxiv.org/abs/1208.4349} {arXiv:1208.4349 [hep-lat]} \BibitemShut {NoStop}%
\bibitem [{\citenamefont {Lepage}(1989)}]{Lepage:1989hd}%
  \BibitemOpen
  \bibfield  {author} {\bibinfo {author} {\bibfnamefont {G.~P.}\ \bibnamefont {Lepage}},\ }in\ \href {https://lib-extopc.kek.jp/preprints/PDF/1990/9003/9003479.pdf} {\emph {\bibinfo {booktitle} {{Theoretical Advanced Study Institute in Elementary Particle Physics}}}}\ (\bibinfo {year} {1989})\BibitemShut {NoStop}%
\bibitem [{\citenamefont {Lahert}\ \emph {et~al.}(2024)\citenamefont {Lahert}, \citenamefont {DeTar}, \citenamefont {El-Khadra}, \citenamefont {Gottlieb}, \citenamefont {Kronfeld},\ and\ \citenamefont {Van~de Water}}]{Lahert:2024vvu}%
  \BibitemOpen
  \bibfield  {author} {\bibinfo {author} {\bibfnamefont {S.}~\bibnamefont {Lahert}}, \bibinfo {author} {\bibfnamefont {C.}~\bibnamefont {DeTar}}, \bibinfo {author} {\bibfnamefont {A.~X.}\ \bibnamefont {El-Khadra}}, \bibinfo {author} {\bibfnamefont {S.}~\bibnamefont {Gottlieb}}, \bibinfo {author} {\bibfnamefont {A.~S.}\ \bibnamefont {Kronfeld}},\ and\ \bibinfo {author} {\bibfnamefont {R.~S.}\ \bibnamefont {Van~de Water}},\ }\Eprint {https://arxiv.org/abs/2409.00756} {arXiv:2409.00756 [hep-lat]}  (\bibinfo {year} {2024})\BibitemShut {NoStop}%
\bibitem [{\citenamefont {Chakraborty}\ \emph {et~al.}(2017{\natexlab{a}})\citenamefont {Chakraborty}, \citenamefont {Davies}, \citenamefont {de~Oliviera}, \citenamefont {Koponen}, \citenamefont {Lepage},\ and\ \citenamefont {Van~de Water}}]{Chakraborty:2016mwy}%
  \BibitemOpen
  \bibfield  {author} {\bibinfo {author} {\bibfnamefont {B.}~\bibnamefont {Chakraborty}}, \bibinfo {author} {\bibfnamefont {C.~T.~H.}\ \bibnamefont {Davies}}, \bibinfo {author} {\bibfnamefont {P.~G.}\ \bibnamefont {de~Oliviera}}, \bibinfo {author} {\bibfnamefont {J.}~\bibnamefont {Koponen}}, \bibinfo {author} {\bibfnamefont {G.~P.}\ \bibnamefont {Lepage}},\ and\ \bibinfo {author} {\bibfnamefont {R.~S.}\ \bibnamefont {Van~de Water}} (\bibinfo {collaboration} {HPQCD}),\ }\href {https://doi.org/10.1103/PhysRevD.96.034516} {\bibfield  {journal} {\bibinfo  {journal} {Phys. Rev. D}\ }\textbf {\bibinfo {volume} {96}},\ \bibinfo {pages} {034516} (\bibinfo {year} {2017}{\natexlab{a}})}\BibitemShut {NoStop}%
\bibitem [{\citenamefont {Neil}\ and\ \citenamefont {Sitison}(2022)}]{Neil:2022joj}%
  \BibitemOpen
  \bibfield  {author} {\bibinfo {author} {\bibfnamefont {E.~T.}\ \bibnamefont {Neil}}\ and\ \bibinfo {author} {\bibfnamefont {J.~W.}\ \bibnamefont {Sitison}},\ }\Eprint {https://arxiv.org/abs/2208.14983} {arXiv:2208.14983 [stat.ME]}  (\bibinfo {year} {2022})\BibitemShut {NoStop}%
\bibitem [{\citenamefont {Neil}\ and\ \citenamefont {Sitison}(2023)}]{Neil:2023pgt}%
  \BibitemOpen
  \bibfield  {author} {\bibinfo {author} {\bibfnamefont {E.~T.}\ \bibnamefont {Neil}}\ and\ \bibinfo {author} {\bibfnamefont {J.~W.}\ \bibnamefont {Sitison}},\ }\href {https://doi.org/10.1103/PhysRevE.108.045308} {\bibfield  {journal} {\bibinfo  {journal} {Phys. Rev. E}\ }\textbf {\bibinfo {volume} {108}},\ \bibinfo {pages} {045308} (\bibinfo {year} {2023})},\ \Eprint {https://arxiv.org/abs/2305.19417} {arXiv:2305.19417 [stat.ME]} \BibitemShut {NoStop}%
\bibitem [{\citenamefont {Davies}\ \emph {et~al.}(2020)\citenamefont {Davies} \emph {et~al.}}]{FermilabLattice:2019ugu}%
  \BibitemOpen
  \bibfield  {author} {\bibinfo {author} {\bibfnamefont {C.~T.~H.}\ \bibnamefont {Davies}} \emph {et~al.} (\bibinfo {collaboration} {Fermilab Lattice, HPQCD, MILC}),\ }\href {https://doi.org/10.1103/PhysRevD.101.034512} {\bibfield  {journal} {\bibinfo  {journal} {Phys. Rev. D}\ }\textbf {\bibinfo {volume} {101}},\ \bibinfo {pages} {034512} (\bibinfo {year} {2020})}\BibitemShut {NoStop}%
\bibitem [{\citenamefont {Lepage}\ \emph {et~al.}(2022)\citenamefont {Lepage}, \citenamefont {Gohlke},\ and\ \citenamefont {Hackett}}]{gvarGitHub}%
  \BibitemOpen
  \bibfield  {author} {\bibinfo {author} {\bibfnamefont {G.}~\bibnamefont {Lepage}}, \bibinfo {author} {\bibfnamefont {C.}~\bibnamefont {Gohlke}},\ and\ \bibinfo {author} {\bibfnamefont {D.}~\bibnamefont {Hackett}},\ }\href@noop {} {}\bibinfo {howpublished} {\href{https://zenodo.org/record/6544356\#.Y6dfuuzMI-S}{\texttt{gplepage/gvar v11.10}}} (\bibinfo {year} {2022})\BibitemShut {NoStop}%
\bibitem [{\citenamefont {Jay}\ and\ \citenamefont {Neil}(2021)}]{Jay:2020jkz}%
  \BibitemOpen
  \bibfield  {author} {\bibinfo {author} {\bibfnamefont {W.~I.}\ \bibnamefont {Jay}}\ and\ \bibinfo {author} {\bibfnamefont {E.~T.}\ \bibnamefont {Neil}},\ }\href {https://doi.org/10.1103/PhysRevD.103.114502} {\bibfield  {journal} {\bibinfo  {journal} {Phys. Rev. D}\ }\textbf {\bibinfo {volume} {103}},\ \bibinfo {pages} {114502} (\bibinfo {year} {2021})}\BibitemShut {NoStop}%
\bibitem [{\citenamefont {Lehner}\ and\ \citenamefont {Meyer}(2020)}]{Lehner:2020crt}%
  \BibitemOpen
  \bibfield  {author} {\bibinfo {author} {\bibfnamefont {C.}~\bibnamefont {Lehner}}\ and\ \bibinfo {author} {\bibfnamefont {A.~S.}\ \bibnamefont {Meyer}},\ }\href {https://doi.org/10.1103/PhysRevD.101.074515} {\bibfield  {journal} {\bibinfo  {journal} {Phys. Rev. D}\ }\textbf {\bibinfo {volume} {101}},\ \bibinfo {pages} {074515} (\bibinfo {year} {2020})}\BibitemShut {NoStop}%
\bibitem [{\citenamefont {Aubin}\ \emph {et~al.}(2020)\citenamefont {Aubin}, \citenamefont {Blum}, \citenamefont {Tu}, \citenamefont {Golterman}, \citenamefont {Jung},\ and\ \citenamefont {Peris}}]{Aubin:2019usy}%
  \BibitemOpen
  \bibfield  {author} {\bibinfo {author} {\bibfnamefont {C.}~\bibnamefont {Aubin}}, \bibinfo {author} {\bibfnamefont {T.}~\bibnamefont {Blum}}, \bibinfo {author} {\bibfnamefont {C.}~\bibnamefont {Tu}}, \bibinfo {author} {\bibfnamefont {M.}~\bibnamefont {Golterman}}, \bibinfo {author} {\bibfnamefont {C.}~\bibnamefont {Jung}},\ and\ \bibinfo {author} {\bibfnamefont {S.}~\bibnamefont {Peris}},\ }\href {https://doi.org/10.1103/PhysRevD.101.014503} {\bibfield  {journal} {\bibinfo  {journal} {Phys. Rev. D}\ }\textbf {\bibinfo {volume} {101}},\ \bibinfo {pages} {014503} (\bibinfo {year} {2020})},\ \Eprint {https://arxiv.org/abs/1905.09307} {arXiv:1905.09307 [hep-lat]} \BibitemShut {NoStop}%
\bibitem [{\citenamefont {Giusti}\ \emph {et~al.}(2018)\citenamefont {Giusti}, \citenamefont {Sanfilippo},\ and\ \citenamefont {Simula}}]{Giusti:2018mdh}%
  \BibitemOpen
  \bibfield  {author} {\bibinfo {author} {\bibfnamefont {D.}~\bibnamefont {Giusti}}, \bibinfo {author} {\bibfnamefont {F.}~\bibnamefont {Sanfilippo}},\ and\ \bibinfo {author} {\bibfnamefont {S.}~\bibnamefont {Simula}},\ }\href {https://doi.org/10.1103/PhysRevD.98.114504} {\bibfield  {journal} {\bibinfo  {journal} {Phys. Rev. D}\ }\textbf {\bibinfo {volume} {98}},\ \bibinfo {pages} {114504} (\bibinfo {year} {2018})},\ \Eprint {https://arxiv.org/abs/1808.00887} {arXiv:1808.00887 [hep-lat]} \BibitemShut {NoStop}%
\bibitem [{\citenamefont {G\'erardin}\ \emph {et~al.}(2019)\citenamefont {G\'erardin}, \citenamefont {C\`e}, \citenamefont {von Hippel}, \citenamefont {H\"orz}, \citenamefont {Meyer}, \citenamefont {Mohler}, \citenamefont {Ottnad}, \citenamefont {Wilhelm},\ and\ \citenamefont {Wittig}}]{Gerardin:2019rua}%
  \BibitemOpen
  \bibfield  {author} {\bibinfo {author} {\bibfnamefont {A.}~\bibnamefont {G\'erardin}}, \bibinfo {author} {\bibfnamefont {M.}~\bibnamefont {C\`e}}, \bibinfo {author} {\bibfnamefont {G.}~\bibnamefont {von Hippel}}, \bibinfo {author} {\bibfnamefont {B.}~\bibnamefont {H\"orz}}, \bibinfo {author} {\bibfnamefont {H.~B.}\ \bibnamefont {Meyer}}, \bibinfo {author} {\bibfnamefont {D.}~\bibnamefont {Mohler}}, \bibinfo {author} {\bibfnamefont {K.}~\bibnamefont {Ottnad}}, \bibinfo {author} {\bibfnamefont {J.}~\bibnamefont {Wilhelm}},\ and\ \bibinfo {author} {\bibfnamefont {H.}~\bibnamefont {Wittig}},\ }\href {https://doi.org/10.1103/PhysRevD.100.014510} {\bibfield  {journal} {\bibinfo  {journal} {Phys. Rev. D}\ }\textbf {\bibinfo {volume} {100}},\ \bibinfo {pages} {014510} (\bibinfo {year} {2019})},\ \Eprint {https://arxiv.org/abs/1904.03120} {arXiv:1904.03120 [hep-lat]} \BibitemShut {NoStop}%
\bibitem [{\citenamefont {Shintani}\ and\ \citenamefont {Kuramashi}(2019)}]{Shintani:2019wai}%
  \BibitemOpen
  \bibfield  {author} {\bibinfo {author} {\bibfnamefont {E.}~\bibnamefont {Shintani}}\ and\ \bibinfo {author} {\bibfnamefont {Y.}~\bibnamefont {Kuramashi}} (\bibinfo {collaboration} {PACS}),\ }\href {https://doi.org/10.1103/PhysRevD.100.034517} {\bibfield  {journal} {\bibinfo  {journal} {Phys. Rev. D}\ }\textbf {\bibinfo {volume} {100}},\ \bibinfo {pages} {034517} (\bibinfo {year} {2019})},\ \Eprint {https://arxiv.org/abs/1902.00885} {arXiv:1902.00885 [hep-lat]} \BibitemShut {NoStop}%
\bibitem [{\citenamefont {DeTar}\ \emph {et~al.}(2019)\citenamefont {DeTar} \emph {et~al.}}]{FermilabLattice:2019dbx}%
  \BibitemOpen
  \bibfield  {author} {\bibinfo {author} {\bibfnamefont {C.~E.}\ \bibnamefont {DeTar}} \emph {et~al.} (\bibinfo {collaboration} {Fermilab Lattice, HPQCD, MILC}),\ }\href {https://doi.org/10.22323/1.363.0070} {\bibfield  {journal} {\bibinfo  {journal} {PoS}\ }\textbf {\bibinfo {volume} {LATTICE2019}},\ \bibinfo {pages} {070} (\bibinfo {year} {2019})}\BibitemShut {NoStop}%
\bibitem [{\citenamefont {Wang}\ \emph {et~al.}(2022)\citenamefont {Wang}, \citenamefont {Draper}, \citenamefont {Liu},\ and\ \citenamefont {Yang}}]{Wang:2022lkq}%
  \BibitemOpen
  \bibfield  {author} {\bibinfo {author} {\bibfnamefont {G.}~\bibnamefont {Wang}}, \bibinfo {author} {\bibfnamefont {T.}~\bibnamefont {Draper}}, \bibinfo {author} {\bibfnamefont {K.-F.}\ \bibnamefont {Liu}},\ and\ \bibinfo {author} {\bibfnamefont {Y.-B.}\ \bibnamefont {Yang}} (\bibinfo {collaboration} {$\chi$QCD}),\ }\Eprint {https://arxiv.org/abs/2204.01280} {arXiv:2204.01280 [hep-lat]}  (\bibinfo {year} {2022})\BibitemShut {NoStop}%
\bibitem [{\citenamefont {Bors{\'a}nyi}\ \emph {et~al.}(2012)\citenamefont {Bors{\'a}nyi}, \citenamefont {D{\"u}rr}, \citenamefont {Fodor}, \citenamefont {Hoelbling}, \citenamefont {Katz}, \citenamefont {Krieg}, \citenamefont {Kurth}, \citenamefont {Lellouch}, \citenamefont {Lippert},\ and\ \citenamefont {McNeile}}]{BMW:2012hcm}%
  \BibitemOpen
  \bibfield  {author} {\bibinfo {author} {\bibfnamefont {S.}~\bibnamefont {Bors{\'a}nyi}}, \bibinfo {author} {\bibfnamefont {S.}~\bibnamefont {D{\"u}rr}}, \bibinfo {author} {\bibfnamefont {Z.}~\bibnamefont {Fodor}}, \bibinfo {author} {\bibfnamefont {C.}~\bibnamefont {Hoelbling}}, \bibinfo {author} {\bibfnamefont {S.~D.}\ \bibnamefont {Katz}}, \bibinfo {author} {\bibfnamefont {S.}~\bibnamefont {Krieg}}, \bibinfo {author} {\bibfnamefont {T.}~\bibnamefont {Kurth}}, \bibinfo {author} {\bibfnamefont {L.}~\bibnamefont {Lellouch}}, \bibinfo {author} {\bibfnamefont {T.}~\bibnamefont {Lippert}},\ and\ \bibinfo {author} {\bibfnamefont {C.}~\bibnamefont {McNeile}} (\bibinfo {collaboration} {BMW}),\ }\href {https://doi.org/10.1007/JHEP09(2012)010} {\bibfield  {journal} {\bibinfo  {journal} {JHEP}\ }\textbf {\bibinfo {volume} {09}},\ \bibinfo {pages} {010}},\ \Eprint {https://arxiv.org/abs/1203.4469} {arXiv:1203.4469 [hep-lat]} \BibitemShut {NoStop}%
\bibitem [{\citenamefont {Dowdall}\ \emph {et~al.}(2013)\citenamefont {Dowdall}, \citenamefont {Davies}, \citenamefont {Lepage},\ and\ \citenamefont {McNeile}}]{Dowdall:2013rya}%
  \BibitemOpen
  \bibfield  {author} {\bibinfo {author} {\bibfnamefont {R.~J.}\ \bibnamefont {Dowdall}}, \bibinfo {author} {\bibfnamefont {C.~T.~H.}\ \bibnamefont {Davies}}, \bibinfo {author} {\bibfnamefont {G.~P.}\ \bibnamefont {Lepage}},\ and\ \bibinfo {author} {\bibfnamefont {C.}~\bibnamefont {McNeile}} (\bibinfo {collaboration} {HPQCD}),\ }\href {https://doi.org/10.1103/PhysRevD.88.074504} {\bibfield  {journal} {\bibinfo  {journal} {Phys. Rev. D}\ }\textbf {\bibinfo {volume} {88}},\ \bibinfo {pages} {074504} (\bibinfo {year} {2013})}\BibitemShut {NoStop}%
\bibitem [{\citenamefont {Chakraborty}\ \emph {et~al.}(2017{\natexlab{b}})\citenamefont {Chakraborty}, \citenamefont {Davies}, \citenamefont {Donald}, \citenamefont {Koponen},\ and\ \citenamefont {Lepage}}]{Chakraborty:2017hry}%
  \BibitemOpen
  \bibfield  {author} {\bibinfo {author} {\bibfnamefont {B.}~\bibnamefont {Chakraborty}}, \bibinfo {author} {\bibfnamefont {C.~T.~H.}\ \bibnamefont {Davies}}, \bibinfo {author} {\bibfnamefont {G.~C.}\ \bibnamefont {Donald}}, \bibinfo {author} {\bibfnamefont {J.}~\bibnamefont {Koponen}},\ and\ \bibinfo {author} {\bibfnamefont {G.~P.}\ \bibnamefont {Lepage}} (\bibinfo {collaboration} {HPQCD}),\ }\href {https://doi.org/10.1103/PhysRevD.96.074502} {\bibfield  {journal} {\bibinfo  {journal} {Phys. Rev. D}\ }\textbf {\bibinfo {volume} {96}},\ \bibinfo {pages} {074502} (\bibinfo {year} {2017}{\natexlab{b}})}\BibitemShut {NoStop}%
\end{thebibliography}%
